\newif\ifmnras
\mnrastrue
\ifmnras
	\documentclass[fleqn,usenatbib]{mnras}
\else
	\documentclass[apj,numberedappendix]{emulateapj}
\fi

\usepackage{natbib}
\usepackage{graphics,epsf}
\usepackage{amsmath}                
\usepackage{amsfonts}               
\usepackage{amssymb}                
\usepackage{epsfig}                 
\usepackage{graphicx}
\usepackage{adjustbox}
\usepackage{subfig}
\usepackage{booktabs}
\usepackage{xfrac}
 \usepackage{hyperref}

\usepackage{tikz}
\usepackage{pgfplots}
\usepackage{pgfplotstable}
\pgfplotsset{compat=1.16} 

\usepackage{xcolor}
\definecolor{plot1}{HTML}{003f5c}
\definecolor{plot2}{HTML}{7a5195}
\definecolor{plot3}{HTML}{ef5675}
\definecolor{plot4}{HTML}{ffa600}

\def \octo{{\sc Octo-Tiger}}

\ifmnras
	\def \aap{A\&A}
	
	\def \apj{ApJ}
	\def \apjl{ApJ}

	\def \mnras{MNRAS}

    \def \jaavso{JAAVSO}
\fi

\definecolor{redak}{rgb}{0.9,0.15,0.05}

\definecolor{gblue}{RGB}{66,133,244}
\definecolor{gyellow}{RGB}{244,160,0}
\definecolor{gred}{RGB}{219,68,55}
\definecolor{ggreen}{RGB}{15,157,88}

\definecolor{seedred}{RGB}{143,86,81}

\definecolor{isoblue}{RGB}{1,142,217}
\definecolor{isoorange}{RGB}{255,170,0}
\definecolor{isogreen}{RGB}{0,199,0}
\definecolor{isoyellow}{RGB}{239,239,53}
\definecolor{isoviolet}{RGB}{170,85,255}
\definecolor{isored}{RGB}{200,0,0}

\newcommand{\coordDotRight}[6]{%
    \coordinate (origin) at (#1,#2);%
    \coordinate (offsetX) at (#1+#3,#2);%
    \coordinate (offsetY) at (#1,#2+#3);%
    \coordinate (offsetZ) at (#1-0.5*#3,#2-0.5*#3);%
    \coordinate (targetX) at (#1+4*#3,#2);%
    \coordinate (targetY) at (#1,#2+4*#3);%
    \draw[->,gblue] (offsetX) -- (targetX);%
    \node[anchor=west] at (targetX) {#4};%
    \draw[->,gblue] (offsetY) -- (targetY);%
    \node[anchor=south] at (targetY) {#5};%
    \draw[black] (origin) circle (#3);%
    \draw[fill,gblue] (origin) circle (0.25*#3);%
    \node[anchor=north east] at (offsetZ) {#6};
}

\newcommand{\coordDotLeft}[6]{%
    \coordinate (origin) at (#1,#2);%
    \coordinate (offsetX) at (#1-#3,#2);%
    \coordinate (offsetY) at (#1,#2+#3);%
    \coordinate (offsetZ) at (#1+0.5*#3,#2-0.5*#3);%
    \coordinate (targetX) at (#1-4*#3,#2);%
    \coordinate (targetY) at (#1,#2+4*#3);%
    \draw[->,gblue] (offsetX) -- (targetX);%
    \node[anchor=east] at (targetX) {#4};%
    \draw[->,gblue] (offsetY) -- (targetY);%
    \node[anchor=south] at (targetY) {#5};%
    \draw[black] (origin) circle (#3);%
    \draw[fill,gblue] (origin) circle (0.25*#3);%
    \node[anchor=north west] at (offsetZ) {#6};
}

\newcommand{\coordCrossRight}[6]{%
    \coordinate (origin) at (#1,#2);%
    \coordinate (circleTL) at (#1-0.707*#3,#2+0.707*#3);%
    \coordinate (circleBL) at (#1-0.707*#3,#2-0.707*#3);%
    \coordinate (circleTR) at (#1+0.707*#3,#2+0.707*#3);%
    \coordinate (circleBR) at (#1+0.707*#3,#2-0.707*#3);%
    \coordinate (offsetX) at (#1+#3,#2);%
    \coordinate (offsetY) at (#1,#2+#3);%
    \coordinate (offsetZ) at (#1-0.5*#3,#2-0.5*#3);%
    \coordinate (targetX) at (#1+4*#3,#2);%
    \coordinate (targetY) at (#1,#2+4*#3);%
    \draw[->,gblue] (offsetX) -- (targetX);%
    \node[anchor=west] at (targetX) {#4};%
    \draw[->,gblue] (offsetY) -- (targetY);%
    \node[anchor=south] at (targetY) {#5};%
    \draw[black] (origin) circle (#3);%
    \draw[ultra thick,gblue] (circleTR) -- (circleBL);%
    \draw[ultra thick,gblue] (circleBR) -- (circleTL);%
    \node[anchor=north east] at (offsetZ) {#6};
}

\newcommand{\colbox}[1]{\protect\tikz{\protect\draw[#1,fill=#1] (0,0) rectangle (0.2,0.2);}}

\renewcommand{\vec}[1]{\mathbf{#1}}

\newcommand{\accretor}{\mathcal{A}}
\newcommand{\donor}{\mathcal{D}}
\newcommand{\other}{\mathcal{O}}

\newcommand{\seconds}{~\text{s}}
\newcommand{\minutes}{~\text{min}}
\newcommand{\gpccm}{~\text{g/cm}^3}
\newcommand{\kelvin}{~\text{K}}
\newcommand{\cmps}{~\text{cm/s}}
\newcommand{\kmps}{~\text{km/s}}

\newif\ifoldimages\oldimagesfalse

\ifmnras
	\title[OctoVis]{Visualizing the mass transfer flow in direct-impact accretion}

	\author[Frank et al.]{Juhan Frank$^{1}$\thanks{E-mail: 
	\href{jfrank@lsu.edu}{jfrank@lsu.edu}},
        Alexander Straub$^{2}$, Sagiv Shiber$^{1,3}$, Parsa Amini$^{4}$, Dominic C. Marcello$^{1,5}$, 
        \newauthor
        Patrick Diehl$^{1,5,6}$, Thomas Ertl$^{2}$, Filip Sadlo$^{7}$, Steffen Frey$^{8}$\\
\\
$^{1}$Department of Physics and Astronomy, Louisiana State University, Baton Rouge, LA 70803, USA \\
$^{2}$Visualization Research Center, University of Stuttgart, Allmandring 19, 70569 Stuttgart, Germany \\
$^{3}$Department of Physics, Florida State University, 77 Chieftan Way, Tallahassee, FL 32306, USA \\
$^{4}$AMD, Santa Clara, CA, USA \\
$^{5}$LSU Center for Computation \& Technology, Louisiana State University, Baton Rouge, LA 70803, USA \\
$^{6}$Applied Computer Science (CCS-7), Los Alamos National Laboratory, Los Alamos, NM 87545, USA \\
$^{7}$Visual Computing Group, IWR, Heidelberg University, Im Neuenheimer Feld 205, 69120 Heidelberg, Germany \\
$^{8}$Bernoulli Institute, University of Groningen, Nijenborgh 9, 9747 AG Groningen, Netherlands
}

\fi

\begin{document}
\label{firstpage}

\pagerange{\pageref{firstpage}--\pageref{lastpage}}
\maketitle

\begin{abstract}
    We use a variety of visualization techniques to display the interior and surface flows in a double white dwarf binary undergoing direct-impact mass transfer and evolving dynamically to a merger. The structure of the flow can be interpreted in terms of standard dynamical, cyclostrophic and geostrophic arguments. We describe and showcase some visualization and analysis techniques of potential interest for astrophysical hydrodynamics. In the context of R Coronae Borealis stars, we find that mixing of accretor material with donor material at the shear layer between the fast accretion belt and the slower rotating accretor body will always result in some dredge-up. We also discuss briefly some potential applications to other
    types of binaries.
\end{abstract}


\section{Introduction}
\label{sec:intro}
In certain subsets of close semi-detached binaries, \emph{e.g.}\ short-period double white dwarf (DWD) binaries, AM Canum Venaticorum (AM CVn) binaries and other low-mass compact binaries, the mass transfer stream may impact the accretor surface in what is known as direct-impact accretion. In the shortest orbital period AM CVn stars direct-impact accretion is secularly stable. However, direct-impact accretion can occur in a transient phase, during the formation of contact binaries or the merger of various types of stars, \emph{e.g.}\ blue stragglers, white dwarf mergers leading to the formation of R Coronae Borealis (RCB) stars \citep{Clayton2012, Shiber2024}, and some progenitors of Type Ia Supernovae (SNe) \citep{Tutukov1981, Iben1997, Dan2015}. Depending on the binary mass ratio, the rate of mass transfer and the ability of the accretor to absorb the transferred mass, an accretion disk may form or, if the transferred matter fills or overfills the Roche lobe of the accretor, a contact binary may result, or the two components may merge into a single object. The purpose of this study is to understand the internal and surface flows induced by the mass transfer during the merger of a DWD, based on numerical simulations of this dynamical phenomenon. In particular, we investigate the case of direct impact in the case of a DWD binary that ultimately merges in a few orbital periods.
With the insights gained in this study, we discuss the applicability to other types of binaries and investigate potential observational consequences.

The AM Canum Venaticorum (AM CVn) binaries are a small sub-class of hydrogen-deficient cataclysmic variables of extremely short orbital periods ranging from 5 minutes to 65 minutes. The donor star is a degenerate or partially degenerate He WD transferring mass by Roche-lobe overflow to a more compact and more massive WD. See \citet{Solheim2010, Ramsay2018} for reviews of AM CVn systems. At least two AM CVn have ultra-short orbital periods below 10 minutes, in which no accretion disk forms, and the mass transfer stream impacts the surface of the accretor: HM Cnc \citep{Roelofs2010} and V407 Vul \citep{Marsh2002}. The recent discovery of three ultra-compact binaries, two with orbital periods below 10 minutes, {\em all} of which are disk accretors, has significantly increased the interest in the origin and evolution of these systems \citep{Chakrabortyetal2024}.

Despite their small number, AM CVn binaries are important for being candidates for progenitors of Type Ia SNe \citep{Shen2009}, and are predicted to be strong sources of milli-Hz gravitational wave emission for the Laser Interferometer Space Antenna
(LISA) or similar space-borne observatories \citep{Roelofs2006, Nelemans2009, Nissanke2012, Amaro-Seoane2023,Kupfer2023}. Finally, because of the potential role in various formation scenarios, and its known role in HM Cnc and V407 Vul, direct impact accretion has been the subject of theoretical studies using analytic approaches \citep{Sepinsky2014, Kramarev2023}. Our goal here is to provide some complementary insights gleaned from our numerical simulations of the mergers of white dwarfs \citep{Marcello2021, Shiber2024}.

In the case of hydrogen-rich cataclysmic variables (CVs; \cite{Warner2003, WarnerWoudt2005}), a white dwarf (WD) accretes matter from a low-mass, near main-sequence star or a giant companion. The angular momentum of the mass transfer stream is high enough to avoid direct impact allowing for the development of an accretion disk. The hydrogen-rich material from the donor in-spirals through the disk and lands on the surface of the white dwarf. The problem of how this material circulating around the WD at near Keplerian speeds 
lands and spreads over the WD surface has been studied in the literature \citep{Kippenhahn1978, Piro2004, PiroDNO}. The concept of {\em accretion belt} was introduced in \citet{Kippenhahn1978} for cataclysmic variables as an equatorial region confined to low latitudes, with chemical composition and velocity distinct from the accretor body.

The plan for the rest of the paper is as follows: in~\autoref{sec:binary} we introduce the two binary simulations that are the subject of this paper and their previous history; in~\autoref{sec:vis} we present the visualization methods to be employed; with these, we then discuss the flow in the surface layers in~\autoref{sec:surfflow}, and the structure of internal flows in~\autoref{sec:internal}; we analyze the equatorial flow in terms of Lagrangian Coherent Structures~(LCS) extracted using Finite-Time Liapunov Exponents~(FTLE) in~\autoref{sec:analysis}; and summarize our results and conclusions in~\autoref{sec:discussion}.

\section{The binary simulation}
\label{sec:binary}

Our goal in this paper is to visualize the surface and internal flows in a mass-transferring binary star taking as our case study the flow in a double white dwarf binary of mass ratio, donor over accretor, $q = M_2/M_1 = 0.7$, and a total mass $M=M_1+M_2=0.9 M_\odot$.
We take the full evolution datasets generated by the 3D adaptive-mesh hydrodynamic code \octo ~\citep{Marcello2021} described in \citet{Shiber2024}.
\octo\ is a high performance code that uses the C\texttt{++} standard library for parallelism and concurrency (HPX) \citep{Kaiser2020} for distributed simulations.
For portability of performance, \octo ~uses Kokkos \citep{trott2021kokkos} and HPX-Kokkos \citep{daiss2021beyond} to support various GPUs, \emph{e.g.}, NVIDIA and AMD. \octo\ was used on ORNL's Summit \citep{diehl2021octo}, Riken's Supercomputer\ Fugaku \citep{10196612,daiss2022merging}, and CSCS's Piz Daint \citep{10.1145/3295500.3356221}.
Recently, \octo\ was ported to RISC-V \citep{10.1145/3624062.3624230,diehl2024preparinghpcriscvexamining} and Intel GPUs \citep{10.1145/3585341.3585354}.

\octo\ evolves the Newtonian compressible fluid in a non-uniform Cartesian grid, where the highest resolution is assigned to regions of interest, such as the binary components and the mass transfer stream, and the lowest resolution is assigned to the low-density ``vacuum" surrounding the binary.
Initially, the vacuum density is set to 
$\rho_{\rm vac}=10^{-7} \gpccm$ everywhere; during evolution, however, the density increases to
$\rho
\approx 10^{-4} \gpccm$ in the vicinity of the binary.
As the evolution proceeds, re-griding can take place as the matter distribution changes and the number of cells in the simulation domain can substantially increase (by a factor of ten or even more). At refinement level $n+1$ the resolution in every direction is twice that of level $n$, where the refinement and de-refinement take place based on a density criterion. This makes sure that
throughout the whole evolution of the simulation, regions of physical interest, that is, both the donor and accretor stars, as well as the mass transfer stream and the accretion belt, are resolved to the maximal level.

The images and insights presented here are based on the datasets generated by two simulations of the evolution of the same DWD binary using 11 (rcb11) and 12 (rcb12) levels of refinement. 
They were generated by simulations L11 and L12, as part of a broader study of WD mergers and the origin of RCB stars. The domain size in both simulations is $\approx 4\times 10^{11}~{\rm cm}$ and, therefore, the smallest cell size with 11 and 12 levels of refinement is $4\times10^{7}~{\rm cm}$ and $2\times10^{7}~{\rm cm}$, respectively. Consequently, the binary is resolved better in rcb12, with 145 cells across the larger donor star's diameter, compared to 73 in rcb11. 
Since the ``vacuum" is not refined to high levels, the total initial number of cells in rcb12, 5.3 million, is only about twice the initial number of cells in rcb11 (see the second column of \autoref{tab:parameters}). The physical parameters of the binary system in rcb11 and rcb12 are otherwise the same. Full details of the simulations can be found in \citet{Shiber2024}.

The initial states at $t=0$ for these simulations are equilibrium states for a synchronous binary on a circular orbit generated by a self-consistent code that finds by iteration the equilibrium configurations and orbital angular velocity \citep{Marcello2021}. The donor star fills its Roche lobe, and all initial velocities vanish in the comoving frame. Mass transfer is started by driving the system for a finite time $t_{\rm dr}$ at a level of 1\% per orbit. This is achieved by artificially removing a fraction $0.01 \Delta t/P_0$ from the angular momentum of all cells, where $\Delta t$ is the time step and $P_0$ is the initial orbital period. For both rcb11 and rcb12, $t_{\rm dr}$ was chosen to be $1.3P_0$, with $P_0=114 \seconds$, and $P_0=113.6 \seconds$ respectively (see \autoref{tab:parameters}).

\begin{figure*}%
    \centering%
    \includegraphics[width=0.8\linewidth,trim=0 250pt 0 570pt,clip]{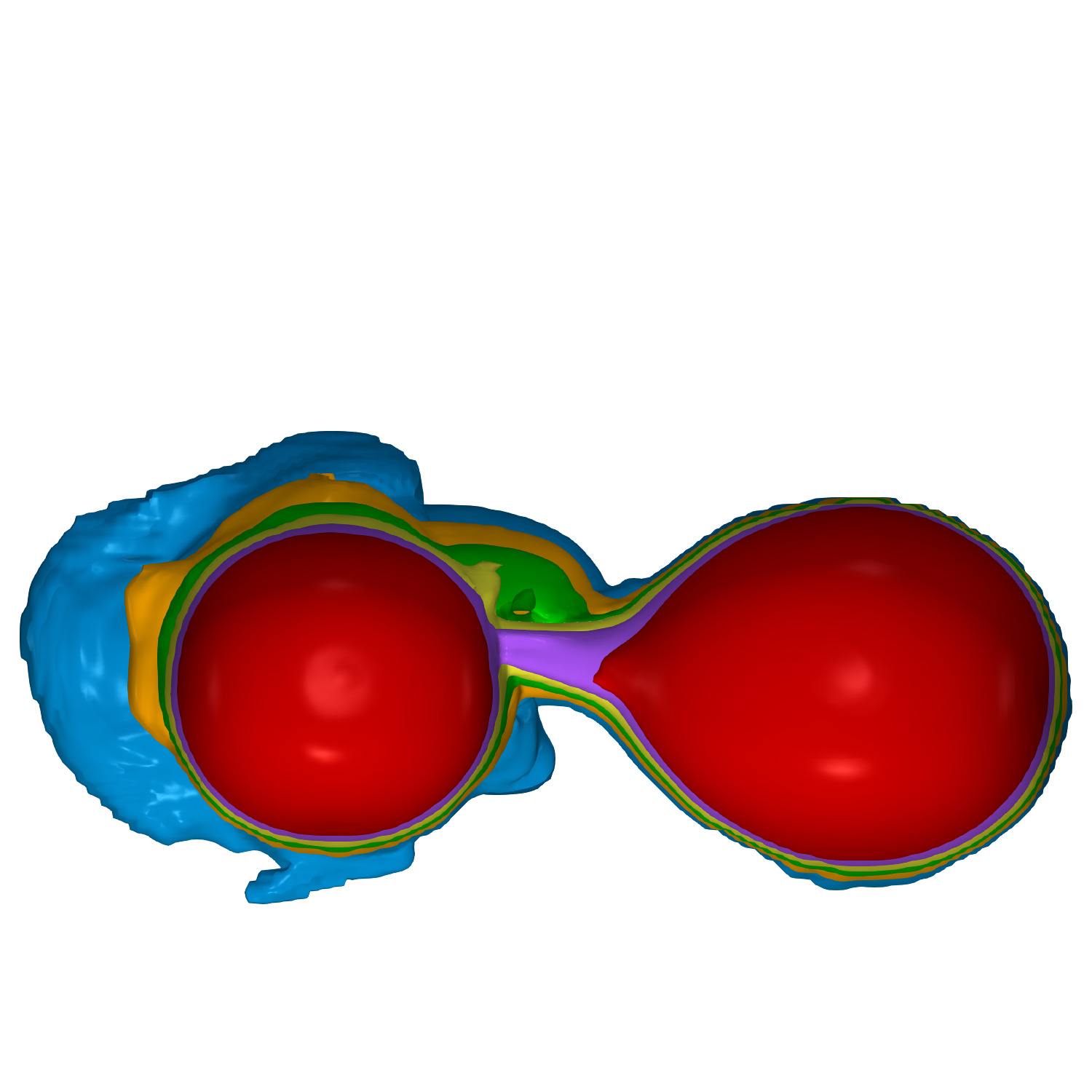}%
    \\%
    \includegraphics[width=0.45\linewidth]{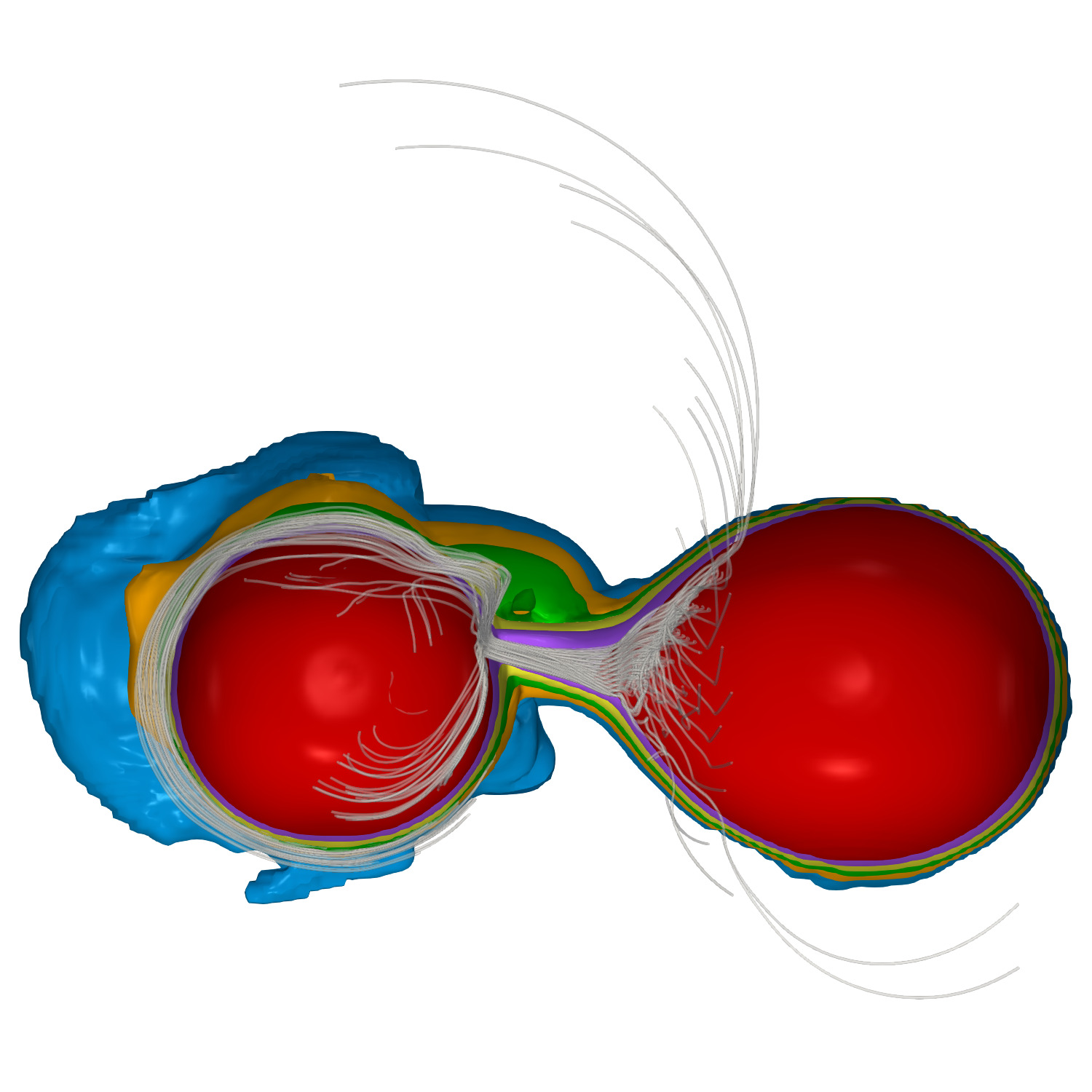}%
    \hspace{2em}%
    \includegraphics[width=0.45\linewidth]{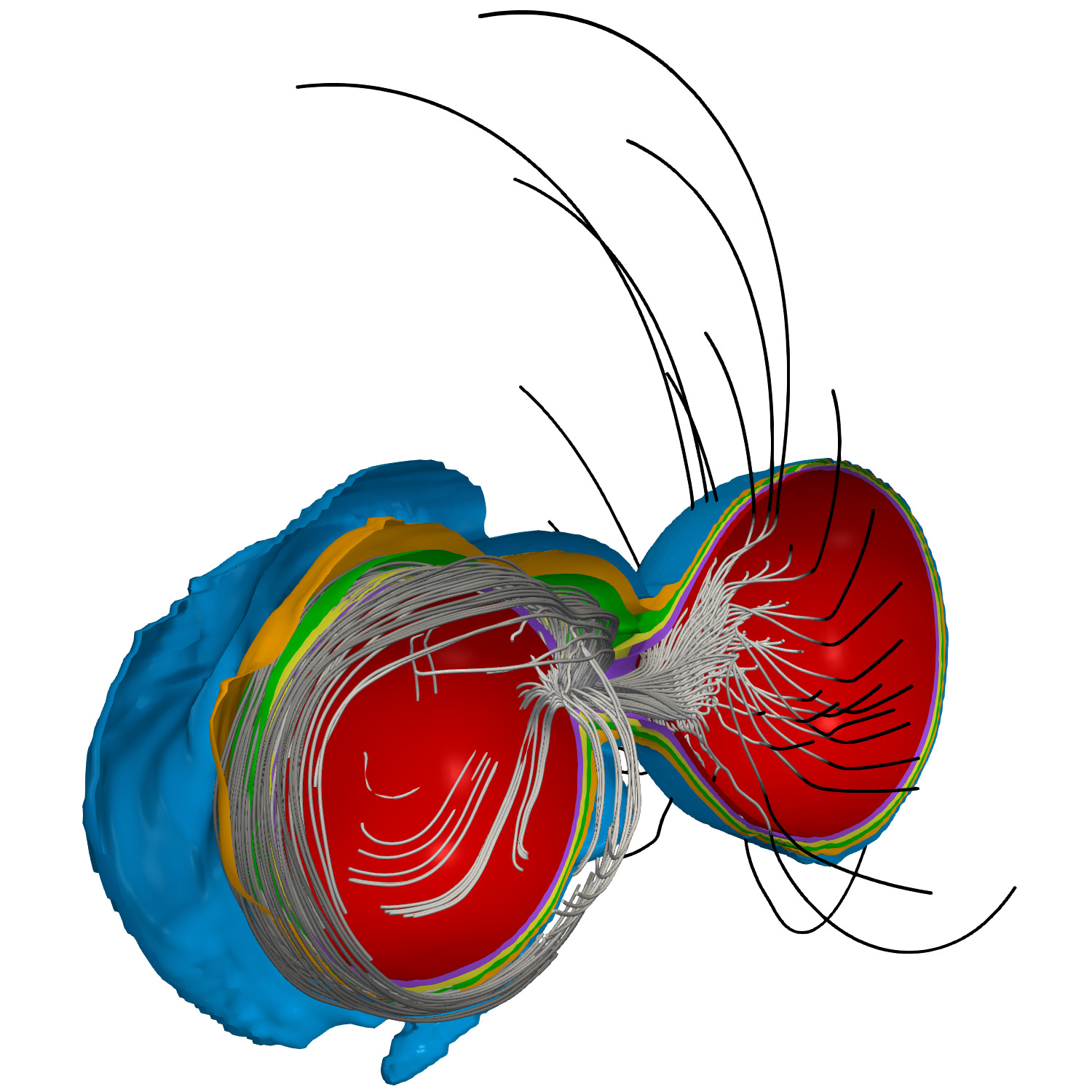}%
    \caption{%
        Surfaces at different isovalues for time step $t=86.83 \seconds$ in $\gpccm$:
        $\rho=0.1$~(\colbox{isoblue}),
        $\rho=1$~(\colbox{isoorange}),
        $\rho=5$~(\colbox{isogreen}),
        $\rho=20$~(\colbox{isoyellow}),
        $\rho=100$~(\colbox{isoviolet}),
        $\rho=500$~(\colbox{isored}) as revealed by slicing through the binary at the orbital plane and viewing it from the side of the removed material.
        The binary rotates counter-clockwise.
        The figure shows an exploded view of the internal structure of the non-degenerate layers of a double white dwarf binary during the mass transfer phase that precedes the eventual merger of the two stars into a single object.
        The star on the right (the donor) is transferring mass through a stream striking the upper layers of the more massive white dwarf on the left (the accretor).
        Several pathlines show how the stream impacts the accretor, with most of the mass flowing around the accretor while some fraction of the stream spreads in all directions around the point of impact.
        A few ``vacuum" particles are swept up in the leading side of the donor, and become part of the stream.
        Similarly a few particles in the trailing side of the donor are swept into the wake and join the surface flow on the donor.
    }%
    \label{fig:overview}%
\end{figure*}

We set the scene in the top frame of \autoref{fig:overview} by showing six nested surfaces of constant density (isopycnic surfaces) at an early stage of the simulation rcb11, when the binary has completed about 3/4 of the first orbital revolution. For the cool WDs involved in these binaries, with core temperatures in the range $10^6-10^7 \kelvin$, the gas at densities $\ge 10^2 \gpccm$ is degenerate, so the lowest 4 density values displayed correspond to the non-degenerate envelope. The lower two frames show two views of the same stratification as above together with the paths of selected tracer particles in the mass transfer stream, integrated forward and backward in time (see details below in \autoref{sec:vis-lines}).

\begin{table} 
\caption{Main parameters of interest for the simulations in this paper}
\label{tab:parameters}
     \begin{tabular}{ccllllll} 
     \toprule
     Dataset & $N_{\rm cells}^{i}$ & $P_0$ & $t_{\rm dr}$ & Length & $t_{\rm merge}$ \\
     \midrule
     rcb11 & $2.5\times10^6$ & 114.0 $\seconds$ & 1.3 $P_0$ &  37 $P_0$ & 24.5 $P_0$ \\
     rcb12 & $5.3\times10^6$ & 113.6 $\seconds$ & 1.3 $P_0$ & 44 $P_0$ & 38.7 $P_0$ \\ [0.5 ex]
     \bottomrule
    \end{tabular}
\end{table}

\section{Visualization Methods and Methodology}
\label{sec:vis}

In the following, the data processing and visualization methods used are introduced.
As a first step, the data, which was stored in an octree in the simulation, was converted to a Cartesian grid to facilitate the application of a variety of visualization methods~(\autoref{sec:vis-grid}).
In addition, the data was transformed to be in a co-rotating frame of reference~(\autoref{sec:vis-frame}) to keep the stars in place for a more intuitive and straightforward representation.
After these pre-processing steps, the visualization methods can be employed:
\begin{itemize}
\item slicing at the equatorial plane in combination with isolines for density  (\autoref{sec:vis-slice}),
\item surface LIC to analyze flow on density isosurfaces (\autoref{sec:vis-isolic}),
\item quantification of inflow and outflow orthogonal to isosurfaces (\autoref{sec:vis-inoutflow}),
\item integral lines (stream- and pathlines) to directly present 3D flow (\autoref{sec:vis-lines}),
\item and forward/backward FTLE to identify regions where flow is separating/converging (\autoref{sec:vis-ftle}).
\end{itemize}

This list roughly orders visualization approaches from simple techniques to more sophisticated methods.
At the same time, it is mostly the same order in which they were applied to gain insight into the data during the exploration phase, where domain experts cooperated closely with visualization researchers.

All stages are implemented as plugins in the visualization framework ParaView~\citep{Ayachit2015}, which is based on the Visualization Toolkit~(VTK).
With its modular approach, available filters can be combined with custom modules programmed in C\texttt{++}.
For this work, some filters in the Two-Phase Flow~(TPF) framework\footnote{\url{https://github.com/UniStuttgart-VISUS/tpf}}~(\autoref{sec:vis-frame}, \autoref{sec:vis-lines}) and general ParaView plugins\footnote{\url{https://github.com/UniStuttgart-VISUS/ParaView-Plugins}}~(\autoref{sec:vis-grid}, \autoref{sec:vis-frame}) were developed and used.
These are already publicly available under the permissive MIT license.
Other visualizations are provided directly by ParaView~(\autoref{sec:vis-slice}, \autoref{sec:vis-isolic}), or by the Visual Computing Group at Heidelberg University\footnote{\url{https://vcg.iwr.uni-heidelberg.de/plugins}}~(\autoref{sec:vis-ftle}).

\subsection{Sampling to a Cartesian grid}
\label{sec:vis-grid}

The data provided by the simulation is stored in an octree data structure, in which data points reside at the cell centers.
This tree structure allows to save space in outer regions, and only provides high resolution in areas of interest and of turbulent flow behavior.
However, most flow visualization techniques are designed and optimized for regular grids.
Therefore, the octree data is resampled to a Cartesian grid before applying the visualizations.
As the depth of the octree is known, and thus the cell size of the highest resolution, a Cartesian grid with this cell size can simply be created.
Further, we limit the grid to the region of interest to save space, where all regions of interest are resolved at the highest resolution.
Hence, nearest neighbor interpolation can be employed for resampling, aligning the grid's nodes with the cell centers of the octree cells.
This guarantees exact matches for the highest resolution cells, and only negligible approximation errors in outside regions.

\subsection{Co-rotating frame of reference}
\label{sec:vis-frame}

The original data is already in a co-rotating grid.
However, this grid rotates steadily at the initial angular frequency of the stars around the common center of mass.
As during the simulation the orbital rotation is accelerating, they begin to move in this frame.
To keep the stars static at their respective positions, the correct angular velocity has to be computed to provide a dynamic co-rotating frame of reference.
\cite{Marcello2021}~already provide such an algorithm for the diagnostics of the simulation.
Here, as a first step, the stars have to be identified.
This is done by simply comparing the densities of the accretor~$\accretor$ and the donor~$\donor$, which are quantities provided by the simulation.
In total, there are four density fields: the actual density $\rho$ of the respective cell, the density $\rho_\accretor$ of the accretor material and $\rho_\donor$ of the donor material, and a density $\rho_\other$ that belongs to neither.
Thus, the initial classification for a cell $i$ is as follows:
\begin{equation}
    \begin{split}
        i \in \accretor, &~~\text{if}~~ \rho_\accretor > \rho_\donor ~\land~ \rho_\accretor > \rho_\other \\
        i \in \donor, &~~\text{if}~~ \rho_\donor > \rho_\accretor ~\land~ \rho_\donor > \rho_\other \\
        i \in \other, &~~\text{else}.
    \end{split}
\end{equation}

Afterwards, the center of mass and the average velocity of both stars can be calculated by accumulating over all cells for the accretor $\accretor$ (analogously for the donor $\donor$):
\begin{align}
    \vec{r}_\accretor
        &= \int_\accretor \rho_\accretor ~\vec{r} ~\text{d}V
        \approx \frac{\sum_\accretor \rho_\accretor V ~\vec{r}}{\sum_\accretor \rho_\accretor V} \\
    \vec{u}_\accretor
        &= \int_\accretor \rho_\accretor ~\vec{u} ~\text{d}V
        \approx \frac{\sum_\accretor \rho_\accretor V ~\vec{u}}{\sum_\accretor \rho_\accretor V} \textbf{.}
\end{align}
From this, the angular frequency can be computed as
\begin{equation}
    \omega = \frac{(\vec{r}_\accretor - \vec{r}_\donor) \times (\vec{u}_\accretor - \vec{u}_\donor)}{||\vec{r}_\accretor - \vec{r}_\donor||^2} \cdot \begin{pmatrix} 0 \\ 0 \\ 1 \end{pmatrix},
\end{equation}
with the assumption that the axis of rotation is the z-axis.

As the accretor and donor material get mixed together due to mass transfer at later times, this serves only as a first approximation, and instead an iterative approach is taken to accurately classify each star and for the calculation of the angular orbital frequency.
To this end, the following steps are repeated to re-classify the cells and thus update the frequency. In \cite{shiber2024b}, this iteration scheme was tested for convergence, showing that except at times closer to the merger, the convergence level after five iterations reaches the level of $10^{-5}$. Therefore, we iterate five times.

The following quantities are calculated for the accretor (analogously for the donor):
\begin{equation}
    g_\accretor = \left( \vec{g} + \hat{\vec{r}} \omega^2 \right) \frac{\vec{r} - \vec{r}_\accretor}{||\vec{r} - \vec{r}_\accretor||},
\end{equation}
with gravitation $\vec{g}$, and $\hat{\vec{r}} = ( r_x ~ r_y ~ 0 )^\top$. 
The classification for a cell $i$ is then as follows:
\begin{equation}
    \begin{split}
        i \in \accretor, &~~\text{if}~~ \min\lbrace g_\accretor, 0\rbrace < \min\lbrace g_\donor, 0\rbrace \\
        i \in \donor, &~~\text{if}~~ \min\lbrace g_\accretor, 0\rbrace > \min\lbrace g_\donor, 0\rbrace \\
        i \in \other, &~~\text{else}.
    \end{split}
\end{equation}
Further optimization can be applied by directly discarding cells with very low density, and by classifying the inner regions of the stars using the Roche lobe radii.
Thus, as a first part, a density cut-off can be defined. 
Since the initial density profiles of the WDs decrease rapidly at their surface from values of $\gtrsim 100 \gpccm$ to values of $\lesssim 1 \gpccm$, we confidently set this cut-off to $1 \gpccm$.
For the second optimization, the Roche lobe radii of the stars have to be calculated, here for the accretor (analogously for the donor):
\begin{equation}
    R_\accretor = \alpha ~ a \frac{m_{\accretor/\donor} ^{\sfrac{2}{3}}}{\beta \left( m_{\accretor/\donor}^{\sfrac{2}{3}} + \ln \left( 1 + m_{\accretor/\donor}^{\sfrac{1}{3}} \right) \right)},
\end{equation}
with mass ratios \smash{$m_{\accretor/\donor} = q^{-1} = \frac{m_\accretor}{m_\donor}$ ($q = m_{\donor/\accretor} = \frac{m_\donor}{m_\accretor}$)}, orbital separation $a = ||\vec{r}_\accretor - \vec{r}_\donor||$, and constants $\alpha = 0.49$ and $\beta = 0.6$ \citep{Eggleton83}.
Cells within the critical region of $\frac{1}{4} R_{\accretor,\donor}$ can now be classified directly as pertaining to the respective star.
For more information, please refer to \cite{Marcello2021}.

\begin{figure}%
    \centering
    \includegraphics[width=\linewidth,trim=0 0.6cm 0 0,clip]{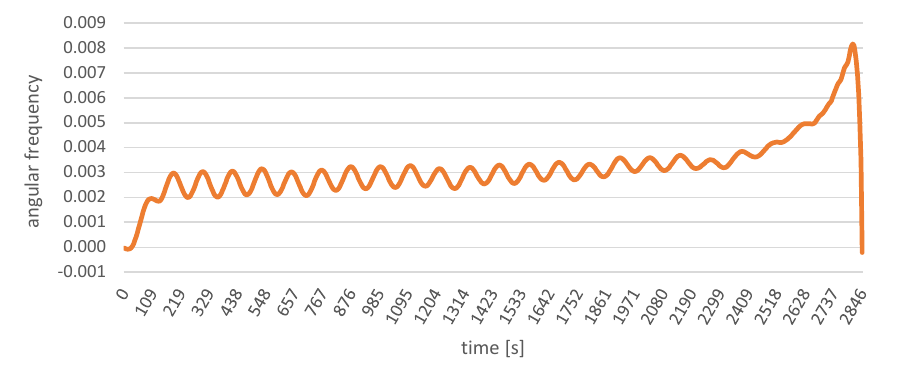}
    \caption{%
        Angular frequency for the rotation of the stars plotted over time for rcb11.
        Note the drastic change close to the merger.
    }
    \label{fig:angular-frequency}
\end{figure}

For the herein presented dataset, the resulting angular frequency is plotted over time in~\autoref{fig:angular-frequency}.
As can be observed, most of the time, the frequency grows gradually as the binary shrinks, exhibiting an epicyclic oscillation induced by the driving.
Nearing the merger, the angular frequency drastically increases before it appears to``reverse''.
However, this last behavior is clearly wrong and results from the fact that the stars have finally merged.
From this point onward, the computation of the angular frequency cannot be conducted, as it requires to identify two separate stars.
We therefore just use the last ``valid'' angular frequency from before the merger, and give up on providing a co-rotating frame of reference.

To remove the rotation of the stars at a time $t$, additional to the current rotational frequency $\omega$, the angle $\phi$, by which the stars have rotated so far, is needed.
This angle can be calculated from the frequencies as
\begin{equation}
    \phi = \int_{\tau = 0}^{t} \omega ~\text{d}\tau \approx \sum_\tau \Delta \tau ~\omega_\tau.
\end{equation}
Then, points $\vec{x}$ of the resampled grid relate to the value at
\begin{equation}
    \tilde{\vec{x}} =
        \begin{pmatrix}
            \cos \phi & -\sin \phi & 0 \\
            \sin \phi &  \cos \phi & 0 \\
            0         & 0          & 1
        \end{pmatrix}
        \vec{x}.
\end{equation}
Additionally, velocities have to be adjusted to point in the correct direction after resampling.
For this, the current angular velocity has to be deducted, and the direction has to be adjusted for the previous rotation:
\begin{equation}
    \tilde{\vec{u}} =
        \begin{pmatrix}
            \cos \phi & -\sin \phi & 0 \\
            \sin \phi &  \cos \phi & 0 \\
            0         & 0          & 1
        \end{pmatrix}^{-1}
        \left( \vec{u} - \omega \begin{pmatrix}
            -y \\ x \\ 0
        \end{pmatrix} \right)
\end{equation}

\subsection{Equatorial slices}
\label{sec:vis-slice}

One of the most interesting phenomena of the binary merger is the mass transfer from the donor to the accretor, and the development of the accretion belt.
As a simple and effective visualization to gain a first insight, the equatorial plane can be extracted.
In our case, this is the plane defined by $z = 0$, which is a 2D Cartesian grid.
In general, the slice can be used to visualize quantities, such as densities and velocity magnitudes by mapping them to color.
Further, density isolines can be extracted to reconstruct the ``surface'' of the stars for a user-defined isovalue.
From this, the mass transfer stream connecting the two stars can be identified.
In addition, the velocities can be visualized as arrow glyphs to also highlight the direction of the flow and thus reveal flow patterns.

\subsection{(Surface) Line Integral Convolution}
\label{sec:vis-isolic}

Similar to the extraction of isolines in the equatorial plane, the isosurface can be extracted directly from the densities in the three-dimensional grid.
This is done using the Marching Cubes algorithm \citep{lorensen1998marching}, which approximates the isosurface as a linear patch for each cell (line in 2D, polygon in 3D).
As the positions of the vertices of such patches are computed along the edges of cells, the resulting surface is a manifold; \emph{i.e.}, it is a closed surface.
To visualize the velocities on the surface, line integral convolution (LIC) 
\citep{Cabral1993}
can be employed to reveal static flow patterns.
The idea behind LIC is to create a dense representation of the vector field.
Intuitively, it can be described as smearing a droplet of ink along the trajectory of a massless particle.
This is repeated for different starting positions until the whole image is covered.
The trajectory for this is computed from the initial value problem
\begin{equation}
    \label{eq:streamline}%
    \frac{\partial}{\partial t} \vec{x}(t) = \vec{u}(\vec{x}(t)),
\end{equation}
for a starting position $\vec{x}(t_0) = \vec{x}_0$.
This means that the trajectory of the particle defined by the position $\vec{x}(t)$ is tangential to the velocity field.
As, in our case, LIC is applied to a surface~\citep{Weiskopf2004}, the velocities have to be projected onto it, thus creating trajectories that are confined to the surface.
In practice, a grayscale image is first initialized with randomly generated noise.
Then, along the trajectories, these values are averaged~(smeared).
This leads to a visualization that neither shows direction (but only orientation) nor velocity magnitude.
The latter can be mitigated by additionally mapping the magnitude to color, and blending the color with the intensity provided by the grayscale values.

\subsection{Inflow and outflow}
\label{sec:vis-inoutflow}

Similar to surface LIC, where the velocity at a point on the surface is projected into tangent space, we can define the in- and outflow at an isosurface by looking at the angle between the original velocity vector and the normal.
For this, we use the analytical normal as the inverse normalized gradient of the density field.
Thus, we compute the angle $\phi$ as
\begin{equation}
    \label{eq:in-outflow}
    \phi = \cos^{-1} \left( \frac{\vec{u} \cdot \nabla \rho}{||\vec{u}|| ~ ||\nabla \rho||} \right) .
\end{equation}
This results in values indicating inflow~($\phi < \frac{\pi}{2}$), tangential flow~($\phi = \frac{\pi}{2}$), and outflow~($\phi > \frac{\pi}{2}$).

\subsection{Stream- and pathlines}
\label{sec:vis-lines}

To visualize the flow directly in three-dimensional space, streamlines for static (time-independent) and pathlines for dynamic (time-dependent) visualization can be employed.
Streamlines follow the same idea of a massless particle as LIC and are generated by the same initial value problem, as stated in~\autoref{eq:streamline}.
This equation is solved using an explicit Euler approach.
Contrary to LIC, however, positions to start the streamlines have to be defined, the so-called seed.
Although there exist various seeding algorithms to only visualize a few representative streamlines, we mostly know our regions of interest.
Particularly, this is the stream and the resulting accretion belt.
But also our previous techniques, \emph{i.e.}, slicing and surface LIC, can be used to find areas of interest.

Pathlines follow the same idea, with the sole difference that the vector field changes over time.
This is reflected in the equation of the initial value problem for pathlines:
\begin{equation}
    \label{eq:pathline}%
    \frac{\partial}{\partial t} \vec{x}(t) = \vec{u}(\vec{x}(t),t),
\end{equation}
again with starting position $\vec{x}(t_0) = \vec{x}_0$.
This equation is again solved using the explicit Euler method, generally using a time step size of $\Delta t=0.1 \seconds$.
This means that, additional to tri-linear spatial interpolation within the grid, the vector fields have to be linearly interpolated temporally between adjacent time steps.
With the chosen time step size, this roughly amounts to $46$ integration steps between two input data files.

The resulting stream- and pathlines are then rendered as tubes.
In contrast to lines, this allows to use lighting for a better depth perception and to distinguish the lines in case of overdrawing.
To highlight the direction of the flow and to distinguish time, the advection step (streamline) or the time (pathline) is mapped to color.

\begin{figure*}
    \centering%
    \ifoldimages%
        \includegraphics[scale=0.66]{figures_old/rho100t100.pdf}%
    \else%
        \begin{minipage}[b]{0.9\linewidth}%
            \centering%
            \subfloat[\label{fig:4views-iso-z}]%
                {%
                    \begin{tikzpicture}%
                        \node[anchor=south west,inner sep=0] at (0,0) {%
                            \includegraphics[width=0.49\linewidth,trim=0 400pt 0 400pt,clip]%
                                {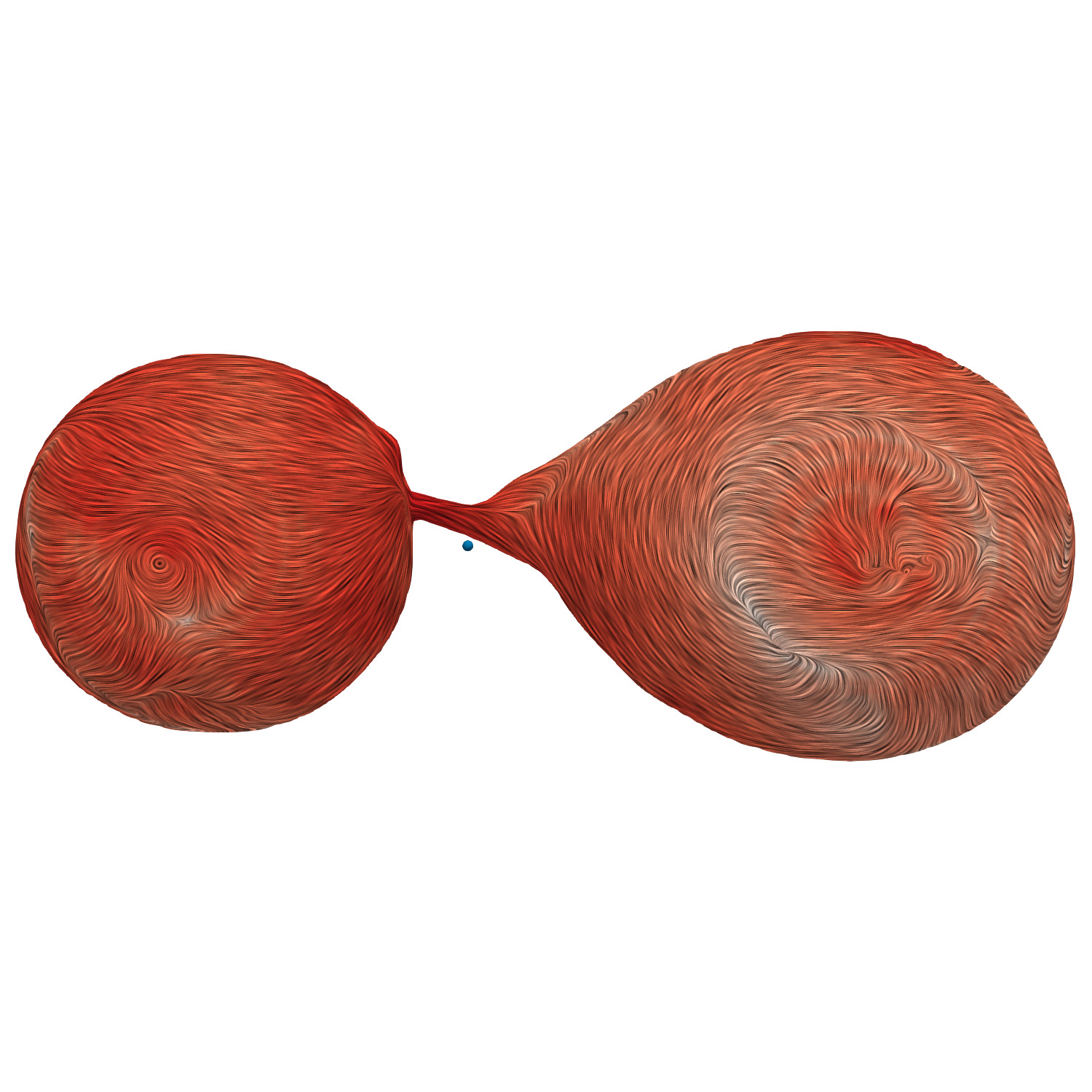}};%
                        \coordDotRight{3.3}{0.5}{0.2}{$x$}{$y$}{$z$}%
                        \draw[black,fill] (3.38,1.854) circle (0.8pt);%
                        \draw[black,line width=0.3mm] (3.38,1.854) circle (2.5pt);%
                    \end{tikzpicture}%
                }%
            \hfill%
            \subfloat[\label{fig:4views-rcb12-iso-z}]%
            {%
                \begin{tikzpicture}%
                    \node[anchor=south west,inner sep=0] at (0,0) {%
                        \includegraphics[width=0.49\linewidth,trim=0 400pt 0 400pt,clip]%
                            {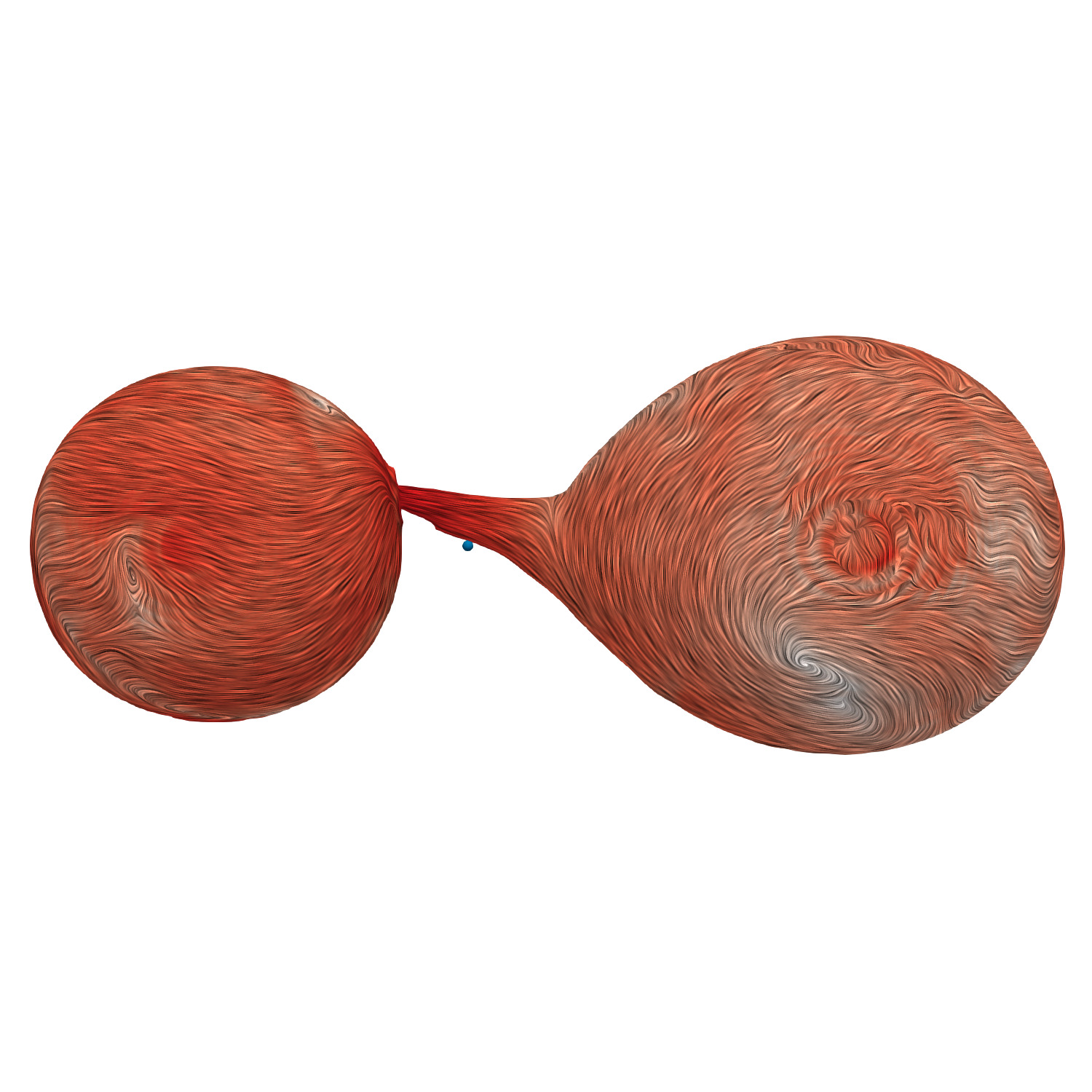}};%
                    \coordDotRight{3.3}{0.5}{0.2}{$x$}{$y$}{$z$}%
                    \draw[black,fill] (3.38,1.854) circle (0.8pt);%
                    \draw[black,line width=0.3mm] (3.38,1.854) circle (2.5pt);%
                \end{tikzpicture}%
            }%
            \\[-1.5em]%
            \subfloat[\label{fig:4views-iso+y}]%
                {%
                    \begin{tikzpicture}%
                        \node[anchor=south west,inner sep=0] at (0,0) {%
                            \includegraphics[width=0.49\linewidth,trim=0 400pt 0 400pt,clip]%
                                {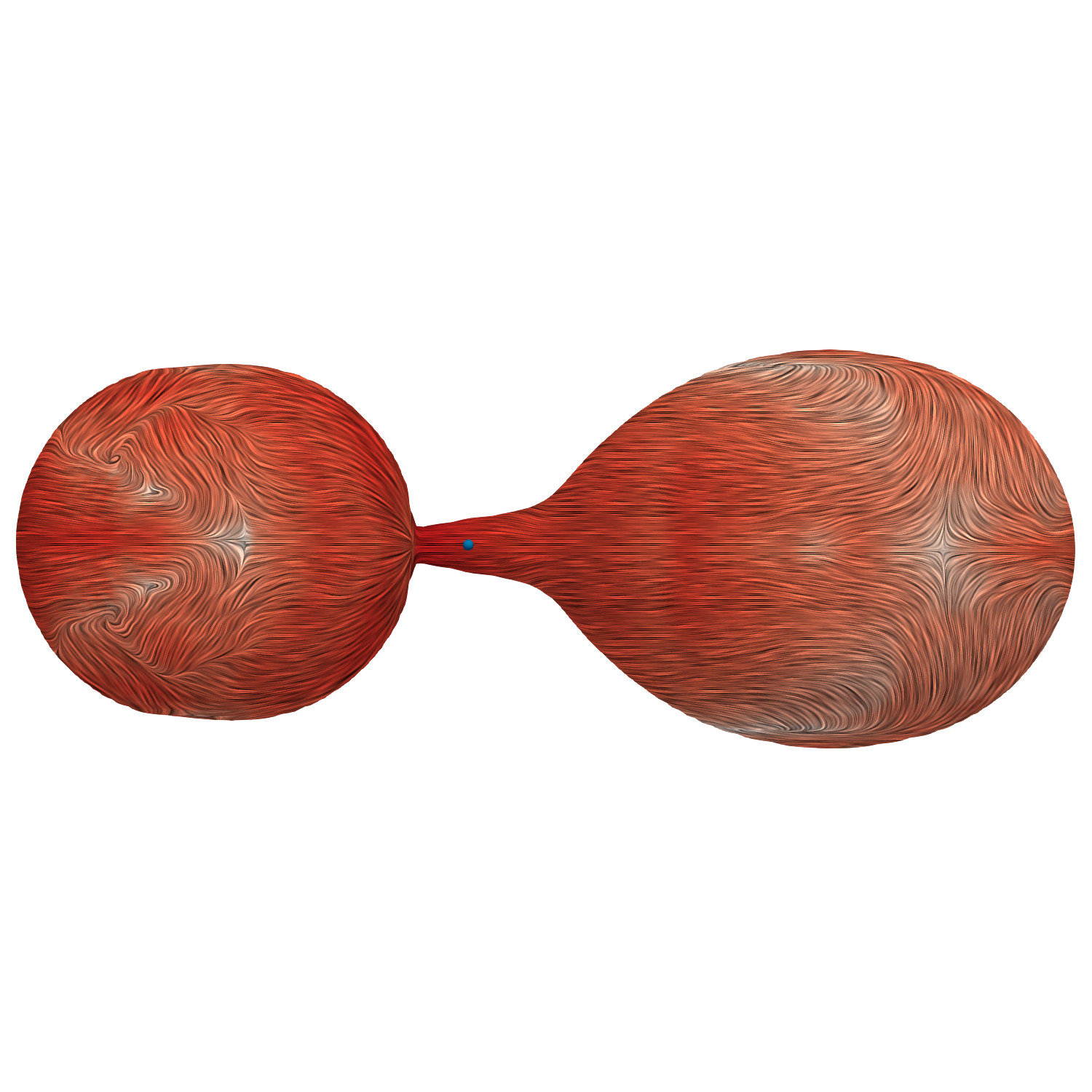}};%
                        \coordCrossRight{3.3}{0.5}{0.2}{$x$}{$z$}{$y$}%
                        \draw[white,fill] (3.381,1.86) circle (0.8pt);%
                        \draw[white,line width=0.3mm] (3.381,1.86) circle (2.5pt);%
                    \end{tikzpicture}%
                }%
            \hfill%
            \subfloat[\label{fig:4views-rcb12-iso+y}]%
            {%
                \begin{tikzpicture}%
                    \node[anchor=south west,inner sep=0] at (0,0) {%
                        \includegraphics[width=0.49\linewidth,trim=0 400pt 0 400pt,clip]%
                            {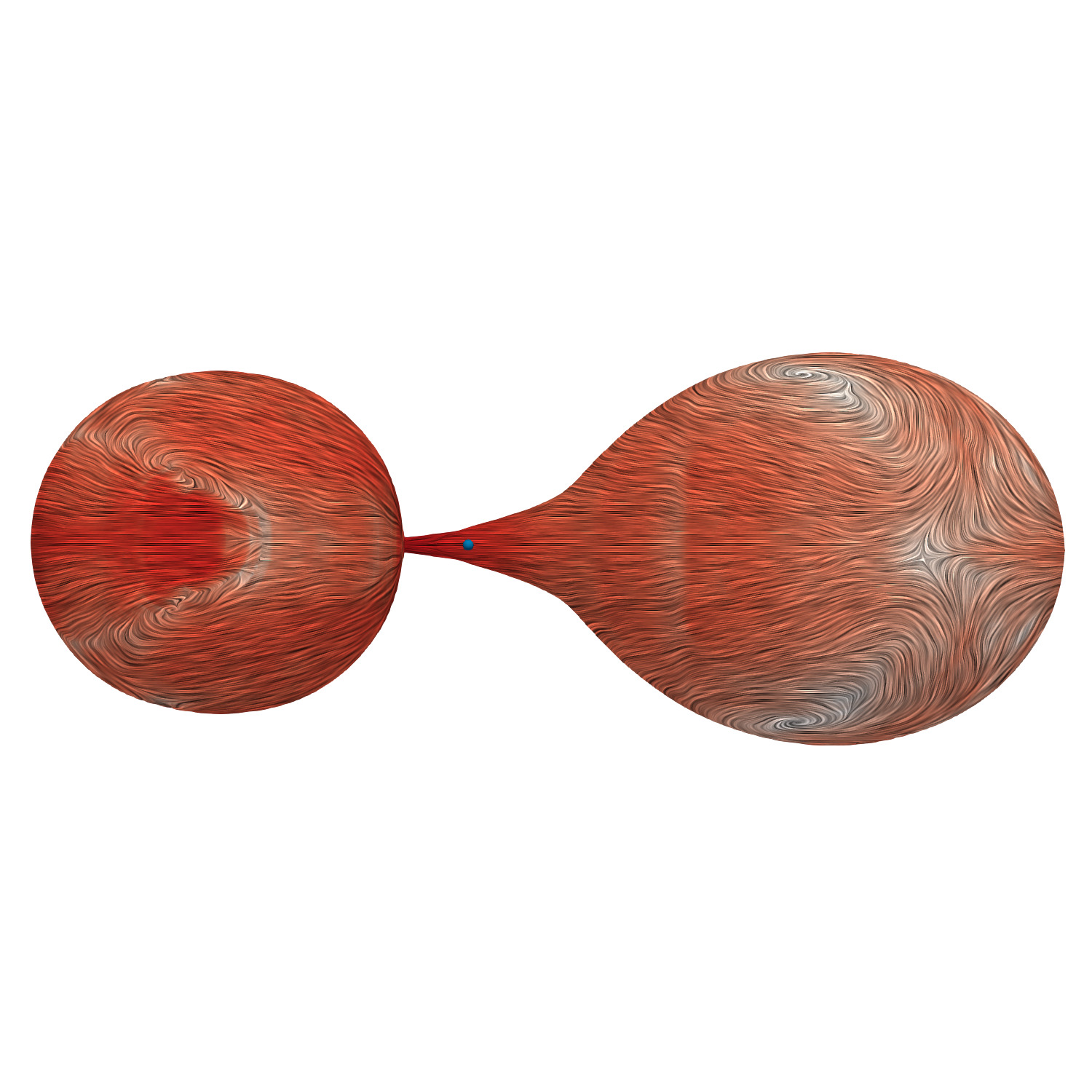}};%
                    \coordCrossRight{3.3}{0.5}{0.2}{$x$}{$z$}{$y$}%
                    \draw[white,fill] (3.381,1.86) circle (0.8pt);%
                    \draw[white,line width=0.3mm] (3.381,1.86) circle (2.5pt);%
                \end{tikzpicture}%
            }%
        \end{minipage}%
    \fi%
    \caption{%
        Two views of the binary for rcb11 at a time $t=100.44 \seconds$ (left) and rcb12 at $t=113.62 \seconds$.
        Velocity magnitudes are mapped logarithmically to color~(\includegraphics[width=6pt,height=32pt,angle=-90,origin=rB]{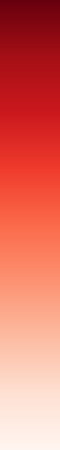}) in the range $[1 \cdot 10^6,~1 \cdot 10^9] \cmps$. The flow lines shown are the result of LIC on the isopycnic surface 
        $\rho =100 \gpccm$. In this and subsequent figures, the donor is on the right and the accretor on the left, unless noted otherwise.
        The triad of axes indicates the viewing direction: \protect\subref{fig:4views-iso-z} and \protect\subref{fig:4views-rcb12-iso-z} are viewed {\em down} the rotation axis, \protect\subref{fig:4views-iso+y} and \protect\subref{fig:4views-rcb12-iso+y} are viewed {\em along} the $y$-axis, i.e. from the trailing side of the donor.
        The common center of mass is shown as black or white dot in a circle.
    }
    \label{fig:4views}
\end{figure*}

\begin{figure*}
    \begin{minipage}[b]{0.9\linewidth}%
        \centering%
        \subfloat[\label{fig:4views-predisruption-iso-z}]%
            {%
                \begin{tikzpicture}%
                    \node[anchor=south west,inner sep=0] at (0,0) {%
                        \includegraphics[width=0.49\linewidth,trim=0 400pt 0 400pt,clip]%
                            {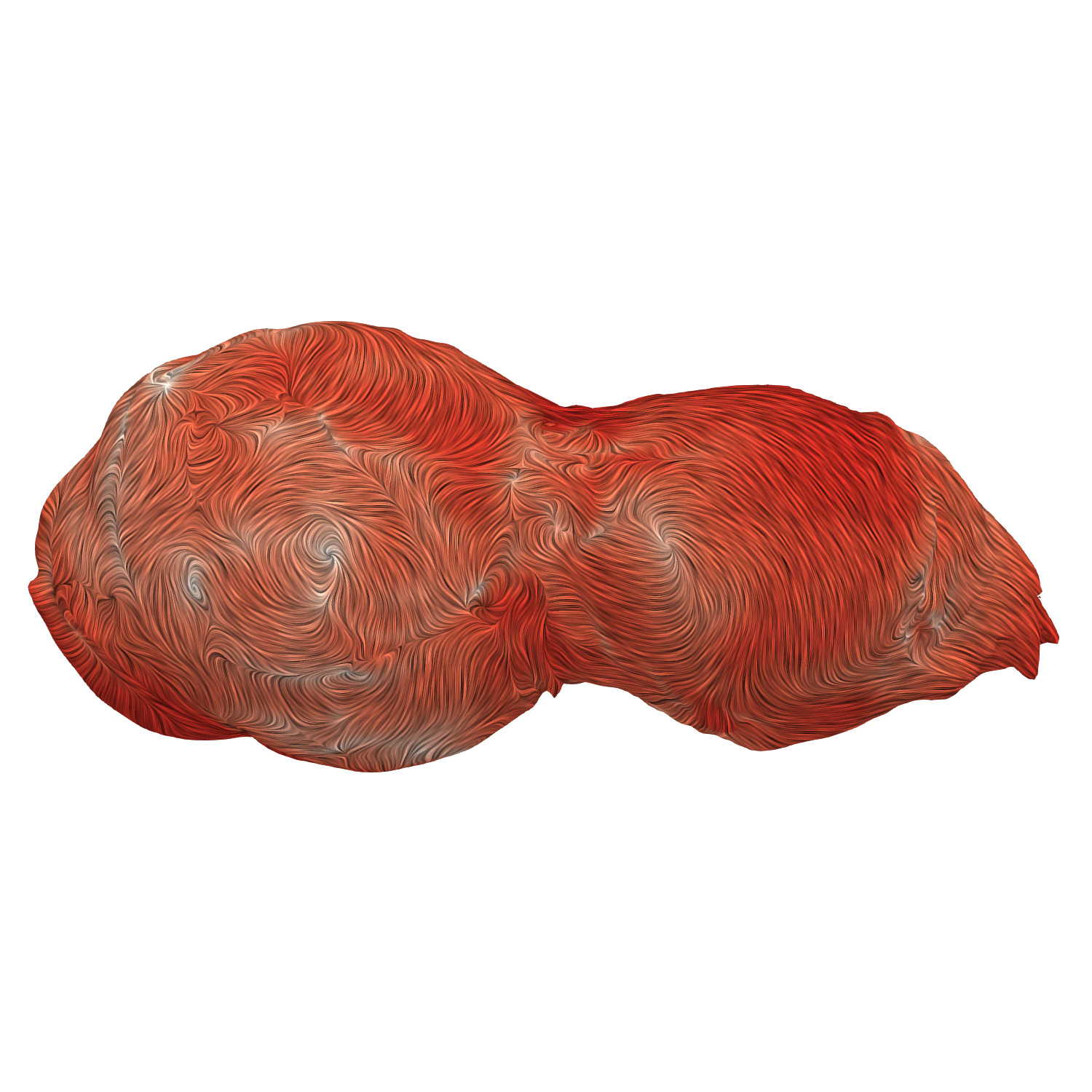}};%
                    \coordDotRight{0.5}{-0.5}{0.2}{$x$}{$y$}{$z$}%
                \end{tikzpicture}%
            }%
        \hfill%
        \subfloat[\label{fig:4views-rcb12-predisruption-iso-z}]%
            {%
                \begin{tikzpicture}%
                    \node[anchor=south west,inner sep=0] at (0,0) {%
                        \includegraphics[width=0.49\linewidth,trim=0 400pt 0 400pt,clip]%
                            {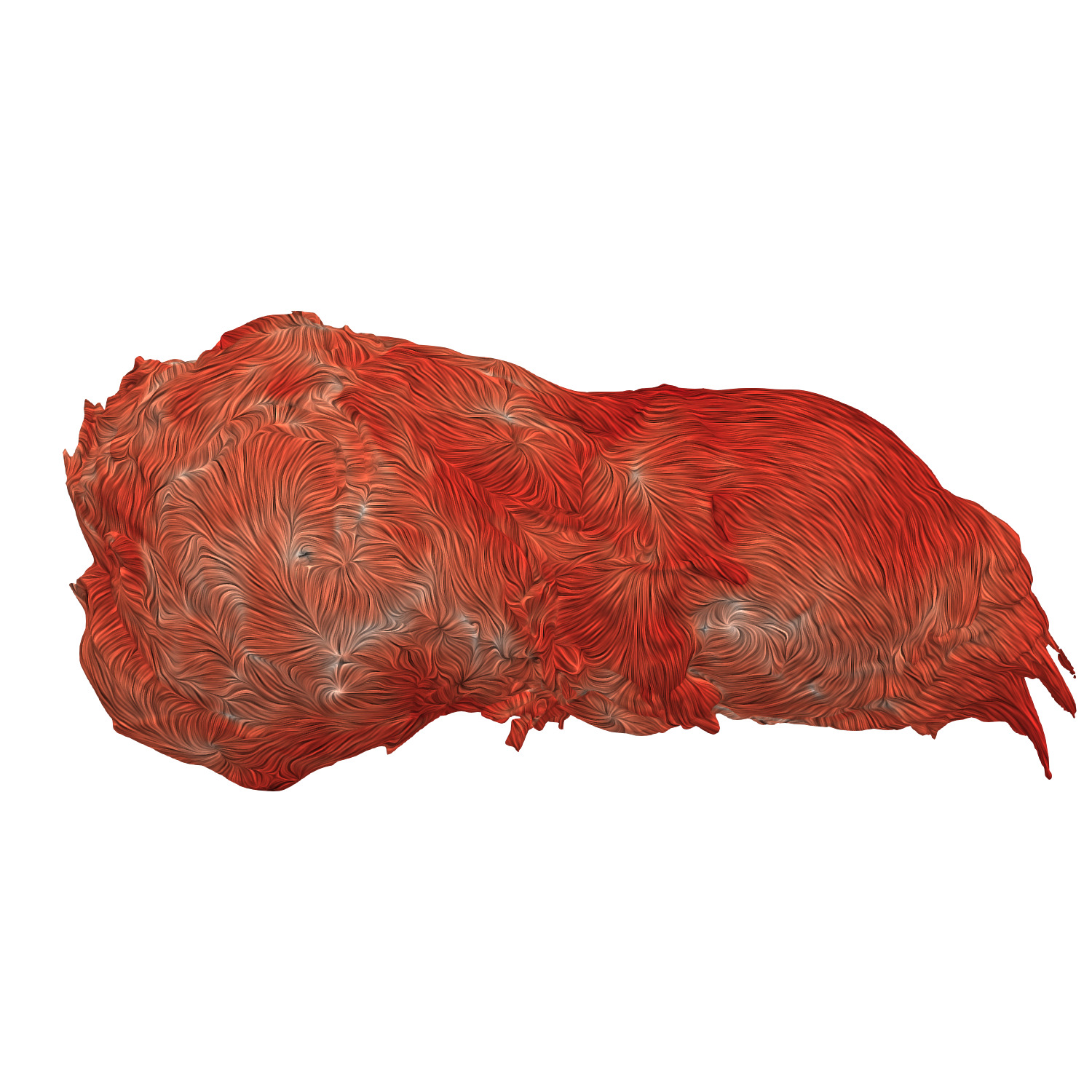}};%
                    \coordDotRight{0.5}{-0.5}{0.2}{$x$}{$y$}{$z$}%
                \end{tikzpicture}%
            }%
        \\[-1.5em]%
        \subfloat[\label{fig:4views-predisruption-iso+y}]%
            {%
                \begin{tikzpicture}%
                    \node[anchor=south west,inner sep=0] at (0,0) {%
                        \includegraphics[width=0.49\linewidth,trim=0 400pt 0 400pt,clip]%
                            {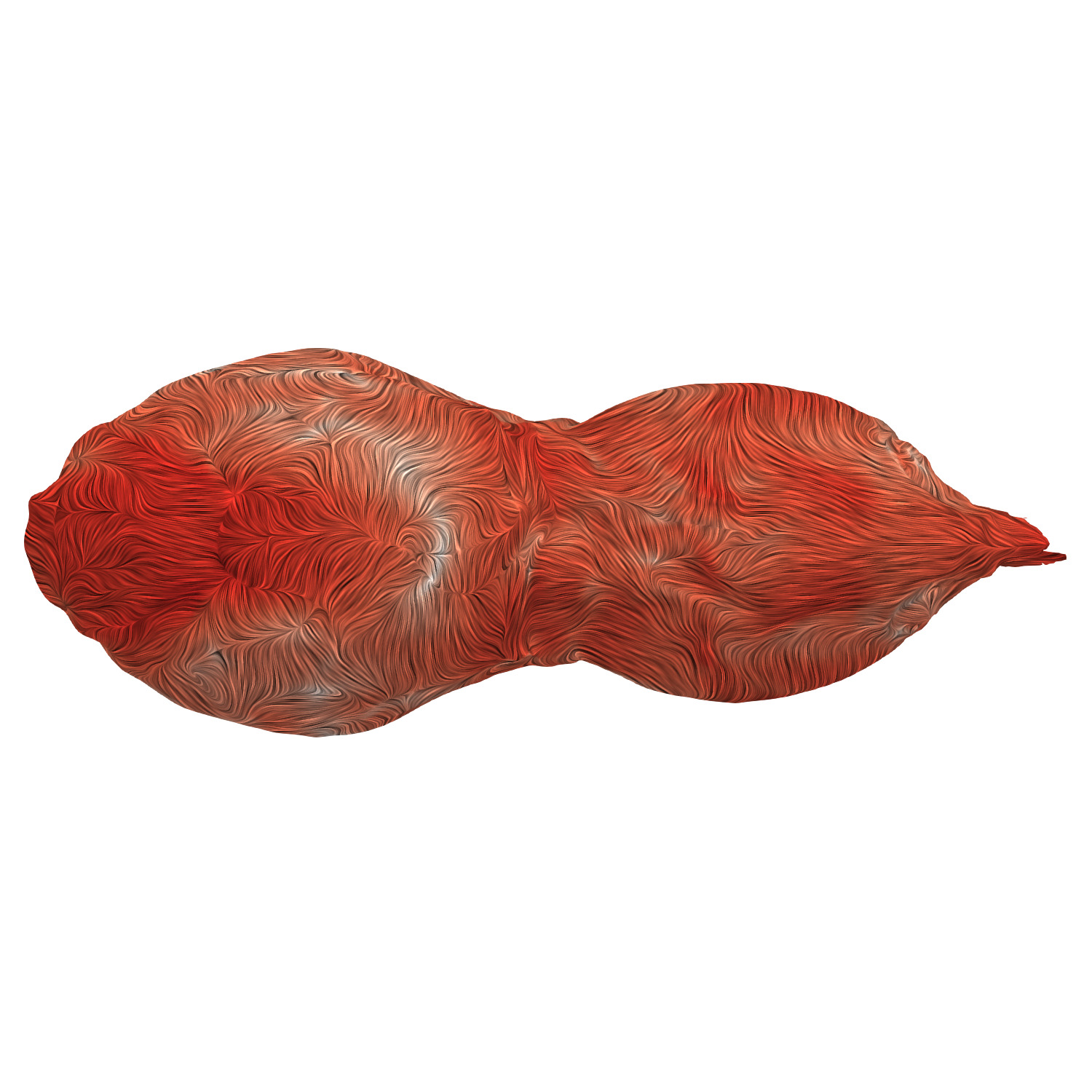}};%
                    \coordCrossRight{0.5}{-0.5}{0.2}{$x$}{$z$}{$y$}%
                \end{tikzpicture}%
            }%
        \hfill%
        \subfloat[\label{fig:4views-rcb12-predisruption-iso+y}]%
            {%
                \begin{tikzpicture}%
                    \node[anchor=south west,inner sep=0] at (0,0) {%
                        \includegraphics[width=0.49\linewidth,trim=0 400pt 0 400pt,clip]%
                            {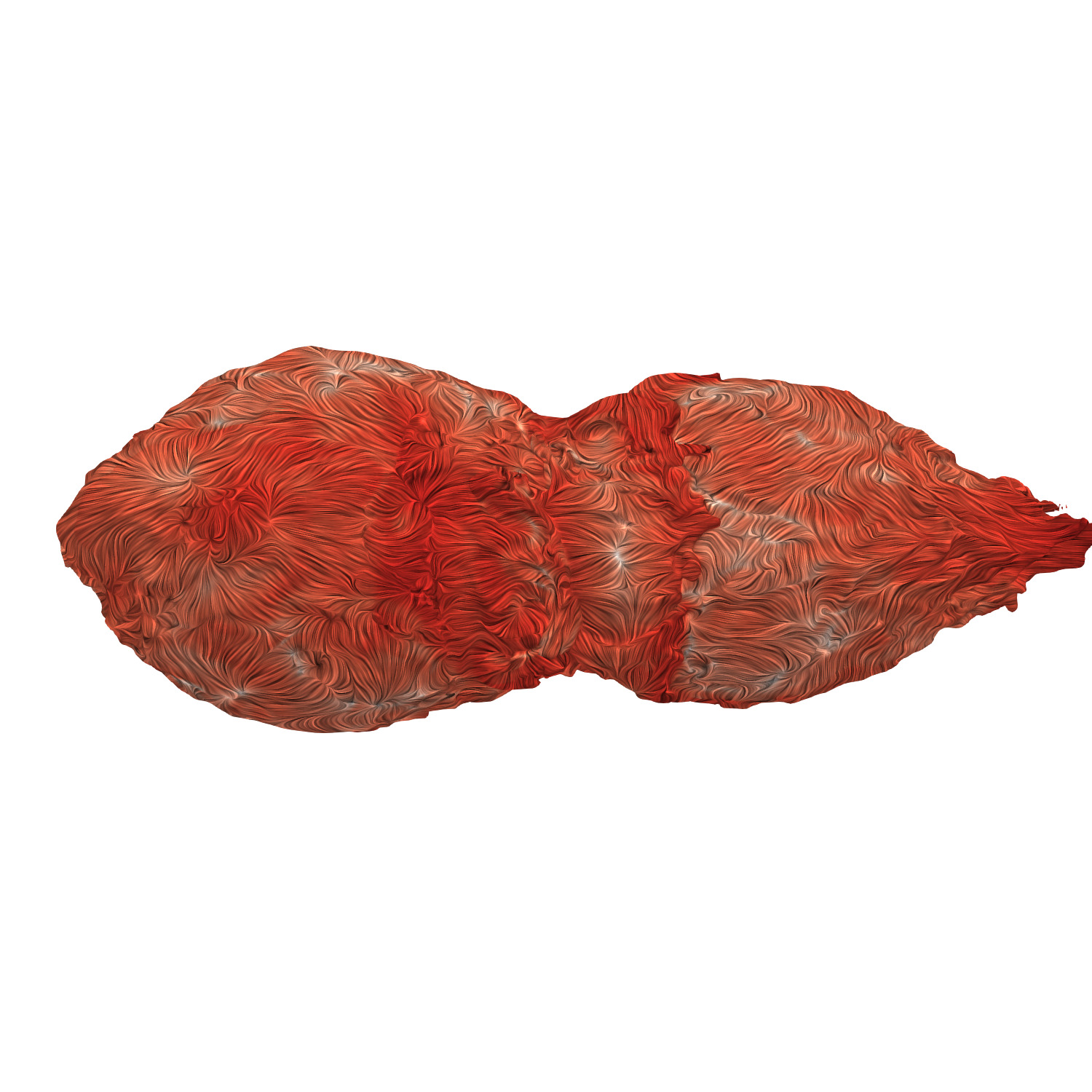}};%
                    \coordCrossRight{0.5}{-0.5}{0.2}{$x$}{$z$}{$y$}%
                \end{tikzpicture}%
            }%
    \end{minipage}%
    \caption{%
        Four views of the binary for rcb11 at a time $t=2759.89 \seconds$ (left) and rcb12 at $t=4306.21 \seconds$, approximately one orbital period before tidal disruption of the donor.
        Velocity magnitudes are mapped logarithmically to color~(\includegraphics[width=6pt,height=32pt,angle=-90,origin=rB]{figures/legend_reds.jpeg}) in the range $[1 \cdot 10^6,~1 \cdot 10^9] \cmps$. The flow lines shown are the result of LIC on the isopycnic surface 
        $\rho =100 \gpccm$. Note how some overflow through $L_2$ on the extreme right is beginning to take place. The bulging on the left of the accretor in \protect\subref{fig:4views-predisruption-iso+y} and \protect\subref{fig:4views-rcb12-predisruption-iso+y} is due to the accretion belt. The center of mass of the system is buried inside the flow, near the center of the images.
    }
    \label{fig:4views-predisruption}
\end{figure*}

\subsection{Finite-time Lyapunov exponent}
\label{sec:vis-ftle}

The finite-time Lyapunov exponent~(FTLE;~\cite{Haller2015}) is a method to provide a scalar value for each point in a grid that represents the separation of two initially very close particles over time.
For the calculation of this value, the same idea as for streamlines (static) and pathlines (dynamic) is employed, virtually seeding two lines close together.
After a user-defined integration time, the distance of the two particles is used to determine the FTLE value $\Lambda_{t_0}^{t_0 + \delta t}$.
This is done by providing a flow map $\mathcal{F}_{t_0}^{t_0 + \delta t}$, which maps the initial particles' positions at a time $t_0$ to their integrated positions at time $t_0 + \delta t$.
From the flow map gradient, the right Cauchy--Green strain tensor can be computed as
\begin{equation}
    \label{eq:cauchy}
    C(\vec{x}_0) = \left[ \nabla \mathcal{F}_{t_0}^{t_0 + \delta t} (\vec{x}_0) \right]^\top \nabla \mathcal{F}_{t_0}^{t_0 + \delta t} (\vec{x}_0),
\end{equation}
whose largest eigenvalue $\lambda_n$ is used to calculate
\begin{equation}
    \label{eq:ftle}
    \Lambda_{t_0}^{t_0 + \delta t} (\vec{x}_0) = \frac{1}{\delta t} \log \sqrt{\lambda_n (\vec{x}_0)}.
\end{equation}
A larger value $\Lambda_{t_0}^{t_0 + \delta t}$ thus means a larger separation, and can be used to find regions where the flow is separating.
One difficulty in employing FTLE is that the resulting value heavily depends on the chosen integration time.
Therefore, it is often used in combination with stream- or pathline visualization to determine an appropriate integration time and to allow the interpretation of the results.
Applying FTLE in reverse time~($\delta t < 0$) further results in a quantity that describes the attachment of the flow.

For the visualization of the scalar FTLE field in two dimensions, the value can be mapped to color, or the field can be interpreted as a height map from which a ridge is extracted.
The latter results in lines that describe where the separation (attachment) of the flow is largest.
This means that particles seeded on one side of the line separate from (converge to) the ones on the other side.
In three dimensions, FTLE is most commonly visualized as a ``height ridge'', which is a surface.
These relatively long-lived ridges are known as Lagrangian coherent structures (LCS;~\cite{Haller2015}).

\begin{figure*}
    \centering%
    \begin{minipage}[b]{\linewidth}%
        \centering%
        \subfloat[\label{fig:4views-postdisruption-iso-z}]%
            {%
                \begin{tikzpicture}%
                    \node[anchor=south west,inner sep=0] at (0,0) {%
                        \includegraphics[width=0.49\linewidth]%
                            {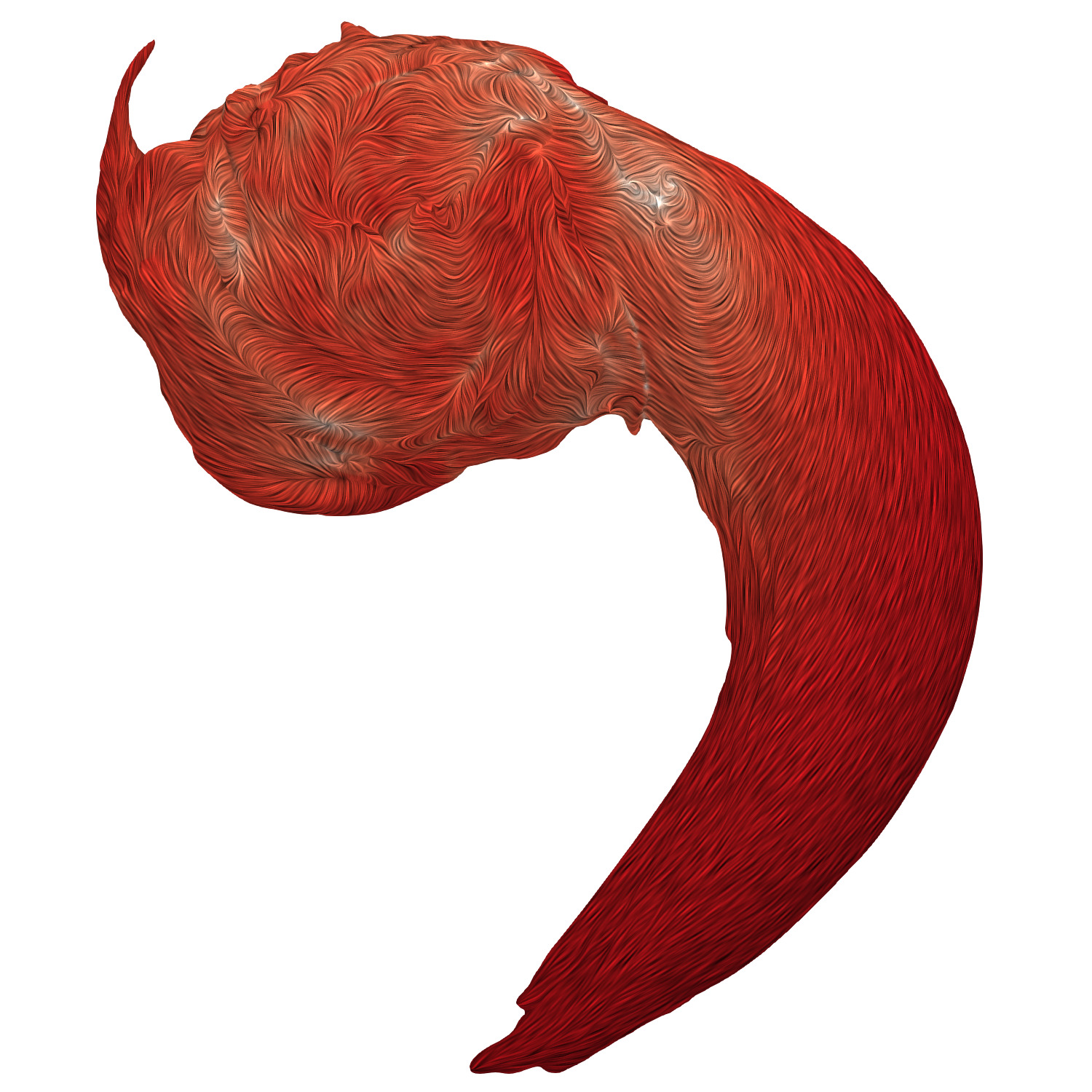}};%
                    \coordDotRight{1.3}{0.5}{0.2}{$x$}{$y$}{$z$}%
                \end{tikzpicture}%
            }%
        \hfill%
        \subfloat[\label{fig:4views-rcb12-postdisruption-iso-z}]%
            {%
                \begin{tikzpicture}%
                    \node[anchor=south west,inner sep=0] at (0,0) {%
                        \includegraphics[width=0.49\linewidth]%
                            {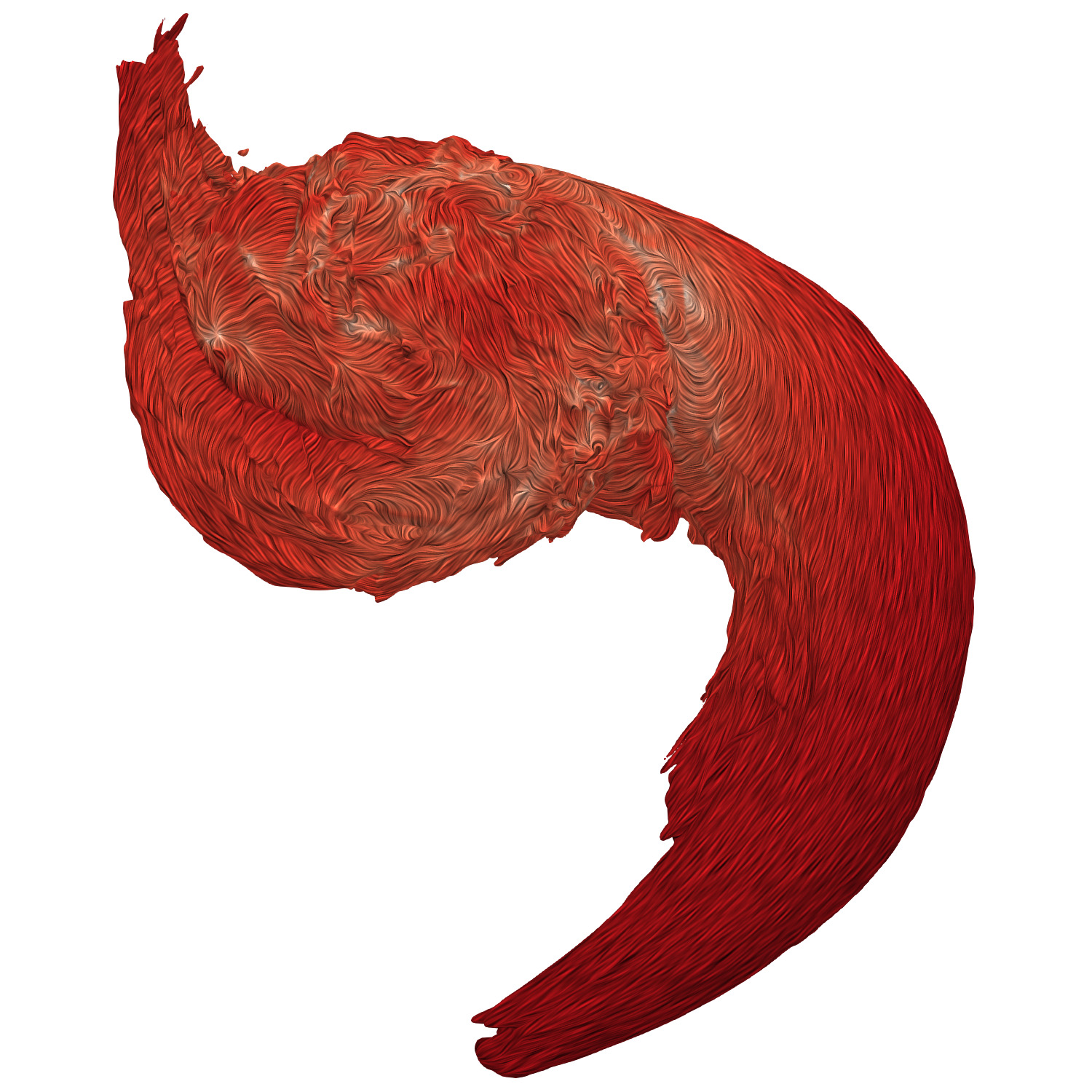}};%
                    \coordDotRight{1.3}{0.5}{0.2}{$x$}{$y$}{$z$}%
                \end{tikzpicture}%
            }%
        \\[-1.5em]%
        \subfloat[\label{fig:4views-postdisruption-iso-y}]%
            {%
                \begin{tikzpicture}%
                    \node[anchor=south west,inner sep=0] at (0,0) {%
                        \includegraphics[width=0.47\linewidth,trim=0 400pt 0 400pt,clip]%
                            {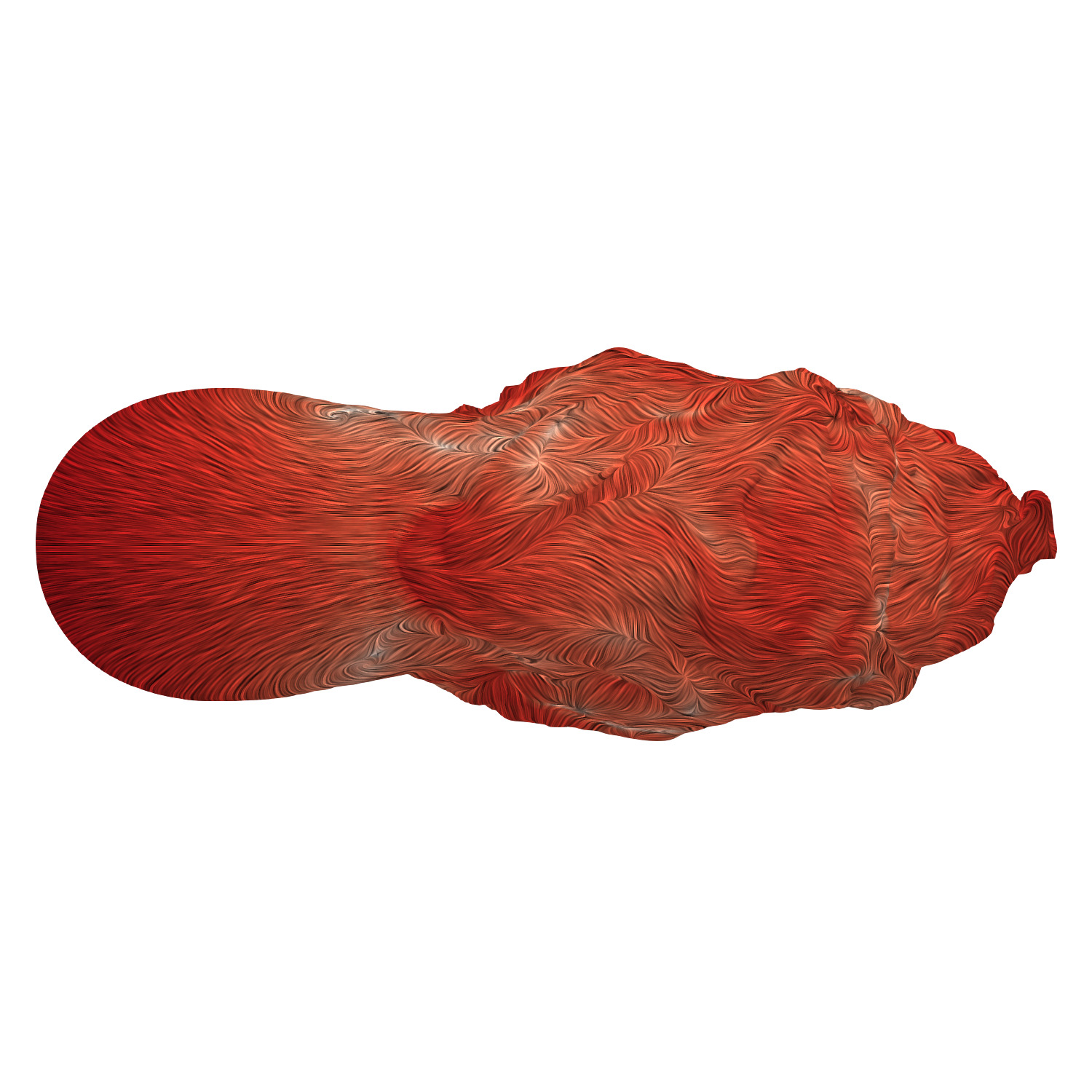}};%
                    \coordDotLeft{8}{0.3}{0.2}{$x$}{$z$}{$y$}%
                \end{tikzpicture}%
            }%
        \hfill%
        \subfloat[\label{fig:4views-rcb12-postdisruption-iso-y}]%
            {%
                \begin{tikzpicture}%
                    \node[anchor=south west,inner sep=0] at (0,0) {%
                        \includegraphics[width=0.47\linewidth,trim=0 400pt 0 400pt,clip]%
                            {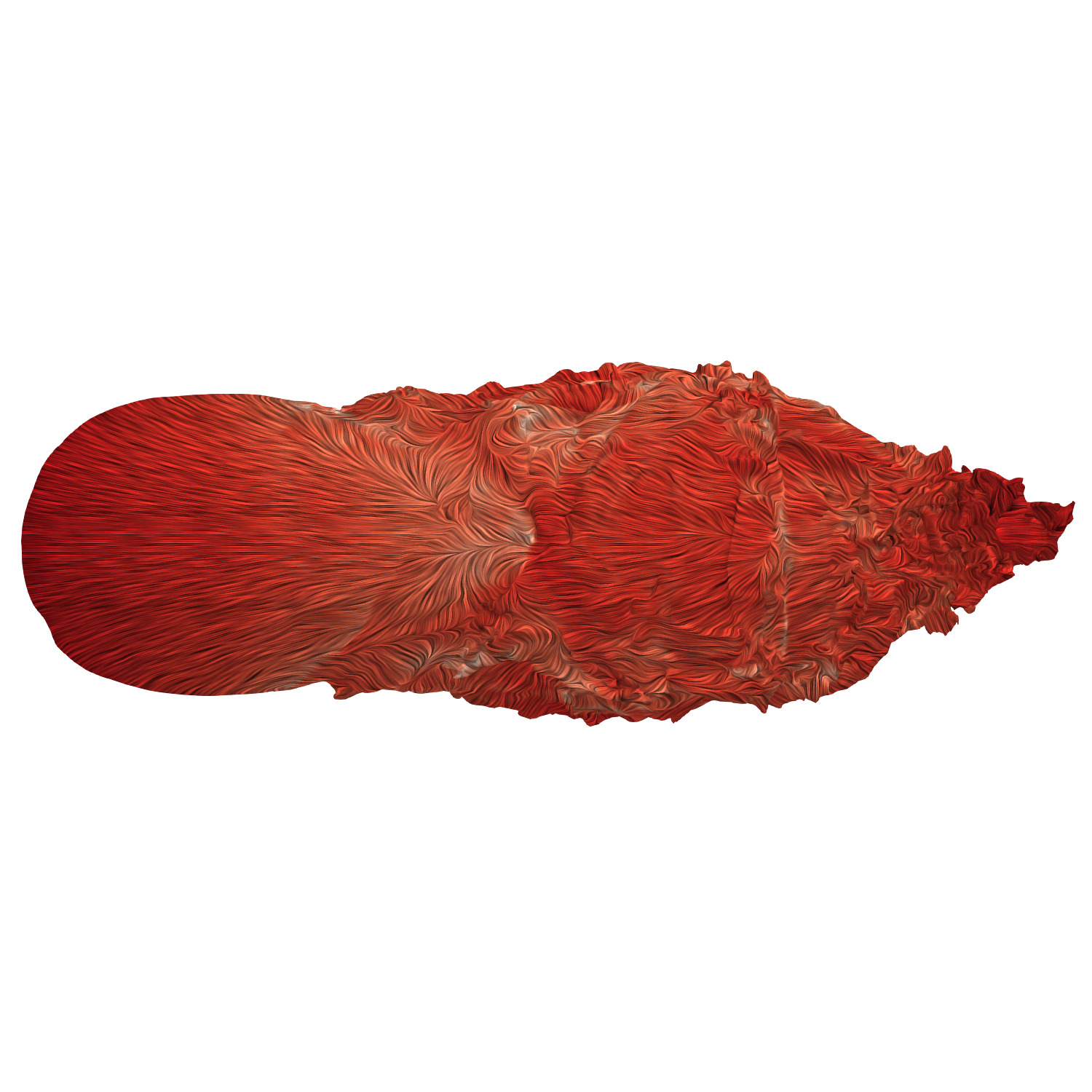}};%
                    \coordDotLeft{8}{0.3}{0.2}{$x$}{$z$}{$y$}%
                \end{tikzpicture}%
            }%
    \end{minipage}%
 \caption{%
        Two views of the binary for rcb 11 at a time $t=2873.81 \seconds$ (left column) and for rcb 12 at a time $t=4419.79 \seconds$ (right column), just as tidal disruption of the donor is occurring. In \protect\subref{fig:4views-postdisruption-iso-z} and \protect\subref{fig:4views-rcb12-postdisruption-iso-z} we see some overflow through $L_3$ on the left of the accretor.
        \protect\subref{fig:4views-postdisruption-iso-z},\protect\subref{fig:4views-rcb12-postdisruption-iso-z}~Surface at $\rho=100 \gpccm$ viewed from $z>0$.
        \protect\subref{fig:4views-postdisruption-iso-y},\protect\subref{fig:4views-rcb12-postdisruption-iso-y}~Surface at $\rho=100 \gpccm$ viewed from $y>0$.
        Projected surface velocity magnitudes are mapped logarithmically to color~(\includegraphics[width=6pt,height=32pt,angle=-90,origin=rB]{figures/legend_reds.jpeg}) in the range $[1 \cdot 10^6,~1 \cdot 10^9] \cmps$. The flow lines shown are the result of LIC on the isopycnic surface 
        $\rho =100 \gpccm$. The view in \protect\subref{fig:4views-postdisruption-iso-y} and \protect\subref{fig:4views-rcb12-postdisruption-iso-y} puts the accretor on the right, showing the incipient overflow through $L_3$ turning toward the observer. The center of mass is invisible, being located inside the fluid.}
    \label{fig:4views-postdisruption}
\end{figure*}

\section{Surface Flows}
\label{sec:surfflow}

We begin by showing external views of the binary 
surface and its surface flows at various times during the evolution for simulations rcb11 and rcb12. The first issue we face here is to specify what we mean by ``surface". The initial equilibrium binary model has an effective boundary where the density drops from $\rho\approx 1 \gpccm$ to $\rho_{\rm vac}$ in one cell. 
During the evolution this boundary is subject to diffusion, rounding errors, and regriding noise. Therefore we choose a higher density isosurface, either 
$\rho_0 =10\; {\rm or}\; 100\, \gpccm$, as the surface for our visualizations, depending on what we wish to emphasize.

\autoref{fig:4views} (left column) shows two views of the rcb11 binary with the fluid flows
obtained by LIC projected onto an isopycnic, constant density $\rho = \rho_0$ surface of the entire binary at an early time $t=100.44 \seconds = 0.88 P_0$, where $P_0=114.0 \seconds$ is the initial binary orbital period for rcb11.
The surface shown corresponds to $\rho_0 = 100 \gpccm$, approximately one five-thousandth of the central density of the donor star. 
The simulation runs from onset of mass transfer to the final merger of the two components into a single, nearly axisymmetric object.
We adopt a right-handed Cartesian coordinate system, with origin at the center of mass of the binary.
We choose the $x$-coordinate axis to lie along the axis of the binary, going through the centers of mass of both donor and accretor, so that the center of mass of the donor lies initially at some positive value of $x$, specifically $x=0.588 a$, where $a$ is the initial separation, while the center of mass of the accretor lies at a negative value $x=-0.412 a$.
The $(x,y)$-plane coincides with the orbital plane of the binary, and the $z$-axis points in the direction of the orbital angular momentum.
The views presented in \autoref{fig:4views} (top row) are from “above”, defined as a view along the $z$-axis, from $z > 0$, so that the binary rotates counterclockwise in the inertial frame.
\autoref{fig:4views} (right column) further shows two views of the projected flows onto the $\rho_0 = 100 \gpccm$ surface for the rcb12 binary, at a slightly later time $t = 113.62 \seconds \approx 1.00 P_0$, where $P_0=113.6 \seconds$.
Note in \autoref{fig:4views-rcb12-iso+y} how the ``head" of the developing accretion belt has advanced further than in 
\autoref{fig:4views-iso+y}.

In the  previous figure, and the rest of the figures described in this section, including the insets in \autoref{fig:pathlines4}, the light-colored areas are regions where the flow is quasi-stagnant in the corotating frame. Paying attention to these helps to identify and characterize physically different flow regimes 
(see also \autoref{sec:analysis}) 

\autoref{fig:4views-predisruption} shows views from the same directions as the previous figures and for the same isopycnic surface $\rho_0 = 100 \gpccm$ as before, at a time just prior to the tidal disruption of the secondary.
The surface flows possess approximate equatorial symmetry, have become generally faster, with small isolated regions of corotation, and dynamical velocities up to 3400 km/s, and the equatorial belt is still visible but no longer the dominant feature.
Similar perspectives for rcb12 
are shown in \autoref{fig:4views-rcb12-predisruption-iso-z} and \ref{fig:4views-rcb12-predisruption-iso+y}
at a later nominal time, selected to be representative of the same evolutionary stage as the views of the rcb11 model.
Qualitatively these are mutually consistent with rcb12 showing more detail.

\autoref{fig:4views-postdisruption} (left column) shows two views of rcb11 from the same directions as the previous figures and for the same isopycnic surface $\rho_0 = 100 \gpccm$ as before, at a time just after the tidal disruption of the secondary. The surface flows possess less  equatorial symmetry than at earlier stages, are generally faster and more chaotic. The tidal tails have become the dominant features, with the trailing tail extending well beyond corotation (light-colored region near the second Lagrangian point $L_2$. 

\autoref{fig:4views-postdisruption} (right column) shows the same views of the rcb12 simulation at a comparable evolutionary stage. The finer detail reveals an even more chaotic flow, with little equatorial symmetry, and a more pronounced leading tidal tail. 
The simulation rcb12 resolves a thinner accretion stream and a lower accretion rate, leading to slower evolution.

\begin{figure*}
    \centering%
    \begin{minipage}{0.47\linewidth}%
        \centering%
        \subfloat[\label{fig:bounce-rcb11}rcb 11, $t= 1072.21 \seconds$]{%
            \begin{tikzpicture}%
                \node[anchor=south west,inner sep=0] at (0,0) {\includegraphics[width=\linewidth]{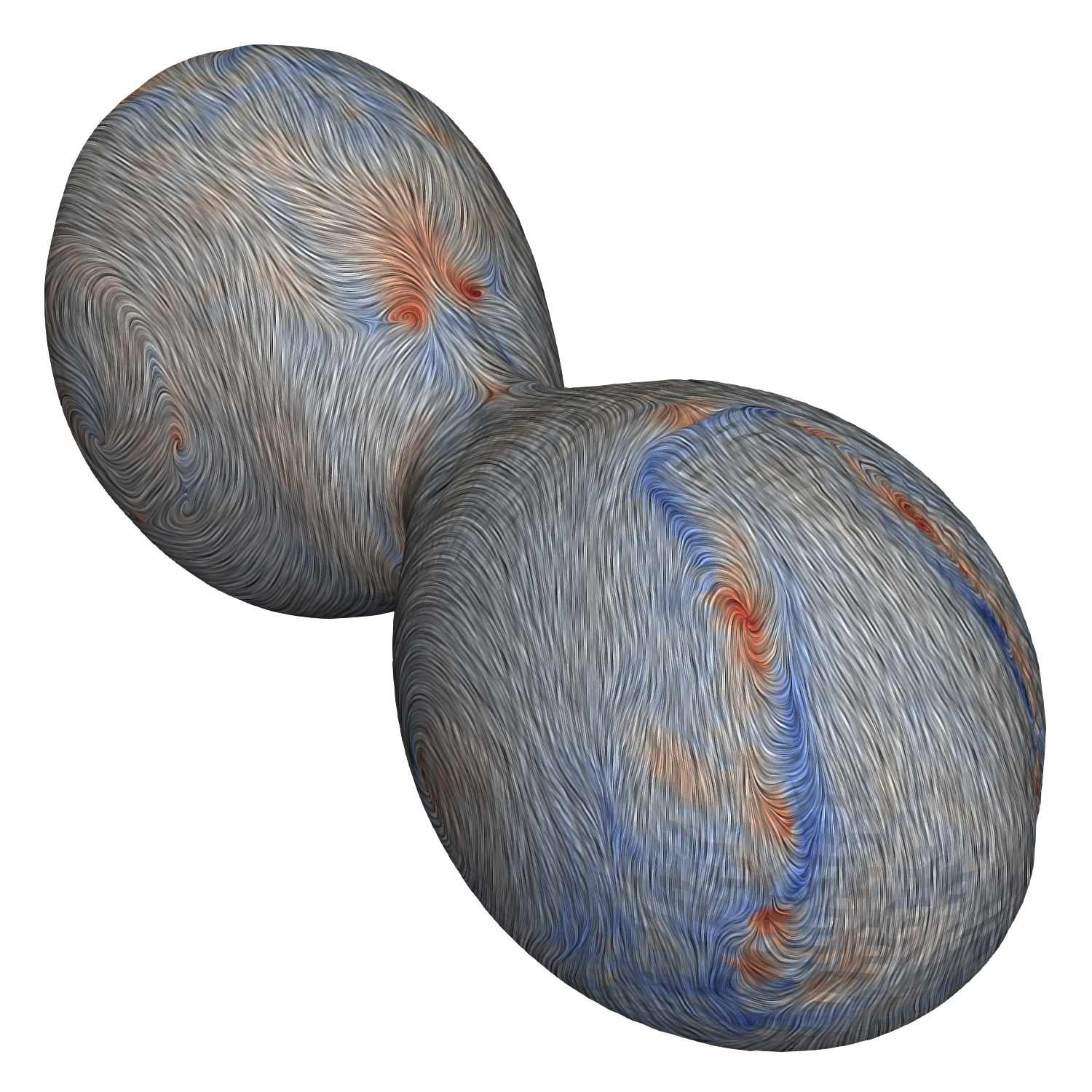}};%
                \node[anchor=south west,inner sep=0] at (0,0) {\fbox{\includegraphics[width=0.4\linewidth]{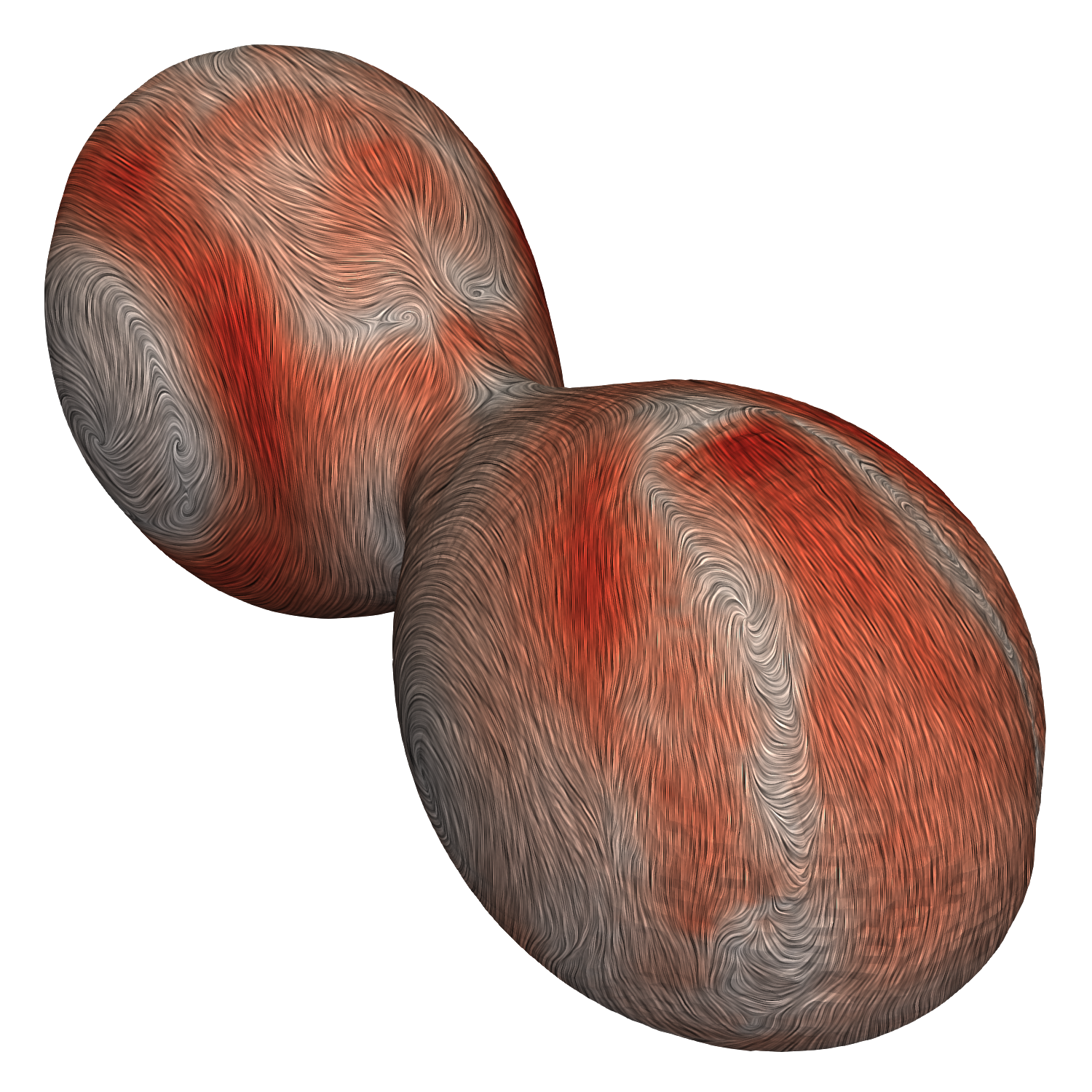}}};%
            \end{tikzpicture}%
        }%
    \end{minipage}%
    \hfill%
    \begin{minipage}{0.47\linewidth}%
        \centering%
        \subfloat[\label{fig:bounce-rcb12}rcb 12, $t= 1067.52 \seconds$]{%
            \begin{tikzpicture}%
                \node[anchor=south west,inner sep=0] at (0,0) {\includegraphics[width=\linewidth]{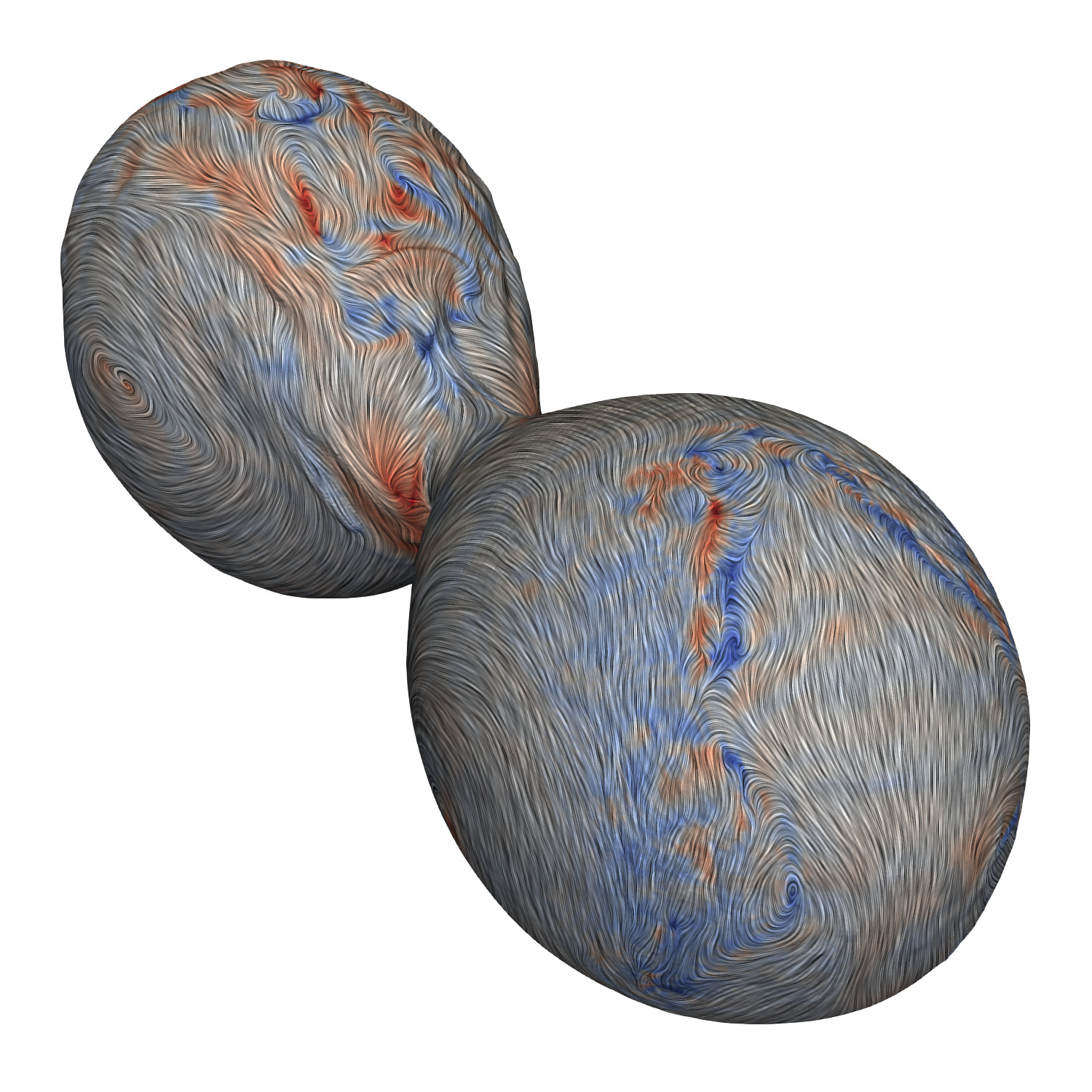}};%
                \node[anchor=south west,inner sep=0] at (0,0) {\fbox{\includegraphics[width=0.4\linewidth]{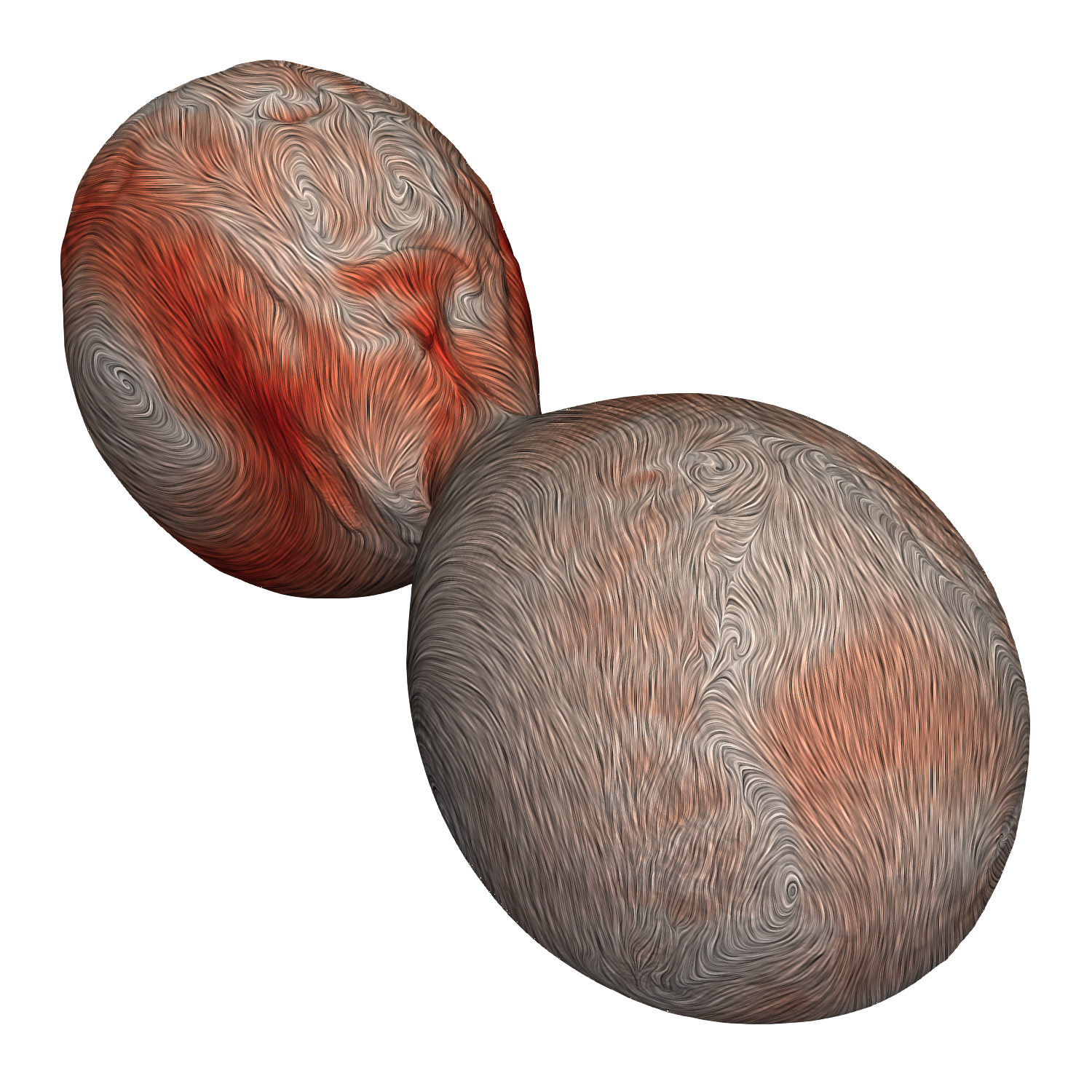}}};%
            \end{tikzpicture}%
        }%
    \end{minipage}%
    \caption{%
        Showing views of the surface flow obtained by LIC for \protect\subref{fig:bounce-rcb11} rcb11 and \protect\subref{fig:bounce-rcb12} rcb12 from above ($z>0$) and behind ($x>0$) the donor for a surface at density $\rho = 100 \gpccm$. The times selected are approximately at the
        same number of orbits after the start $t=9.4 P_0$.
        The angle between the surface normal and the velocity is mapped to color~(\includegraphics[width=6pt,height=32pt,angle=-90,origin=rB]{figures/legend_pv}) in the range $[0,\pi]$, where a value $< \frac{\pi}{2}$~(blue) indicates inflow, a value $> \frac{\pi}{2}$~(red) indicates outflow, and a value of $\frac{\pi}{2}$~(gray) indicates tangential flow
        (see \autoref{eq:in-outflow}). Note the bounce of the stream: the red (outward) spots where the stream meets the accretor indicate splashing or bounce of the stream material.
        The smaller figures~(black rectangle) show the {\em magnitude} of the total velocity~(\includegraphics[width=6pt,height=32pt,angle=-90,origin=rB]{figures/legend_reds}) in the range $[0,~1 \cdot 10^8] \cmps$.
    }%
    \label{fig:bounce}%
\end{figure*}

The image shown in \autoref{fig:bounce} shows a perspective view of the binary from a point of view located on the donor side, above the orbital plane, showing the point of impact of the stream and the equatorial belt.
The red regions show where the component normal to the surface shown is positive (outward), while in the blue regions the normal component is negative (inward).
This view shows the binary at a  time $t\approx 9.4 P_0$ when the stream flow, equatorial belt, and return flow over the back of the donor are well developed.
After the stream bounces upon impact on the accretor, the flow at first spreads mostly forward and sideways---and even backward---before converging and resuming the direction of the equatorial accretion belt.
The flow of the accretion belt goes around the accretor and impacts the stream from behind.
Some of this material splits around the stream and converges joining the main stream behind the point of impact and the bounce region.
As evolution proceeds, an increasing amount of material is leaking around the outer Lagrangian point $L_2$.

We expect the surface flow to be driven by the pressure gradients that arise as mass flows over the inner Lagrange point to be subject to gravity, pressure gradients, and Coriolis forces.
Our goal is to understand the flow pattern based on integrating forward and backward in time the paths of tracer particles placed in suitable initial locations in the flow. 
Given the fluid velocity fields in the corotating frame (\autoref{sec:vis-frame}), pathlines can be constructed as seen by observers in this frame, by placing seed particles at strategically chosen locations and following their motion forward and backward in time by integration.
When choosing the integration time to be a fraction of the orbital period such that the flow field is in a quasi-steady state, flow lines and stream lines are approximately the same, and one learns where the particles came from and where they are headed. 


\begin{figure*}%
\scalebox{0.5}{
    \centering
    \begin{tikzpicture}%
        \node[anchor=south west,inner sep=0] at (0,0)%
            {\includegraphics[width=\linewidth]%
            {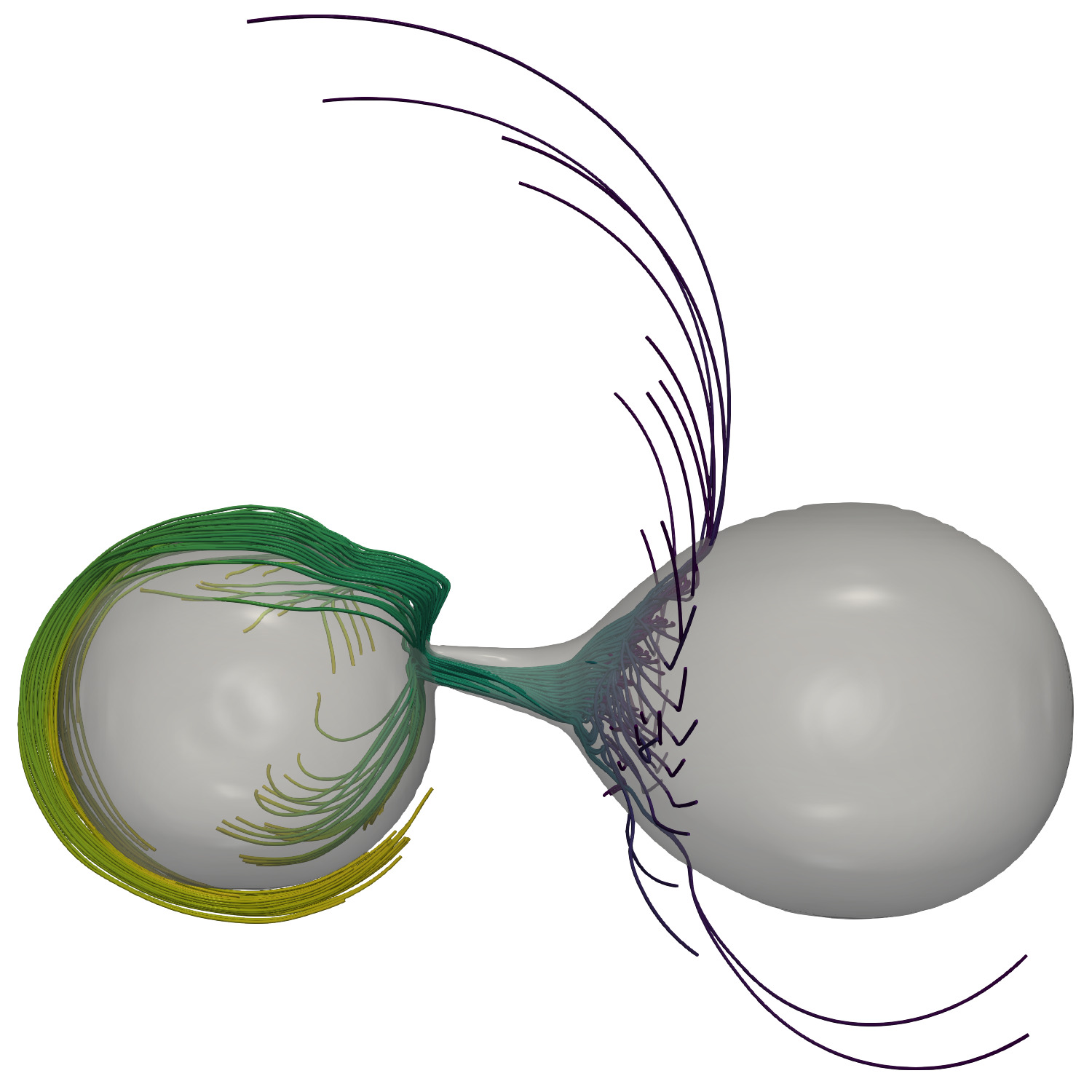}};%
        %
        \node[anchor=south west,inner sep=0] at (0.2,10.2)%
            {\fbox{\includegraphics[width=0.5\linewidth,trim=0 325pt 0 375pt,clip]%
            {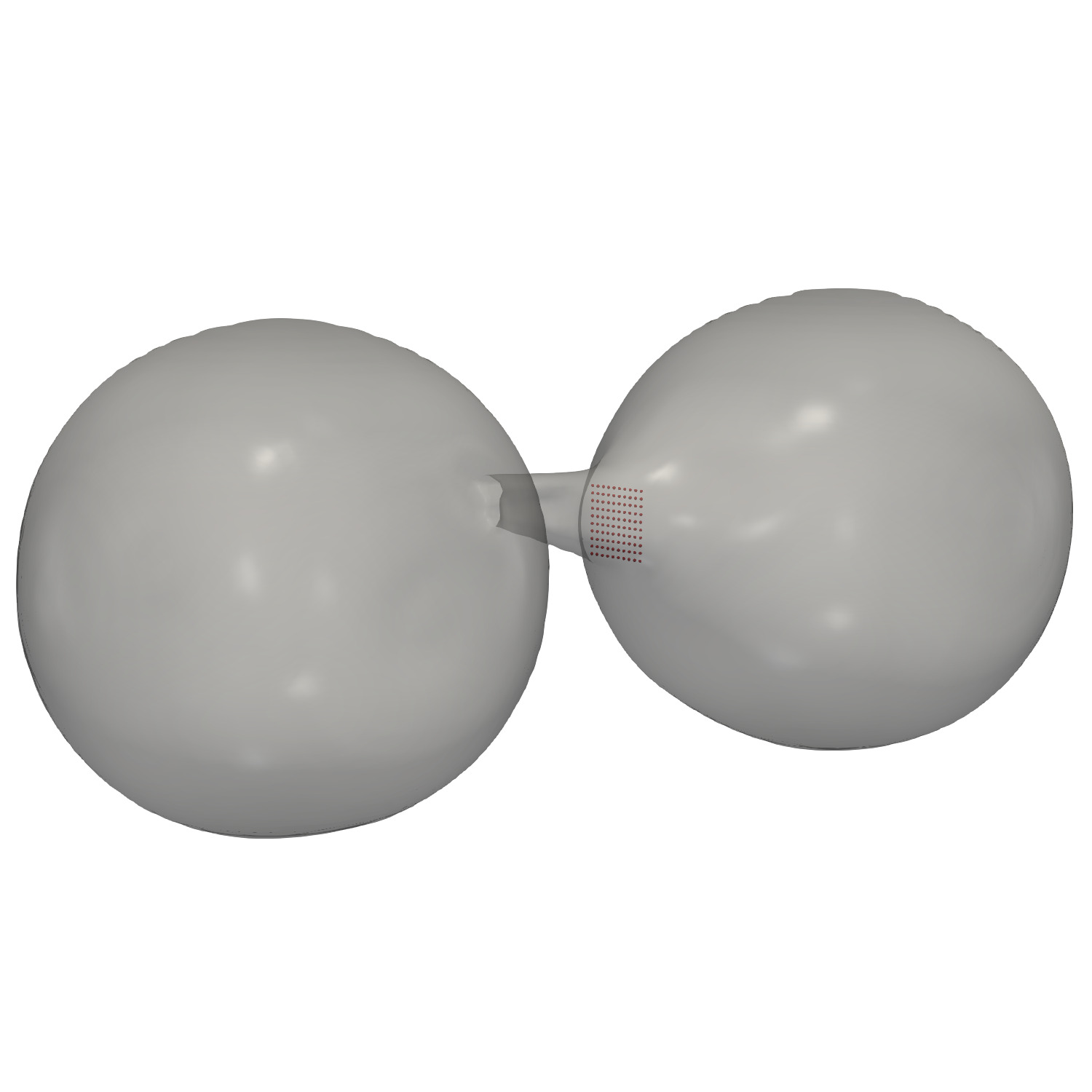}}};%
        \node[anchor=south east] at (9.3,10.2) {pathline seed};%
        %
        \draw[seedred,ultra thick] (4.8,12.4) rectangle (6,13.7);%
        \draw[seedred,ultra thick] (8.8,5.8) rectangle (9.7,7.2);%
        \draw[seedred,thick] (4.8,12.4) -- (8.8,5.8);%
        \draw[seedred,thick] (6,13.7) -- (9.7,7.2);%
        %
        \node[anchor=south west,inner sep=0] at (15,11)%
            {\includegraphics[height=160pt]%
            {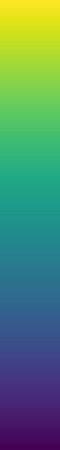}};%
        \node[anchor=west] at (15.6,11.01) {--- 6.0 \seconds};%
        \node[anchor=west] at (15.6,13.83) {--- 86.83 \seconds};%
        \node[anchor=west] at (15.6,16.65) {--- 166.0 \seconds};%
    \end{tikzpicture}%
    }
    \caption{%
        Pathlines seeded at $t=86.83 \seconds$ integrated forward in time until $t=166.0 \seconds$ (\includegraphics[width=6pt,height=32pt,angle=-90,origin=rB,trim=0 150pt 0 0,clip]{figures/legend_viridis.jpeg}) and integrated backward in time until $t=6.0 \seconds$ (\includegraphics[width=6pt,height=32pt,angle=90,origin=lB,trim=0 0 0 150pt,clip]{figures/legend_viridis.jpeg}). The backward-integrated paths extending outside of the binary are due to vacuum particles swept up as the binary rotates.
        }
    \label{fig:pathlines4}
\end{figure*}

An example of such pathlines is shown in \autoref{fig:pathlines4}.
The inset shows a grid of the ``seed" particles covering a cross-section of the stream.
The pathlines obtained by forward integration (green to yellow) show the stream impacting the accretor, with most of the flow bouncing and joining the fast counter-clockwise (CCW) motion of the equatorial belt.
The rest of the stream splashing to higher latitudes and moving backward (CW: clockwise in the corotating frame).
The backward pathlines (green to blue) show that most of the stream material originates in the surface layers of the donor, and moves slowly toward the inner Lagrangian point at the base of the stream.
A few of those particles in the leading side of the donor (recall the binary rotates CCW in the inertial frame) appear to be swept up from the ``vacuum" by the rotating binary.
Relative to the binary frame these particles move with orbital velocities toward the donor.
On the trailing side, even fewer vacuum particles are drawn into the wake of the donor, make it to the surface and join the trailing side of the stream.

\begin{figure*}%
    \centering
    \subfloat[\label{fig:3pathlines-y}]%
        {\includegraphics[width=0.33\linewidth]%
        {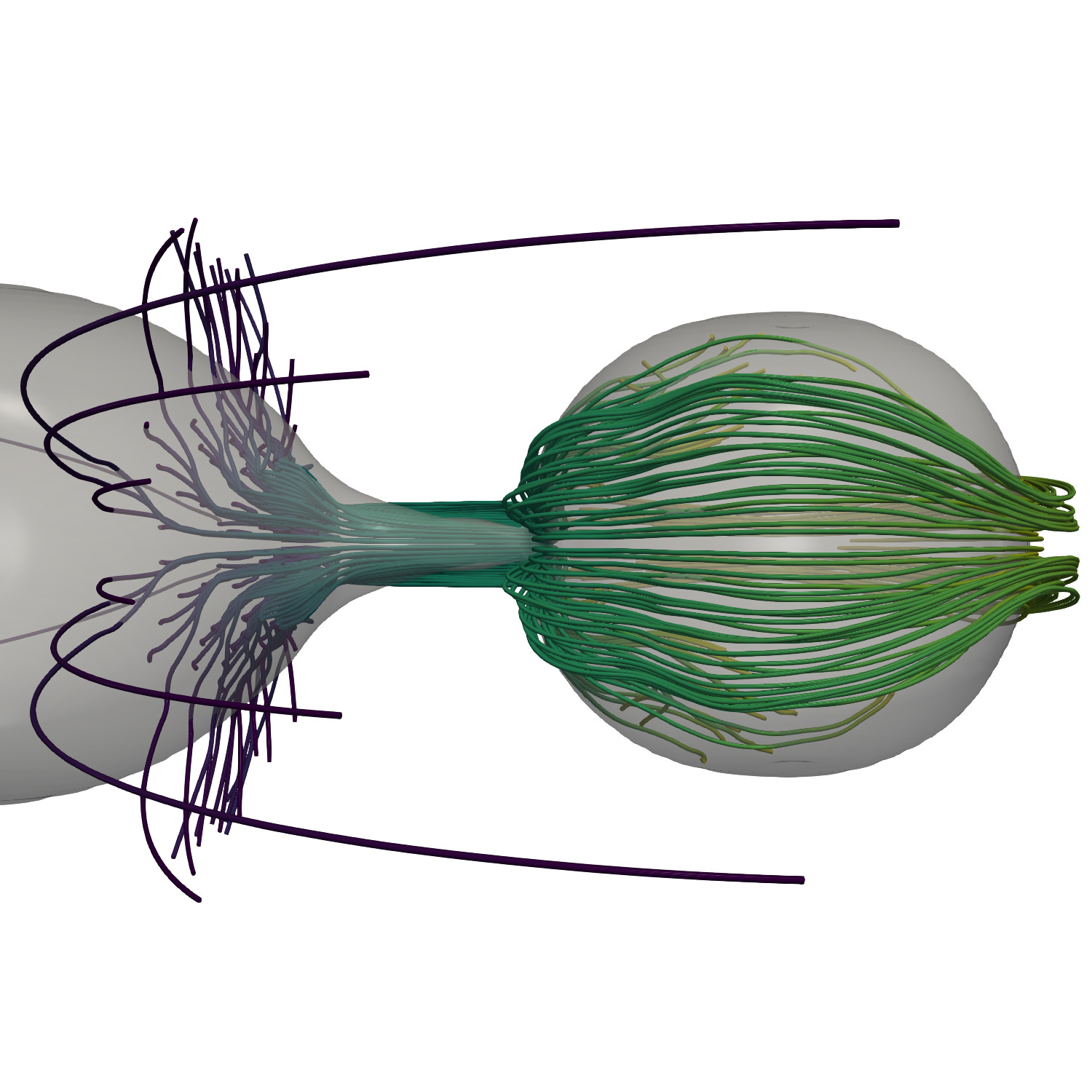}}%
    \hfill%
    \subfloat[\label{fig:3pathlines+x}]%
        {\includegraphics[width=0.33\linewidth]%
        {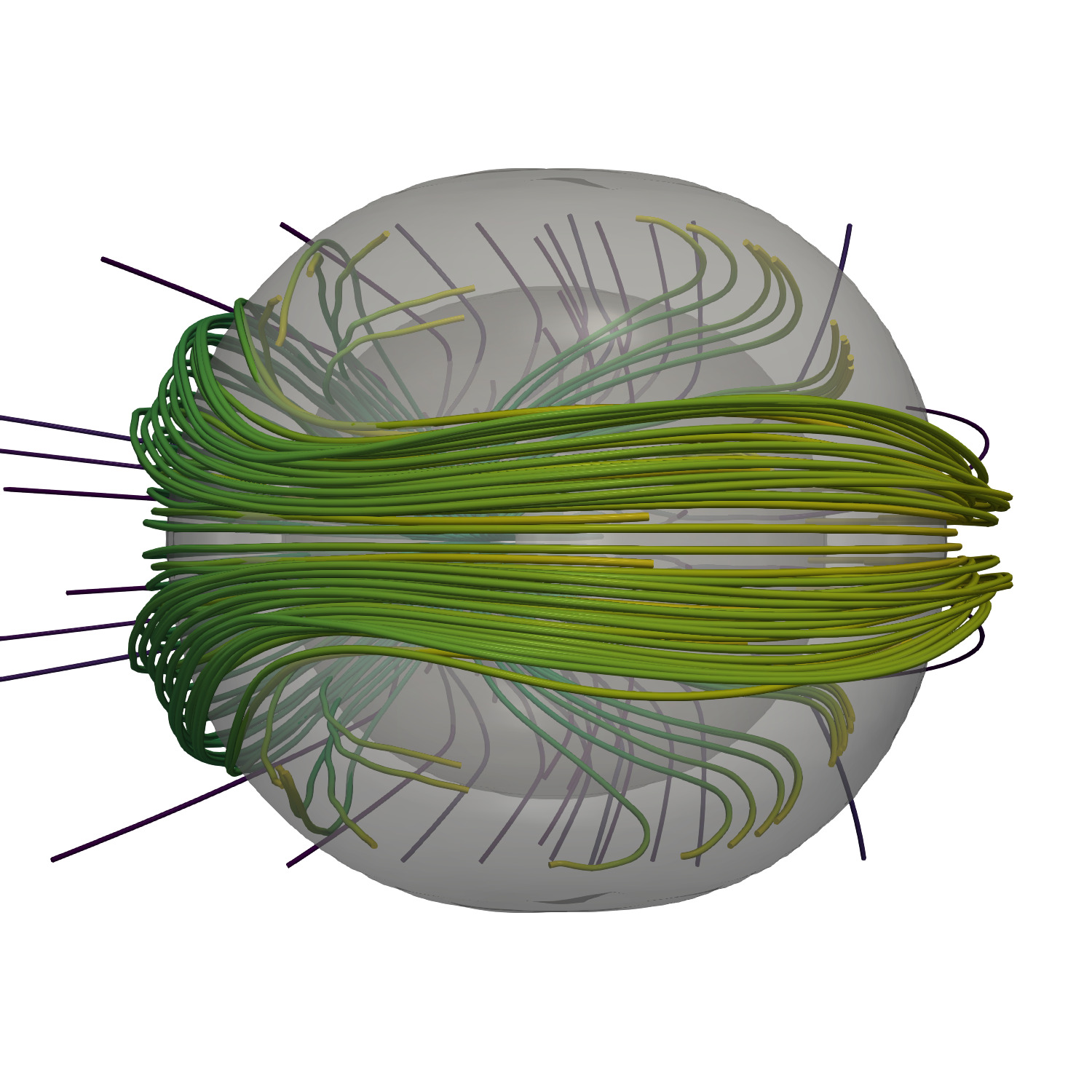}}%
    \hfill%
    \subfloat[\label{fig:3pathlines+y}]%
        {\includegraphics[width=0.33\linewidth]%
        {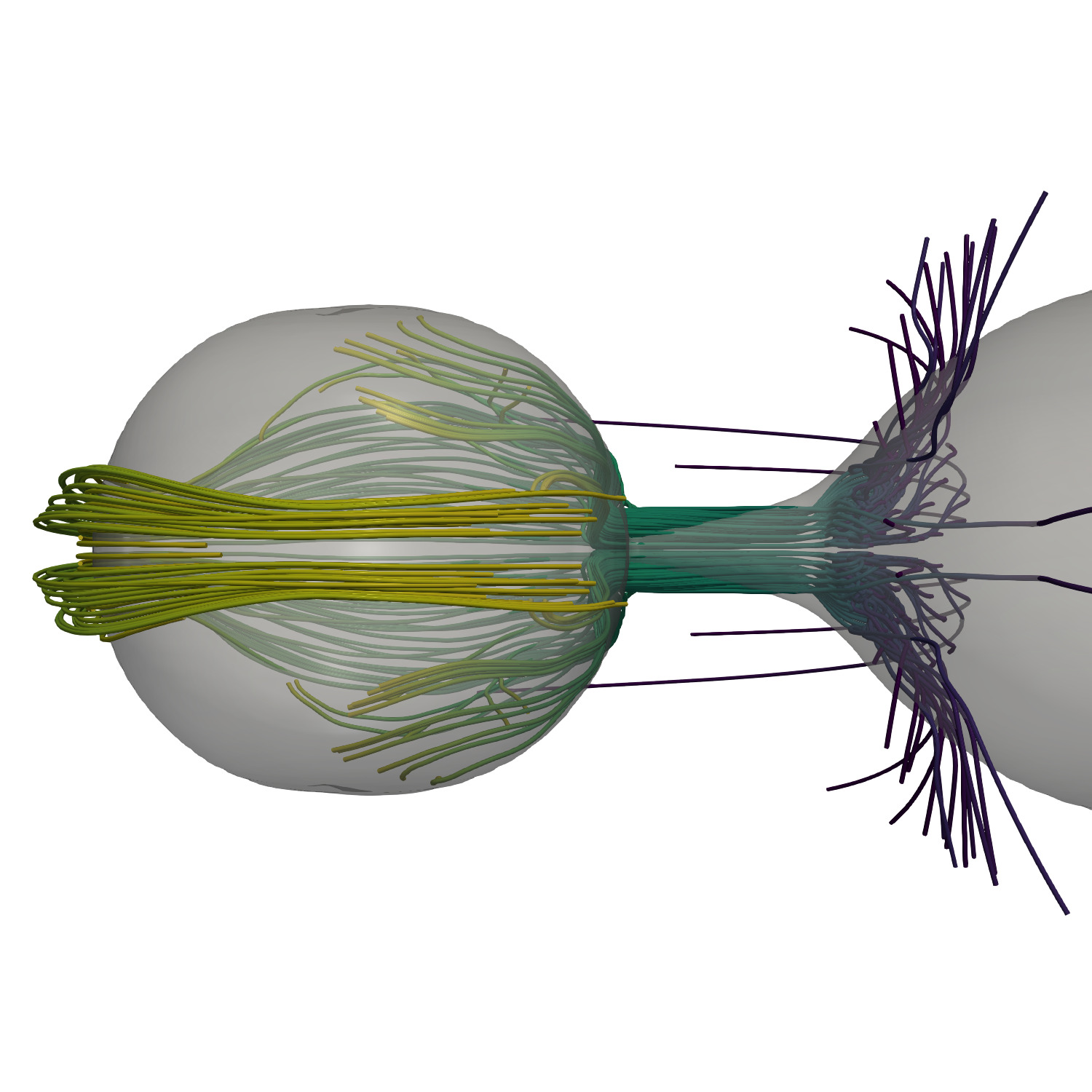}}%
    \caption{%
        Three views show the flow looking in \protect\subref{fig:3pathlines-y}~the negative y-direction, \protect\subref{fig:3pathlines+x}~the positive x-direction, and \protect\subref{fig:3pathlines+y}~the positive y-direction.
        In the left panel we are looking at the stream moving toward the observer, out of the plane of the image.
        In the central panel we see the flow rounding the back of the accretor, moving from left to right, and on the right panel we are looking at the back of the stream.
        Pathline time from $t=6.0 \seconds$ to $t=166.0 \seconds$ mapped to color~(\includegraphics[width=6pt,height=32pt,angle=-90,origin=rB]{figures/legend_viridis.jpeg}).}
    \label{fig:3pathlines}
\end{figure*}

Completing the picture of the early mass transfer flow, \autoref{fig:3pathlines} shows three 
sideways views of the flow showing clearly that most of the mass transfer forms a so-called accretion belt, swirling around the accretor at low latitudes above and below the equator, only a small fraction of the mass is splashed toward higher latitudes.
As the accreted fluid carries a higher specific angular momentum than the accretor’s surface, the accretion belt flows around faster than corotation and hits the stream impact region from behind.
At this point the flow interacts with the splashing of the stream and begins to flow to higher latitudes. Note that the effective accretion rate in our simulations is orders of magnitude higher than in CVs, where the latitudinal spread may depend on the accretion rate \citep{Piro2004}.

\section{Development of Internal Flows}
\label{sec:internal}

The images presented in this section display the velocity fields in two of the principal cross sections of the binary: the orbital (equatorial) plane $(x,y)$ and the meridional plane $(x,z)$ of the binary containing the centers of mass of both components and the axis of rotation. 

The flow on the equatorial plane is viewed from ``above" ($z>0$), including the equatorial slices of the accretor (left) and the donor (right) and the mass transfer stream.
The color scale spans the range of velocities from subsonic $0.1 \kmps$ to supersonic $3400 \kmps$. The higher end velocities are typical dynamical velocities of the corotating binary, and are mainly encountered in the mass transfer stream and the accretion belt.
Large clockwise velocities are also seen in the ``vacuum", at rest in the inertial frame, resulting from the flow of gas past the body of the binary as a reflection of the counter-clockwise orbital motion. 
\begin{figure*}
    \centering%
    \subfloat[\label{fig:EqSlice-rcb11}rcb 11, $t= 114.19 \seconds$]%
        {\includegraphics[width=0.49\linewidth,trim=0 400pt 0 400pt,clip]%
        {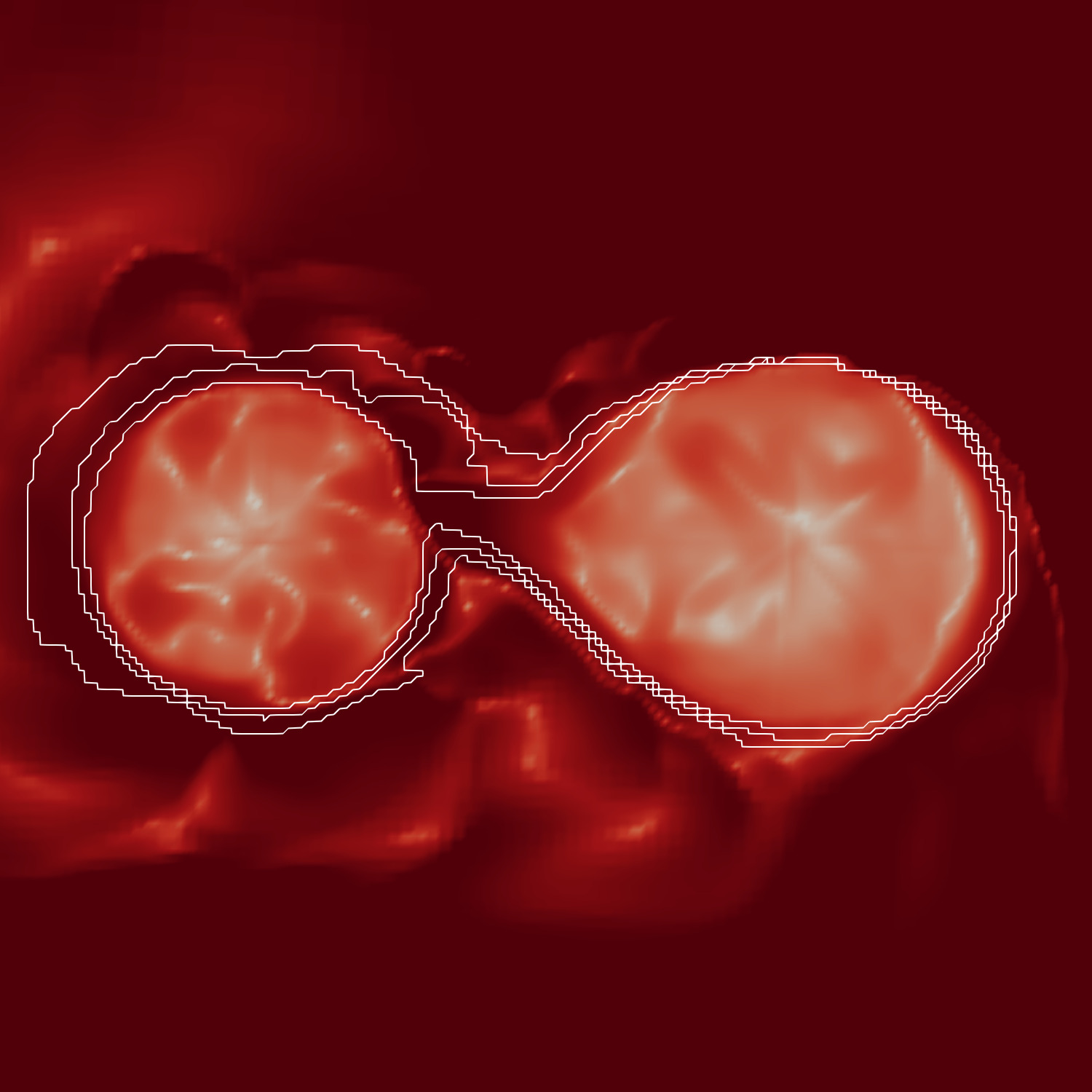}}%
    \hfill%
    \subfloat[\label{fig:EqSlice-rcb12}rcb 12, $t= 113.62 \seconds$]%
        {\includegraphics[width=0.49\linewidth,trim=0 400pt 0 400pt,clip]%
        {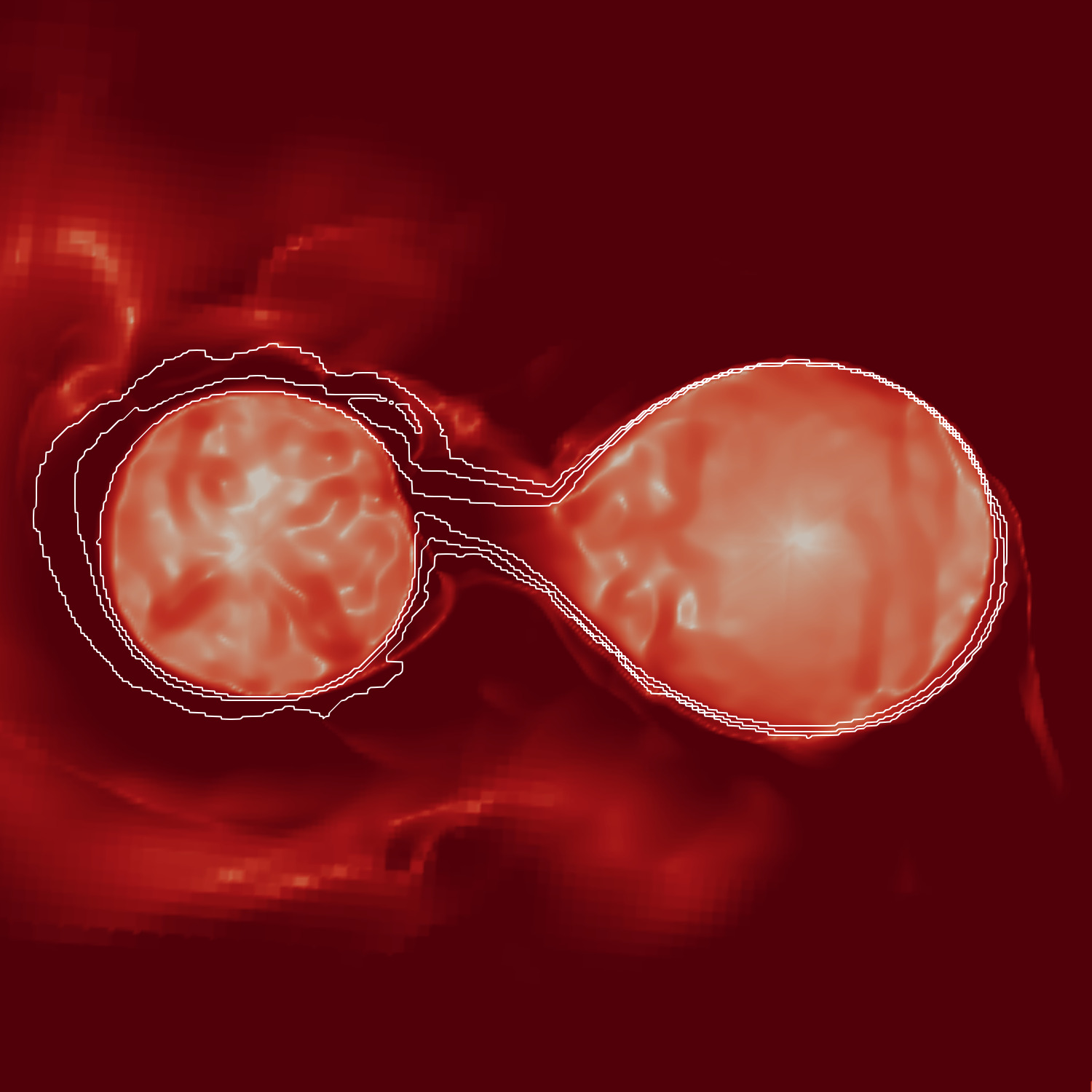}}%
    \\%
    \subfloat[\label{fig:EqSlice-velocity-rcb11}rcb 11, $t= 114.19 \seconds$]%
        {\includegraphics[width=0.49\linewidth,trim=0 450pt 0 450pt,clip]%
        {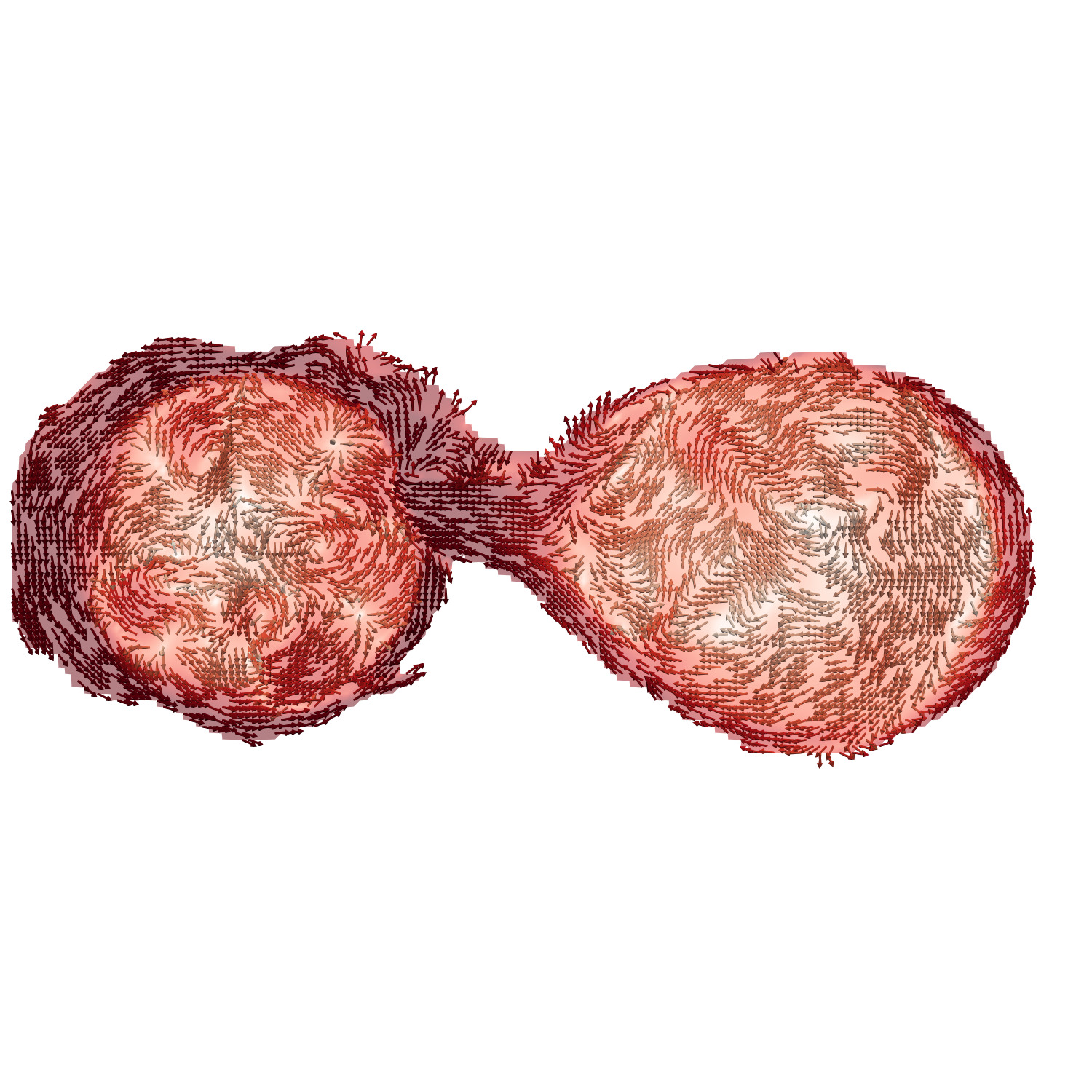}}%
    \hfill%
    \subfloat[\label{fig:EqSlice-velocity-rcb12}rcb 12, $t= 113.62 \seconds$]%
        {\includegraphics[width=0.49\linewidth,trim=0 450pt 0 450pt,clip]%
        {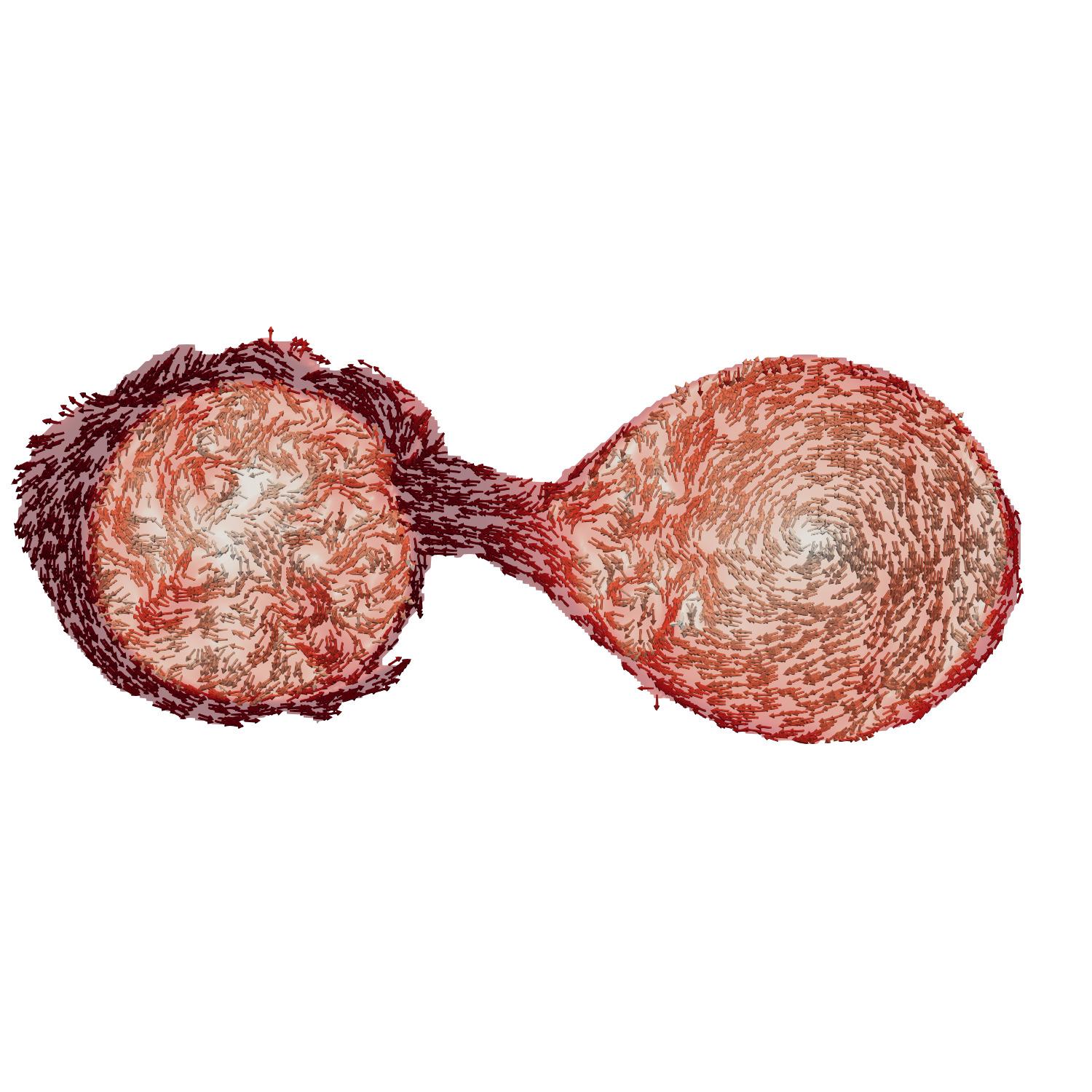}}%
    \\%
    \subfloat[\label{fig:EqSlice-lic-cutout-rcb11}rcb 11, $t= 114.19 \seconds$]%
        {\includegraphics[width=0.49\linewidth,trim=0 450pt 0 450pt,clip]%
        {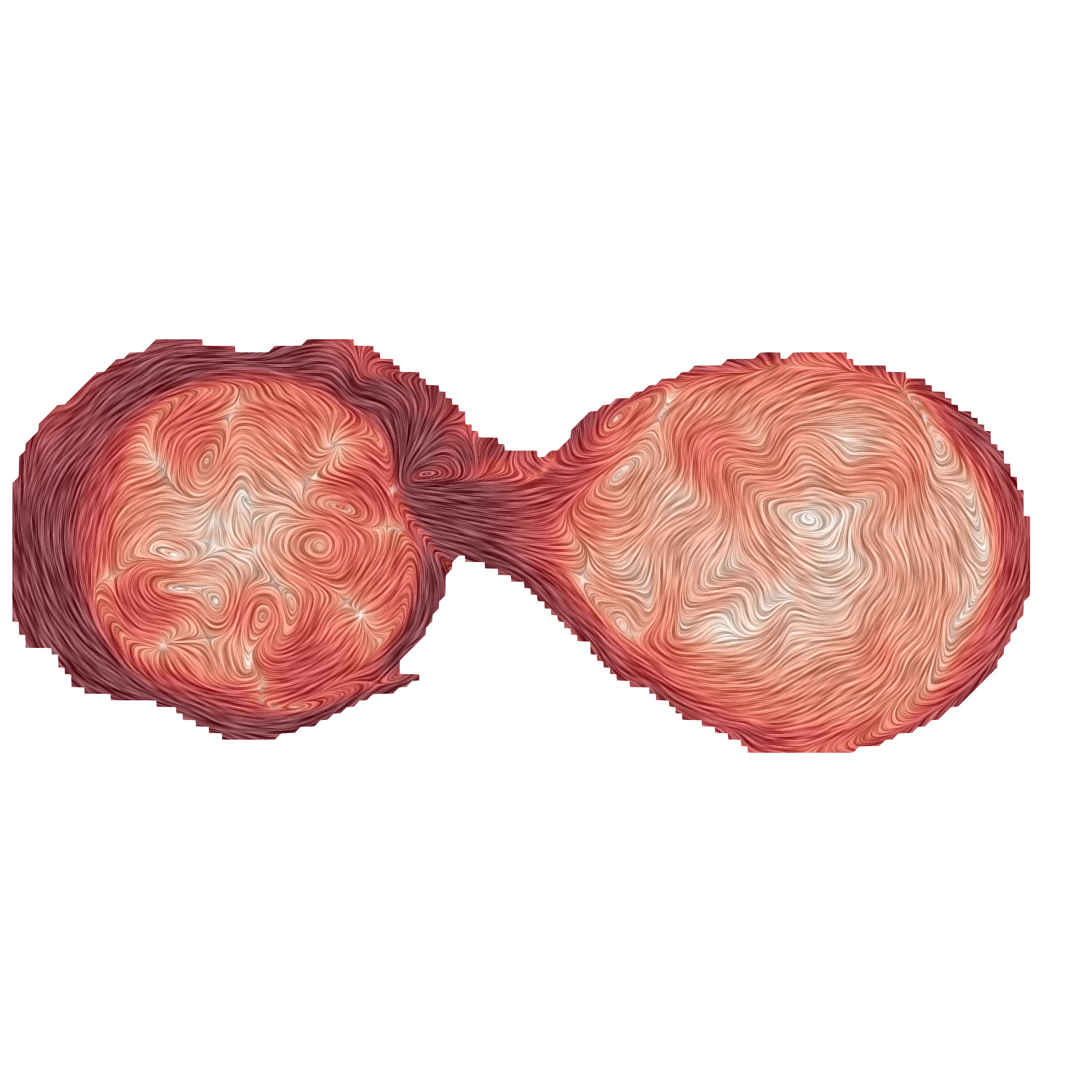}}%
    \hfill%
    \subfloat[\label{fig:EqSlice-lic-cutout-rcb12}rcb 12, $t= 113.62 \seconds$]%
        {\includegraphics[width=0.49\linewidth,trim=0 450pt 0 450pt,clip]%
        {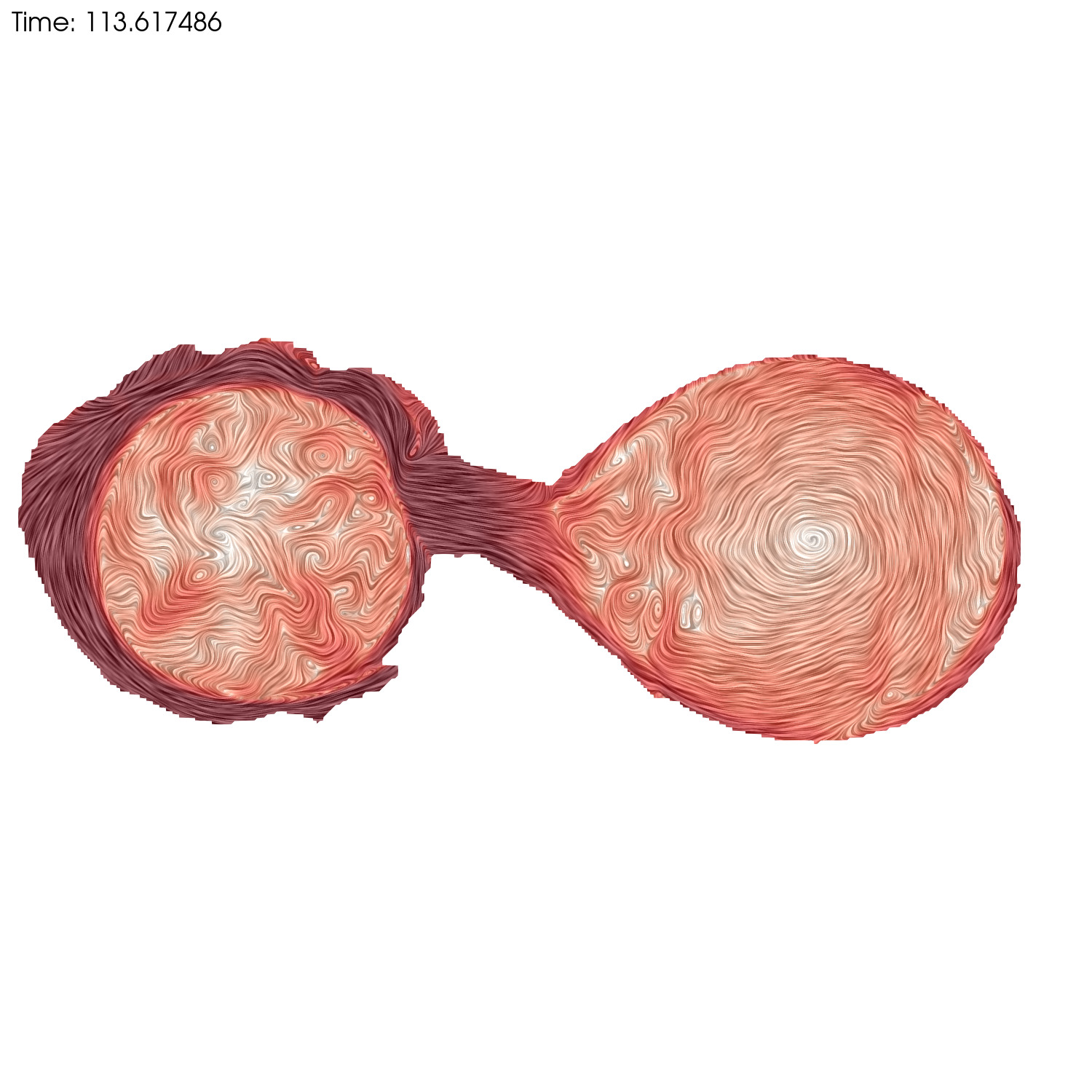}}%
    \caption{%
        Equatorial slices for rcb11 (left column) and rcb12 (right column) viewed from above ($z>0$). The position of the center of mass is the same as in~\autoref{fig:4views}. 
        We do not show it since it would obscure part of the flow near $L_1$.
        \protect\subref{fig:EqSlice-rcb11},\protect\subref{fig:EqSlice-rcb12}~Velocity magnitudes mapped to color.
        The white lines are (from inside to outside) the isosurfaces at $\rho=100, 10, \rm{and}\, 1 \gpccm$.
        \protect\subref{fig:EqSlice-velocity-rcb11},\protect\subref{fig:EqSlice-velocity-rcb12}~Velocities mapped to arrows and color for densities $\rho>1 \gpccm$.
        \protect\subref{fig:EqSlice-lic-cutout-rcb11},\protect\subref{fig:EqSlice-lic-cutout-rcb12}~LIC with velocity magnitudes mapped to color with a cutout at density $\rho<1 \gpccm$.
        Frame~\protect\subref{fig:EqSlice-lic-cutout-rcb11} shows the impact of the mass transfer stream on the accretor and the splash in both forward and backward directions. The downstream flow displays the characteristic undulating bore and trapping flow similar to the downstream waves in a waterfall.
        A few near stagnation regions can be seen on the accretor: near its center and the centers of two rolls, one below the point of impact and one as the accretion belt approaches the stream from behind.
        Velocity magnitudes are mapped logarithmically to color~(\includegraphics[width=6pt,height=32pt,angle=-90,origin=rB]{figures/legend_reds.jpeg}) in the range $[1 \cdot 10^5,~1 \cdot 10^8] \cmps$.
    }%
    \label{fig:EqSlice}%
\end{figure*}

We begin by examining \autoref{fig:EqSlice}, which shows equatorial slices through rcb11 (left column) and rcb12 (right column) at two times close to $1P_0$. The magnitude of the velocities of the fluid with respect to the corotating frame is displayed on the top row, whereas the glyphs showing the direction of the velocity vectors are plotted in the middle row. As before, the color scale indicates the magnitude of the velocity. As can be clearly seen, the high velocity features are confined to the mass transfer stream, its bounce off the accreting WD surface, and the accretion belt. In the bottom row, we employ LIC to display the flow patterns corresponding to the velocity fields in the equatorial plane. In the middle and bottom rows, we limit the display to regions with density $\rho> 1 \gpccm$. This allows us to see the impact of the stream, the bounce, the undulating bore downstream, and the trapping flow between the stream and the bounce.
These features are analogous to those commonly observed in the flow downstream of a waterfall. There is general agreement between rcb11 and rcb12, with the latter showing some finer details. Finally, we note that all the frames in \autoref{fig:EqSlice}
show light, near-white, stagnation regions near the centers of mass of both the donor and the accretor, as expected.

Focusing now on the donor star, the figures show that the internal motions in the equatorial plane of the donor are in the range of velocities from $1 \kmps$  to $100 \kmps$. 
By carefully examining the flow patterns revealed by LIC in the bottom row
(\autoref{sec:vis-isolic}), together with the glyphs shown in the middle row, it is possible to assign orientation to all flow features in the equatorial plane of the donor. Three regions of distinct flows are present: 1) most of the body of the donor gently rotates
CW spiraling outward, joining the stream at the trailing edge of the donor.
2) Near $L_1$, a transition region with a few eddies separates the internal flow from the stream. 3) At the far edge of the donor, as we approach $L_2$, an elongated banana-shaped eddy rotates CCW, and at later times feeds the flow that leaks over $L_2$. This flow pattern persists as the mass transfer increases right up to the tidal disruption of the donor. Most of the material in the mass transfer stream comes from the trailing side of the donor. The sense of the CW undulating flow spiraling away from the center of the donor is due to Coriolis forces. 

In addition, we provide links to two movies showing the velocity field on the equatorial/orbital plane 
for rcb11\footnote{Movie of simulation rcb11 can be obtained via \href{https://youtube.com/watch/NHWKZ6O-jL0?}{this link}}, and
rcb12\footnote{Movie of simulation rcb12 can be obtained via \href{https://youtube.com/watch/v7yBvVqzczY?}{this link}}. 

These movies
reveal how internal flows evolve from the onset of mass transfer to the tidal disruption of the donor.
Very early on, as already discussed, a pattern of flows appears in the donor, and a fast
equatorial accretion belt of donor material is distinguishable.
As evolution proceeds, the counterclockwise (CCW) accretion belt is gaining strength in mass and radial extent.
At approximately 1/4 of the time to merger
($t\approx 6P_0\approx 680 \seconds$ for rcb11, and
$t\approx 10P_0\approx 1140 \seconds$ for rcb12) 
the accretion belt is well defined. The hot spot where the stream impacts the accretor shows a splashing flow with some stream penetration, visible as a dimple near the point of impact. 
A large-scale layer of CCW internal flow develops in the accretor, adjacent to the accretion belt, driven by the radial gradient of the flow. A CW eddy is present near the foot of the stream impact. In the donor. a general outward flow from high pressure to lower pressure, is expected to develop a CW anti-cyclonic flow.
At about 1/3 of the total evolution, a strong CCW accretion belt is present, driving a growing CCW internal flow. A CW eddy is still present trailing the impact point.
Nearly half-way through merger, the accretion belt has become the dominant flow feature. The interior flow in the accretor is also mostly CCW with some very weak CW eddies present. On the donor, the flow is evolving toward an overall CW flow with one or two weak CCW eddies. 

Below we list times for selected stages of both simulations. The approximate times cited for each stage correspond to simulation rcb11, followed in parentheses by the time observed in simulation rcb12 for the same phase.
The accretion belt is completed, fully surrounding the accretor at $t_{\rm ab}\approx 173 {\rm \seconds}\approx 1.5 P_0$
($t_{\rm ab}\approx 182 {\rm \seconds}\approx 1.6 P_0$).
After some time, $t_1\approx 2035 {\rm \seconds}\approx 17.2 P_0$ ($t_1\approx 2275 {\rm \seconds}\approx 20. P_0$), the interior flow settles to a relatively stable configuration that lasts for several orbital periods $5.4 P_0$ ($7.4 P_0$). During this quasi-steady phase the binary is in overcontact, the density contour $\rho = 10 \gpccm$ extends beyond the Roche lobes, with the ``neck" containing the mass transfer stream having a thickness on the order of the stellar radius of the components.

The fluid near the centers of both components appears to turn slowly CW as it lags behind the orbital rotation.
In the comoving frame, a closed corotation region in the accretor separates the accretion belt from the interior. In the donor the flow in the CW interior is accelerating and spiraling out to larger radii to join the mass transfer stream. The corotation regions are arcs with a flow that turns around near the ends.
 Finally, the flow in the outermost layers in both components is highly variable, generally in the CCW direction. Near the far-side of the donor, in the $L_2$ region, a fraction of the CCW flow in the donor spills out of the binary.

The quasi-steady phase comes to an end at $t_2 \approx 2648 {\rm \seconds}\approx 23 P_0$
($t_2 \approx 3115 {\rm \seconds}\approx 27.4 P_0$)
as the outermost layers become wobbly and unstable. As time passes, the rate of flow through $L_2$ increases, leading to the tidal disruption of the donor and the merger at
$t_{\rm merge} \approx 2787 {\rm \seconds}\approx 24.5 P_0$
($t_{\rm merge} \approx 4400 {\rm \seconds}\approx 38.7 P_0$).

\begin{figure*}
    \centering%
    \subfloat[\label{fig:EqSlice-rcb12-disruption-pre}rcb 12, $t= 4306.21 \seconds$]%
        {\includegraphics[width=0.49\linewidth]%
        {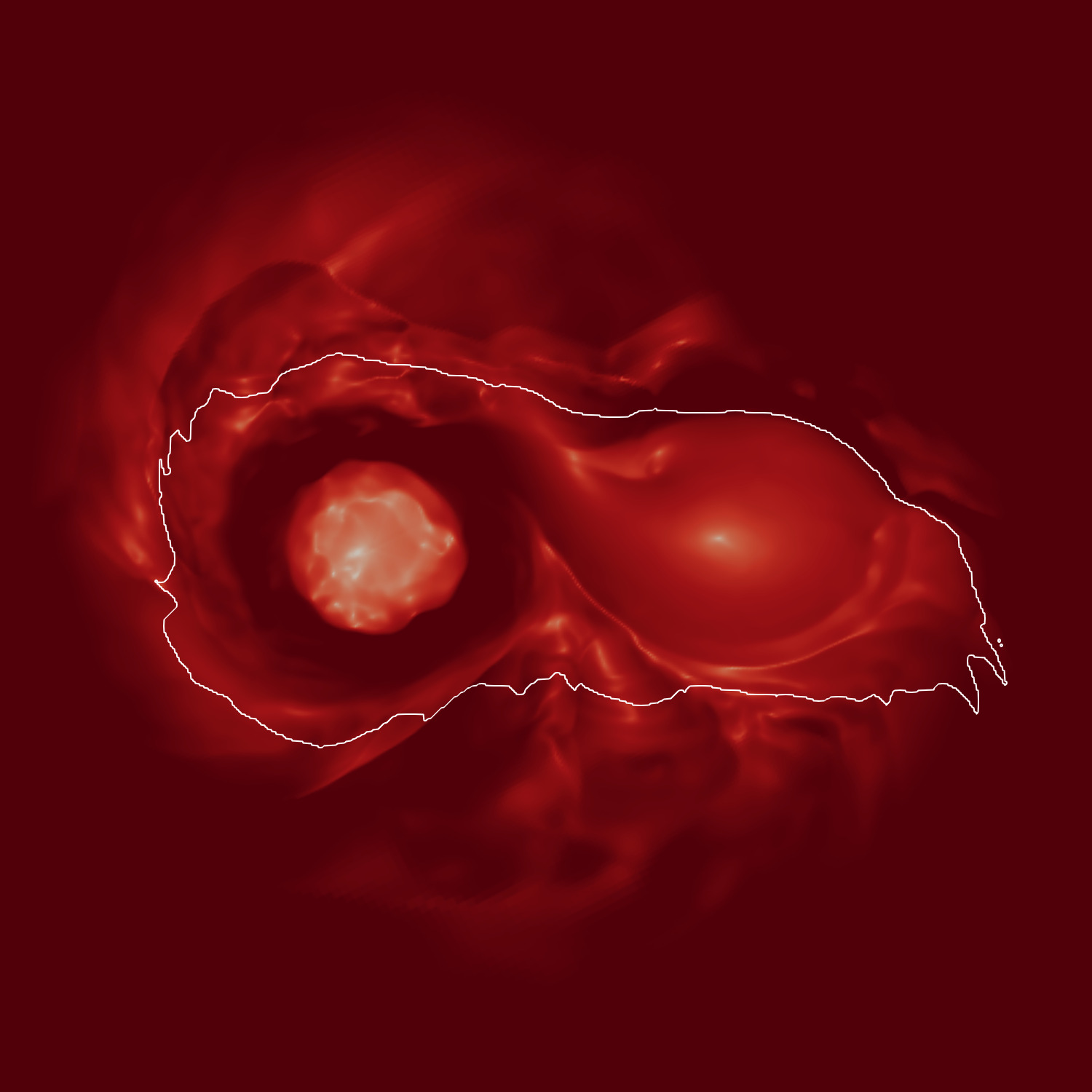}}%
    \hfill%
    \subfloat[\label{fig:EqSlice-rcb12-disruption-post}rcb 12, $t= 4419.79 \seconds$]%
        {\includegraphics[width=0.49\linewidth]%
        {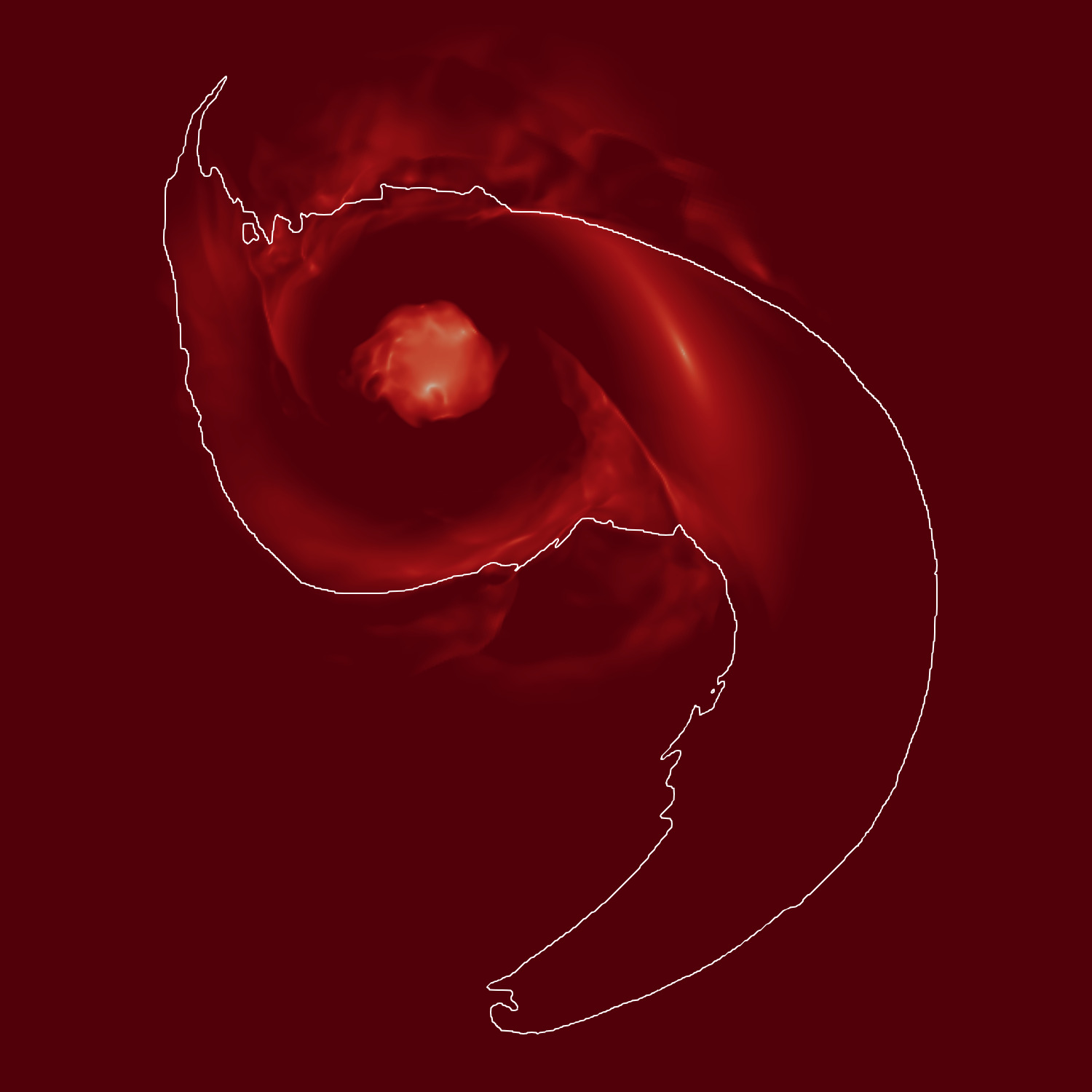}}%
    \\[-1.5em]%
    \subfloat[\label{fig:EqSlice-rcb12-disruption-velocity-pre}rcb 12, $t= 4306.21 \seconds$]%
        {\includegraphics[width=0.49\linewidth]%
        {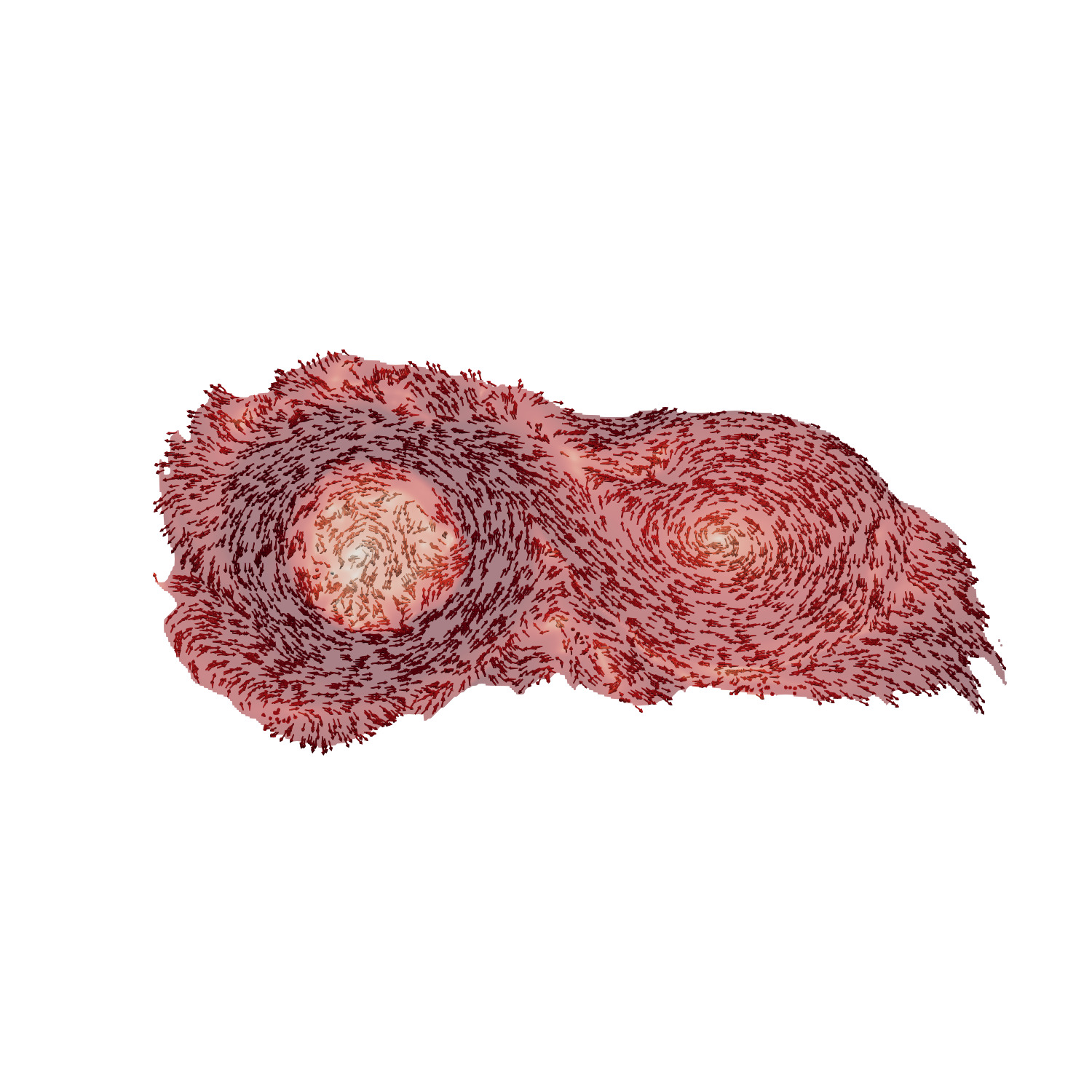}}%
    \hfill%
    \subfloat[\label{fig:EqSlice-rcb12-disruption-velocity-post}rcb 12, $t= 4419.79 \seconds$]%
        {\includegraphics[width=0.49\linewidth]%
        {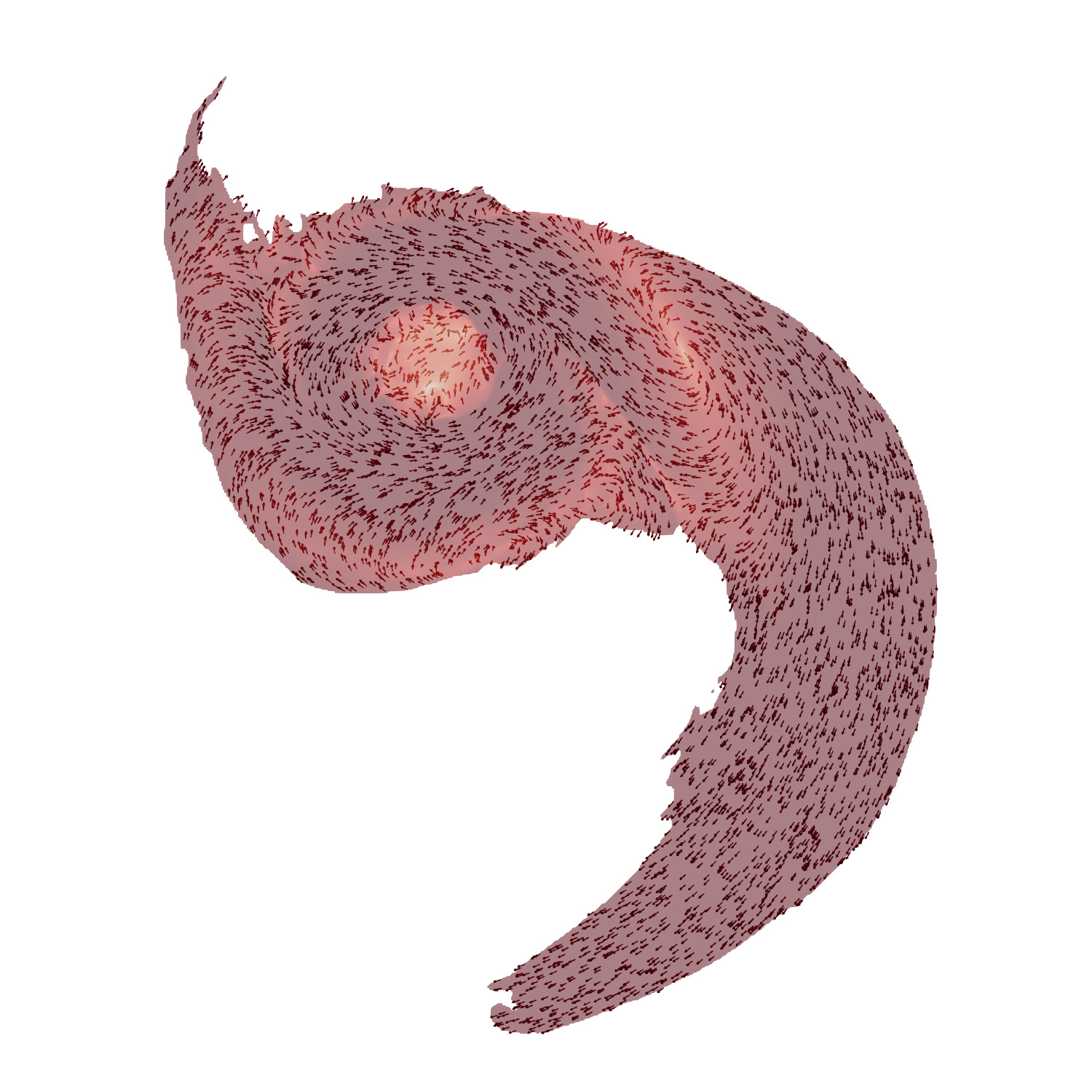}}%
    \caption{%
        Equatorial slices viewed from above ($z>0$).
        \protect\subref{fig:EqSlice-rcb12-disruption-pre},\protect\subref{fig:EqSlice-rcb12-disruption-post}~Velocity magnitudes mapped to color.
        The white line is the isosurface at $\rho=100 \gpccm$.
        \protect\subref{fig:EqSlice-rcb12-disruption-velocity-pre},\protect\subref{fig:EqSlice-rcb12-disruption-velocity-post}~Velocities mapped to arrows and color for densities $\rho>100 \gpccm$ at later times.
        A few near stagnation regions can be seen on the accretor: near its center and the centers of two rolls, one below the point of impact and one as the accretion belt approaches the stream from behind.
        Velocity magnitudes are mapped logarithmically to color~(\includegraphics[width=6pt,height=32pt,angle=-90,origin=rB]{figures/legend_reds.jpeg}) in the range $[1 \cdot 10^5,~1 \cdot 10^8] \cmps$.
    }
    \label{fig:EqSlice-rcb12-disruption}
\end{figure*}

\autoref{fig:EqSlice-rcb12-disruption} shows snapshots just before and just after tidal disruption for rcb12. These figures show clearly the fastest motions occurring in the accretion belt and the stream near the point of impact. The interior of the donor is unrolling CW as it disrupts, while the cores of both accretor and donor are still relatively undisturbed. Just after the disruption one can barely see the remnant donor lobe, and the trailing tidal tail extending outward and backward beyond what was the $L_2$ point. 

\begin{figure*}%
    \centering%
    \subfloat[\label{fig:merid1-rcb11-0}$t= 0.0 \seconds$]%
        {\includegraphics[width=0.245\linewidth,trim=0 600pt 0 600pt,clip]%
        {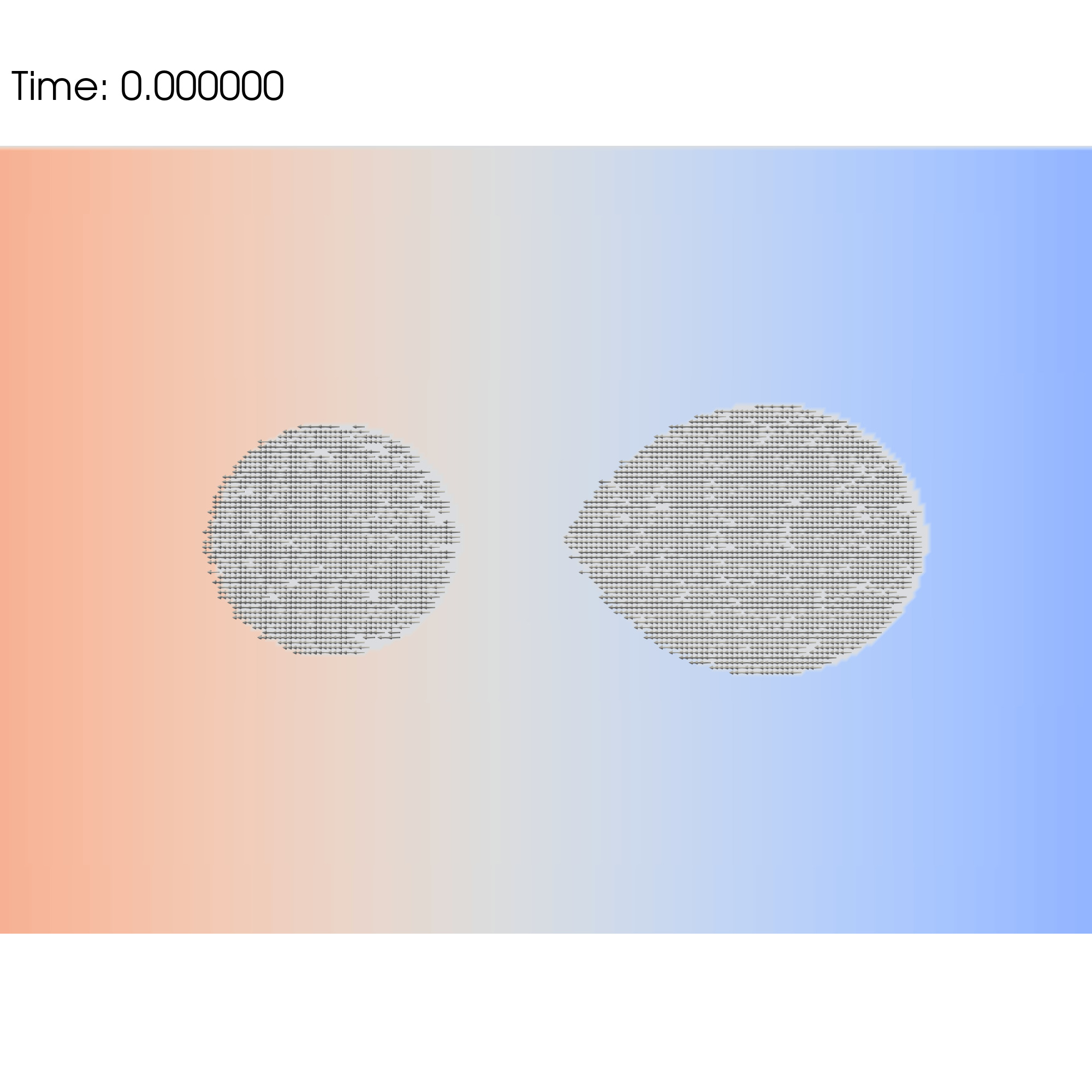}}%
    \hfill%
    \subfloat[\label{fig:merid1-rcb11-10}$t= 45.67 \seconds$]%
        {\includegraphics[width=0.245\linewidth,trim=0 600pt 0 600pt,clip]%
        {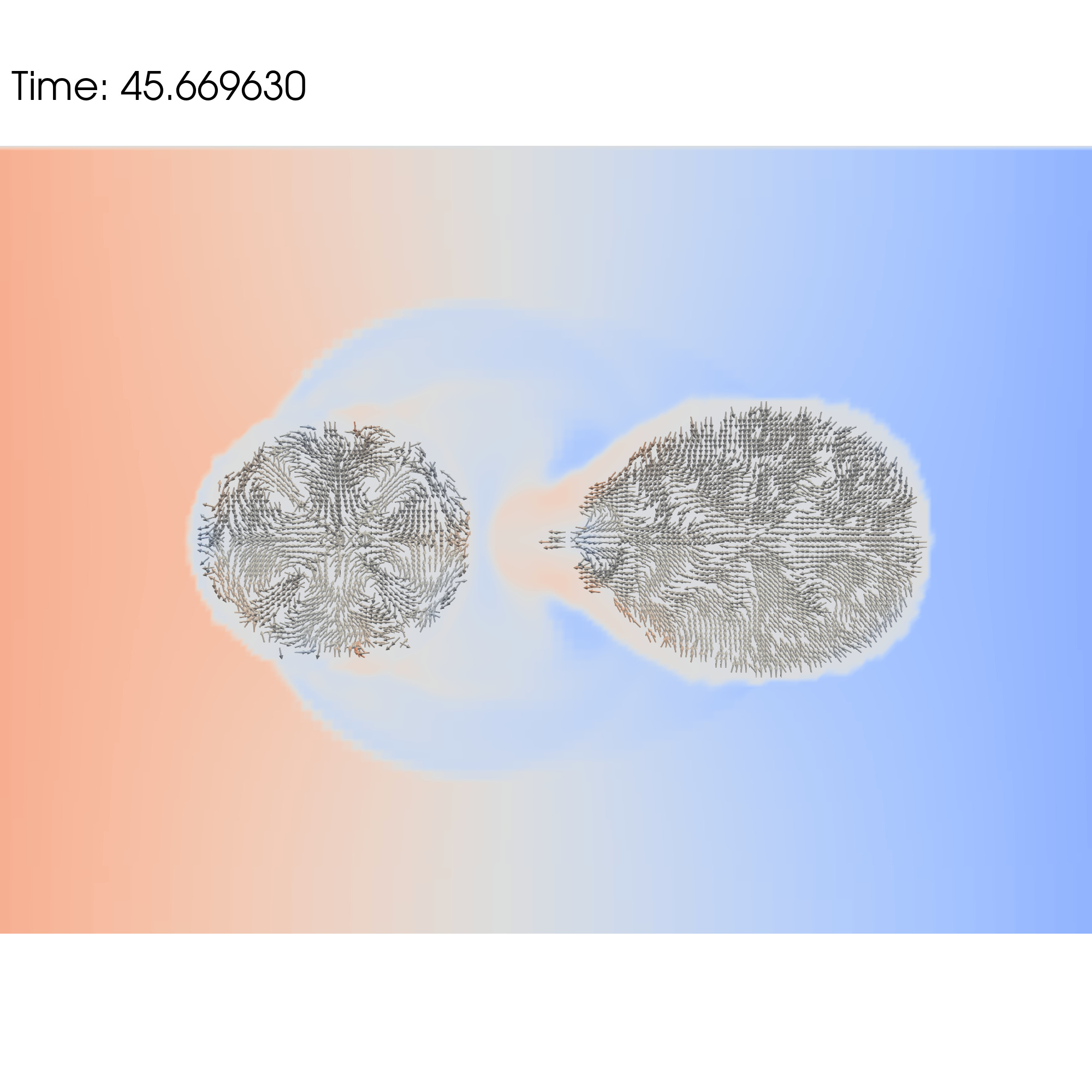}}%
    \hfill%
    \subfloat[\label{fig:merid1-rcb11-113}$t= 515.57 \seconds$]%
        {\includegraphics[width=0.245\linewidth,trim=0 600pt 0 600pt,clip]%
        {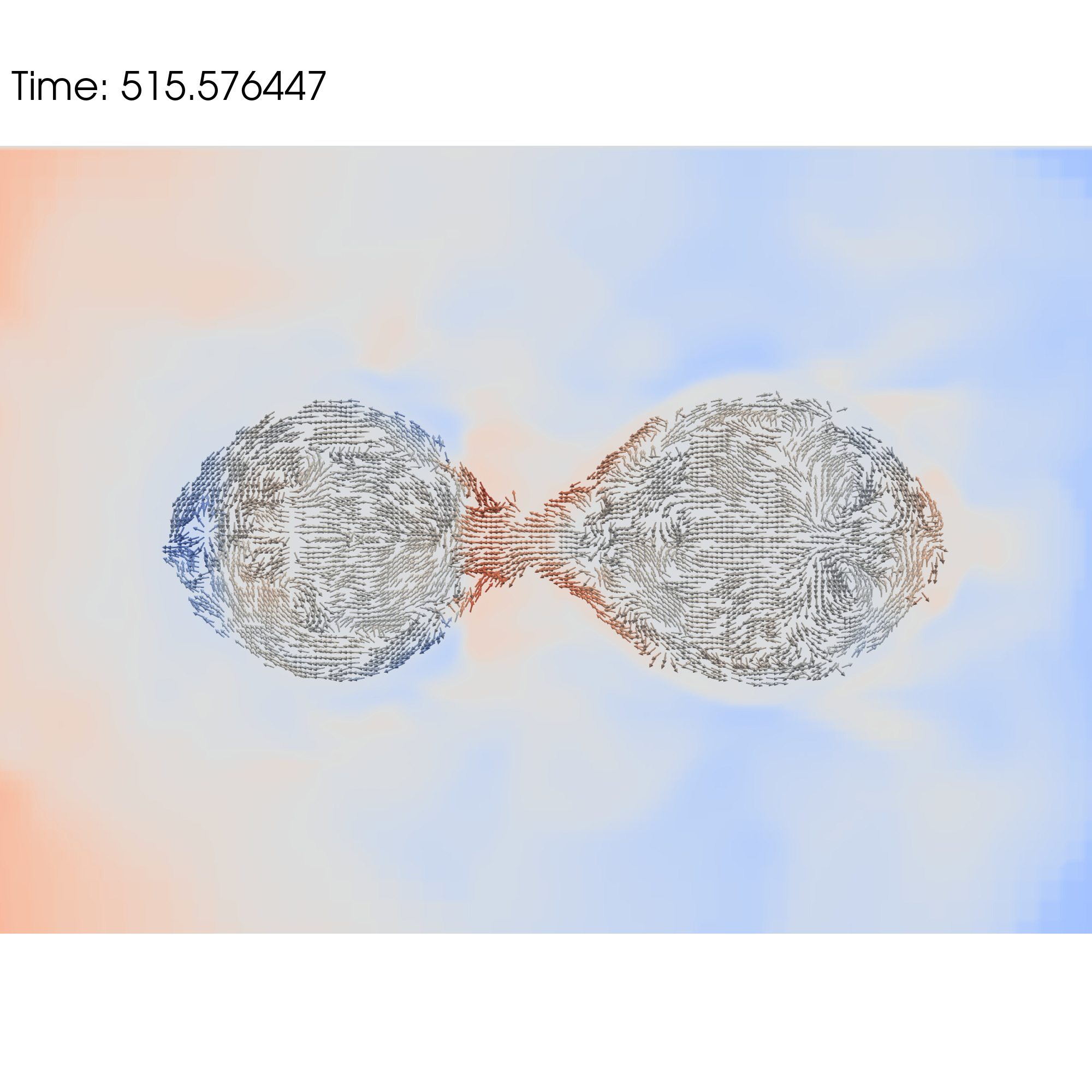}}%
    \hfill%
    \subfloat[\label{fig:merid1-rcb11-288}$t= 1313.84 \seconds$]%
        {\includegraphics[width=0.245\linewidth,trim=0 600pt 0 600pt,clip]%
        {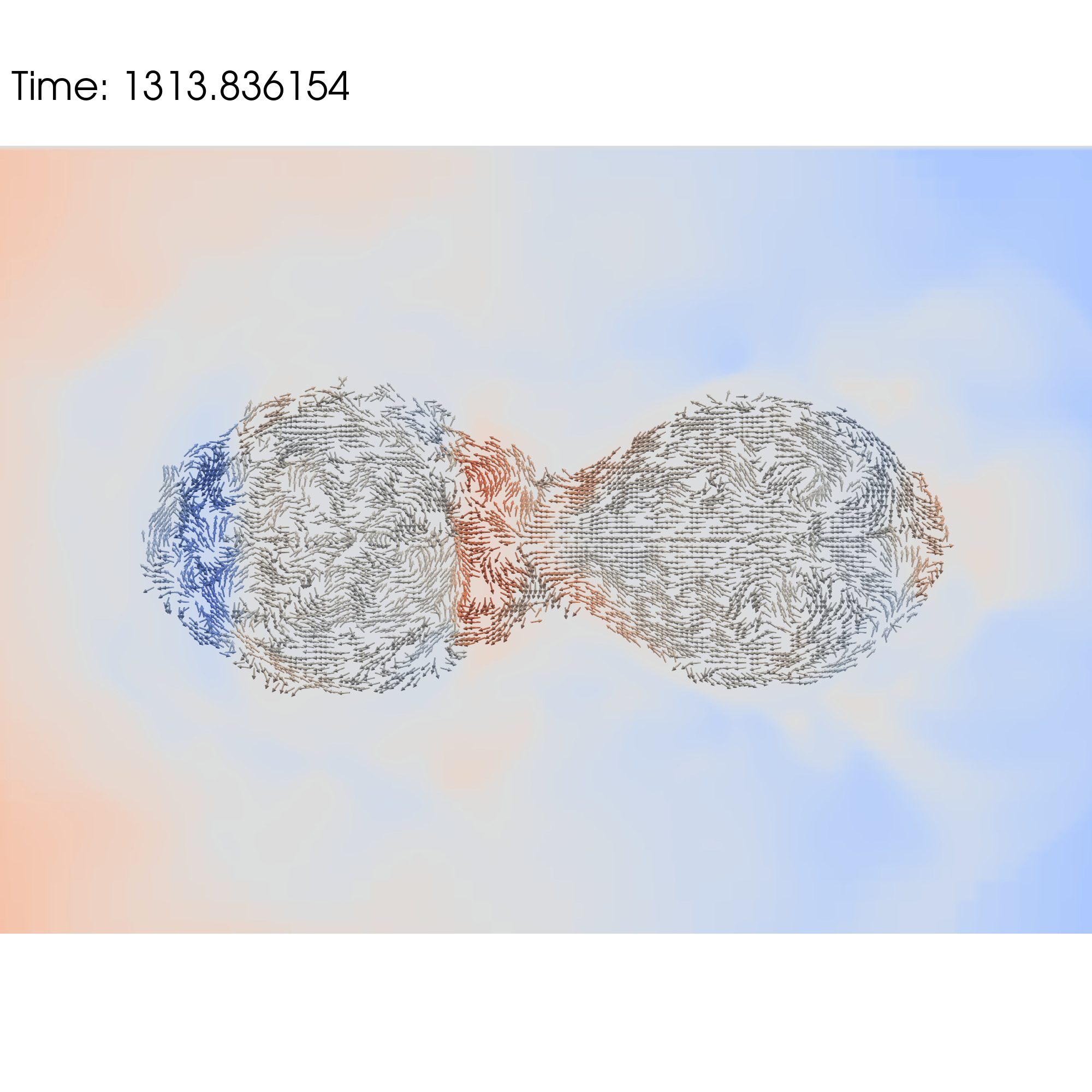}}%
    \\%
    \subfloat[\label{fig:merid1-rcb11-570}$t= 2600.12 \seconds$]%
        {\includegraphics[width=0.245\linewidth,trim=0 400pt 0 400pt,clip]%
        {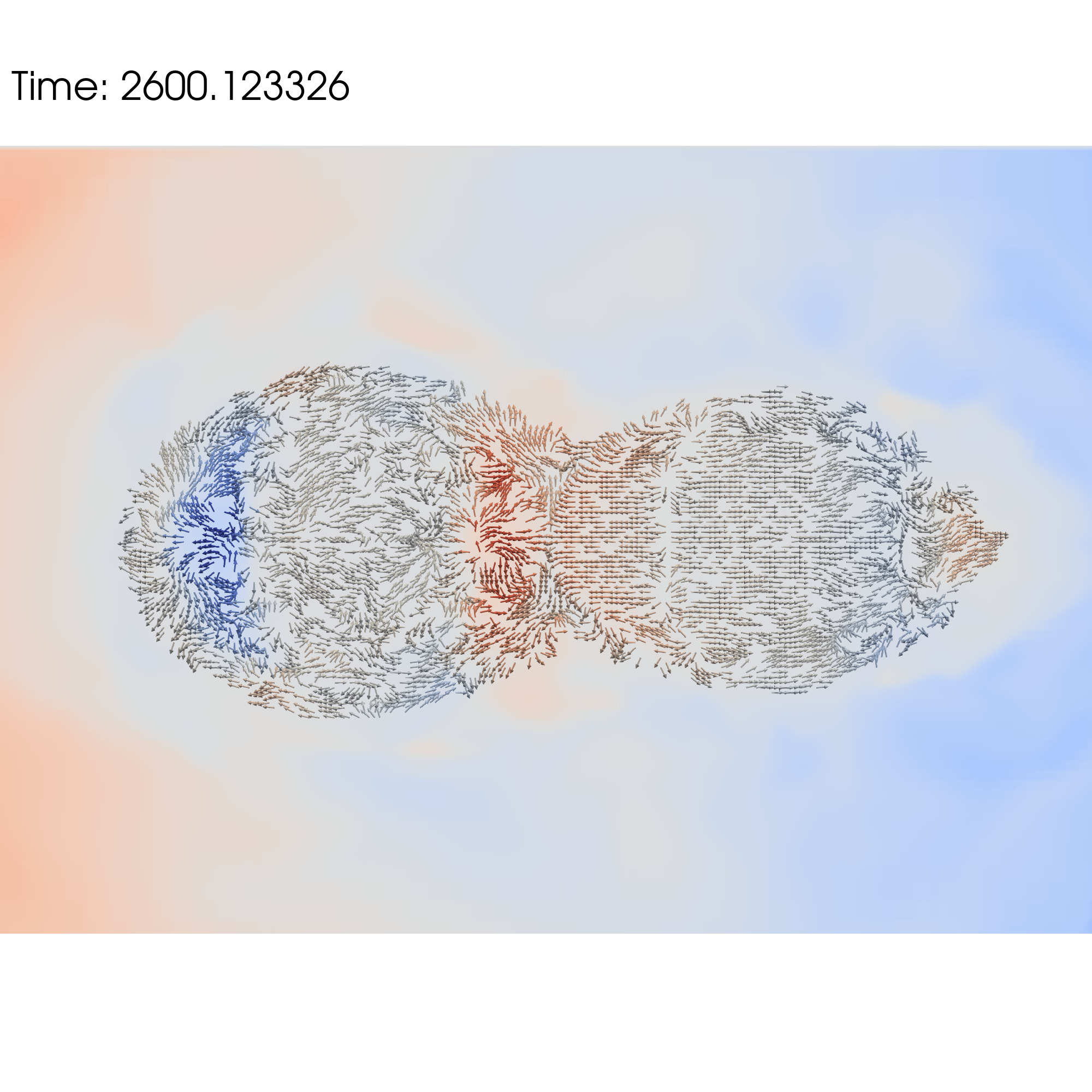}}%
    \hfill%
    \subfloat[\label{fig:merid1-rcb11-645}$t= 2942.28 \seconds$]%
        {\includegraphics[width=0.245\linewidth,trim=0 400pt 0 400pt,clip]%
        {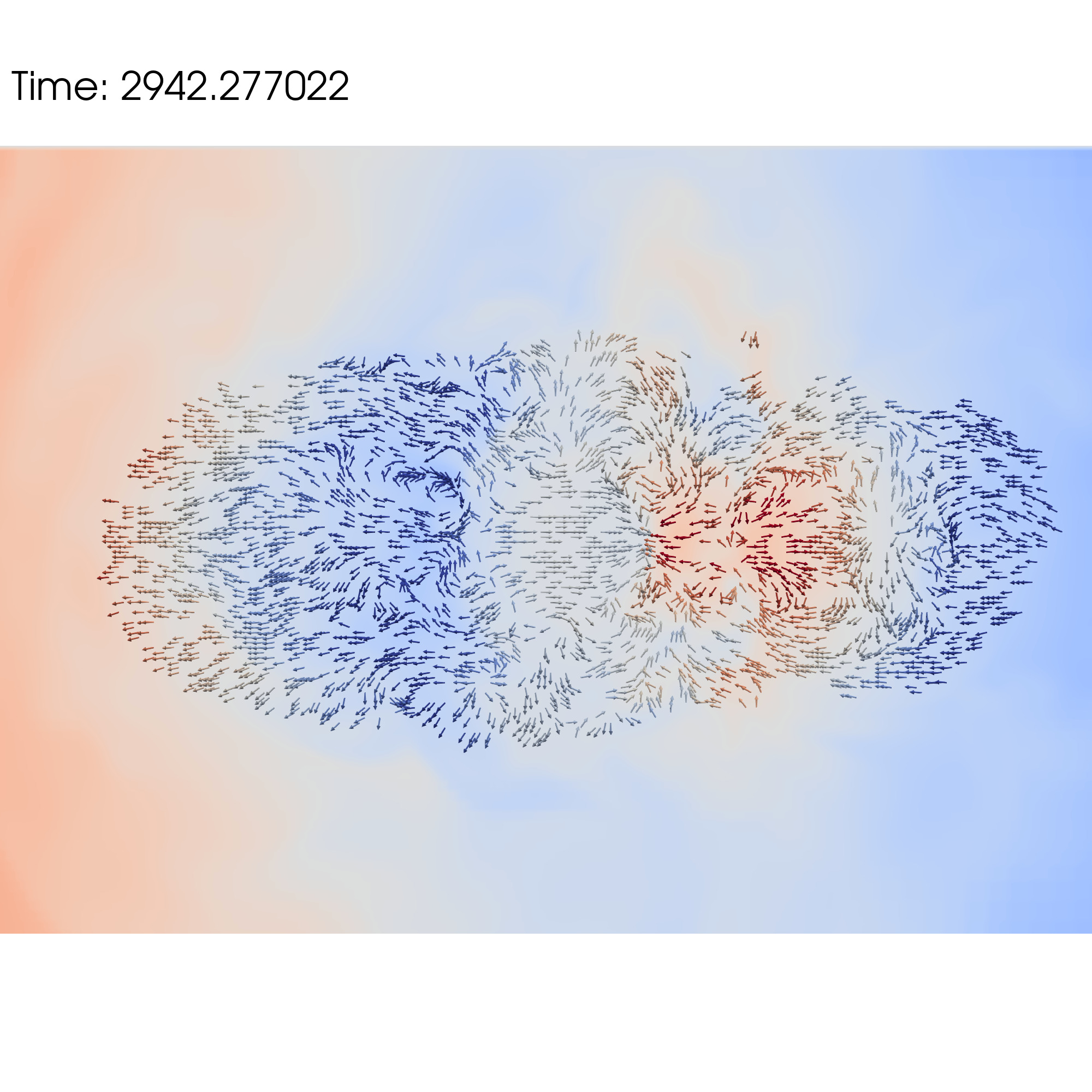}}%
    \hfill%
    \subfloat[\label{fig:merid1-rcb11-767}$t= 3498.74 \seconds$]%
        {\includegraphics[width=0.245\linewidth,trim=0 400pt 0 400pt,clip]%
        {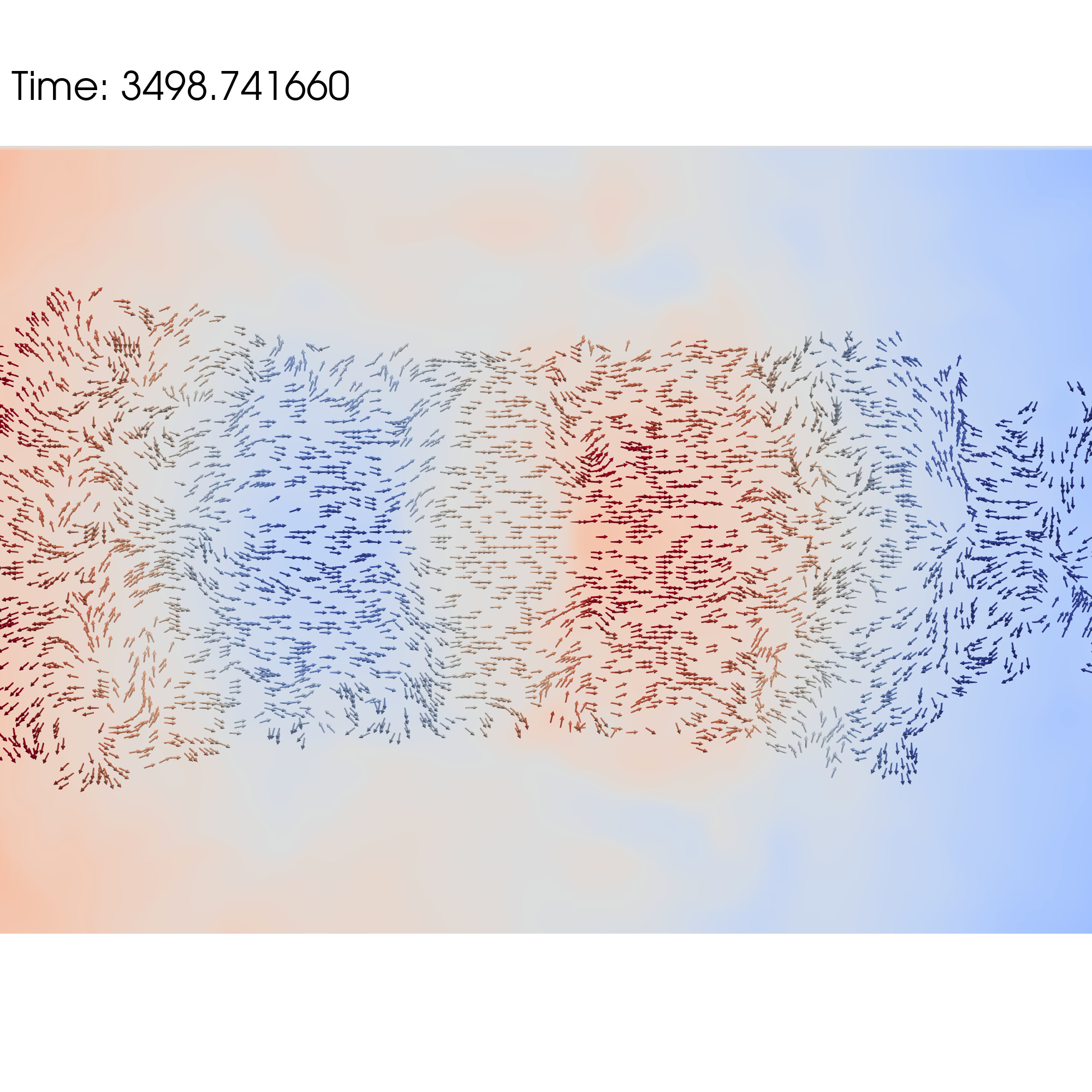}}%
    \hfill%
    \subfloat[\label{fig:merid1-rcb11-971}$t= 4429.30 \seconds$]%
        {\includegraphics[width=0.245\linewidth,trim=0 400pt 0 400pt,clip]%
        {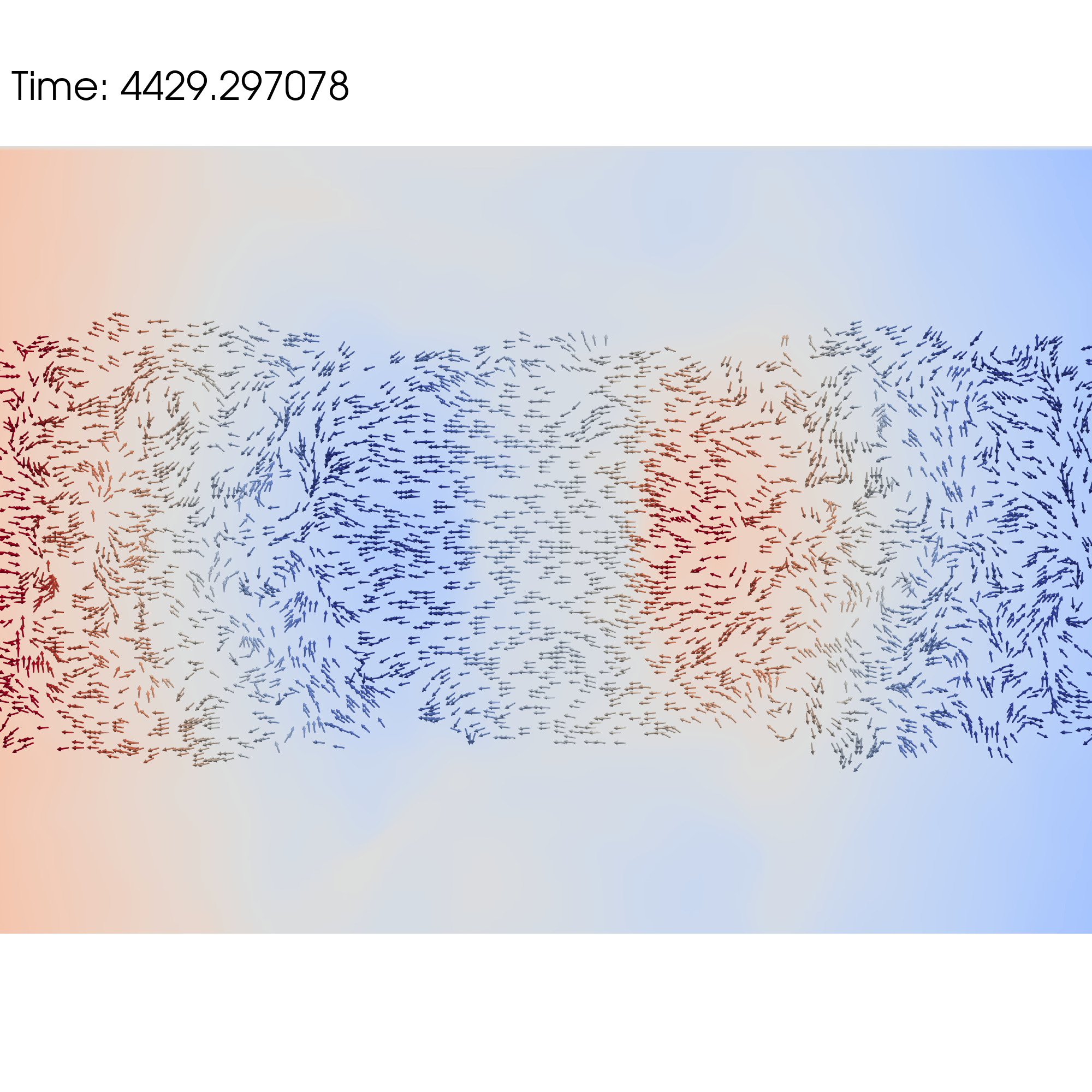}}%
    \caption{%
        Flow on the principal meridional plane $(x,z)$ at different times for rcb11, as seen
        from a point
        $(0,y<0,0)$.
        Only the direction 
        of the meridional velocity $(v_x, v_z)$ is shown.
        The value of the $v_y$ velocity is mapped linearly to color~(\includegraphics[width=6pt,height=32pt,angle=-90,origin=rB]{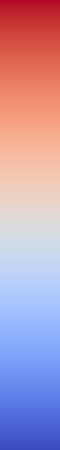}) in the range $[-5.7 \cdot 10^8,~5.7 \cdot 10^8] \cmps$. The position of the center of mass of the system is at the center of the frames. Careful examination of the post-merger frames in the lower row, shows the merged object spinning counterclockwise, while orbiting slowly slightly off the center of mass.
    }%
    \label{fig:merid1}%
\end{figure*}

To complete the description of internal flows in the merging binary, in \autoref{fig:merid1} we present several views of the meridional cross section through the principal axis, \emph{i.e.}, the $(x,z)$ plane viewed along the $y$-axis, from $y<0$ (sideways view). 
The display is limited to densities $\rho>100\gpccm$. For clarity, the arrows only indicate the {\em direction} of the $(v_x, v_z)$ velocity.
The value of the $v_y$ component along the viewer's line of sight is encoded by color from blue ($v_y<0$) to red ($v_y>0$).
The more intense the color, the greater the magnitude of the $v_y$-velocity. A gray-white color indicates $v_y\approx 0$, and indicates a region of corotation with the observer's frame. 

\autoref{fig:merid1-rcb11-0} shows the initial situation: Since the velocities are measured in the corotating frame, there is no internal flow, while the external medium, stationary in the inertial frame, is in solid-body rotation with maximum positive $v_y$ on the left edge and maximum negative $v_y$ on the right edge. In this and all subsequent frames, the $(v_x, v_z)$ patterns are symmetric above and below the orbital plane $z=0$. Note that only the directions of the
$(v_x, v_z)$ velocities are shown with glyphs. During the evolution the magnitudes of the meridional motions remain small $v_{xz}\leq 50 \kmps$, while $|v_y|\leq 5700 \kmps$. 

\autoref{fig:merid1-rcb11-10} shows that
soon after the start of the simulation,
a layer of the external medium around the donor is being dragged along, and similarly the accretor has stirred the surrounding layers into a motion approaching the viewer. Note also the motion $v_y>0$ near $L_1$ at the base of the mass transfer stream. 

\autoref{fig:merid1-rcb11-113} and \ref{fig:merid1-rcb11-288} show the strengthening of the stream and the accretion belt. The growth of the CW circulation inside the donor is evidenced by the development of a $v_y<0$ region on its right side.

Once the stream and the equatorial belt are well developed, we see in \autoref{fig:merid1-rcb11-570} the high blue-shifted motion on the equatorial accretion belt on the left edge of the accretor, and the strong $v_y>0$ motions near $L_1$ and the accretion belt impacting the stream and overflowing it above and below. 

There is already a hint of positive motion $v_y$ on the far edge of the donor, near the $L_2$ point. This is some material that has gone around the donor and is moving faster than the body of the donor. Note also how the surface layers of the donor and the base of the stream have acquired $v_y>0$ as the gas approaches $L_1$. 

Some time later, as the disruption of the donor is under way, in \autoref{fig:merid1-rcb11-645} the accretion belt has widened to higher latitudes, as can be seen both left and right of the accretor, and the body of the donor moves with $v_y>0$ as it is being pulled apart. 

Finally, \autoref{fig:merid1-rcb11-767} and \ref{fig:merid1-rcb11-971} show the meridional flow at late times, after the merger.
Before $t_{\rm merge}$ the observer rotates with the binary orbital period. After the merger, the observer maintains the rotation rate at the time of the merger, so the material beyond the apparent corotation radius will be rotating slower. Thus, from the point of view of the observer, the outer material will appear to approach (blue-shifted) on the right side and to recede (red-shifted) on the left side.

\section{Analysis of the Mass Transfer Flow and Merger}
\label{sec:analysis}

\begin{figure*}
    \centering
    \subfloat[\label{fig:ftle-1072-slice}]{%
        \begin{tikzpicture}%
            \node[anchor=south west, inner sep=0] at (0,0)%
                {\includegraphics[width=0.49\linewidth,trim=50pt 400pt 50pt 400pt,clip]%
                {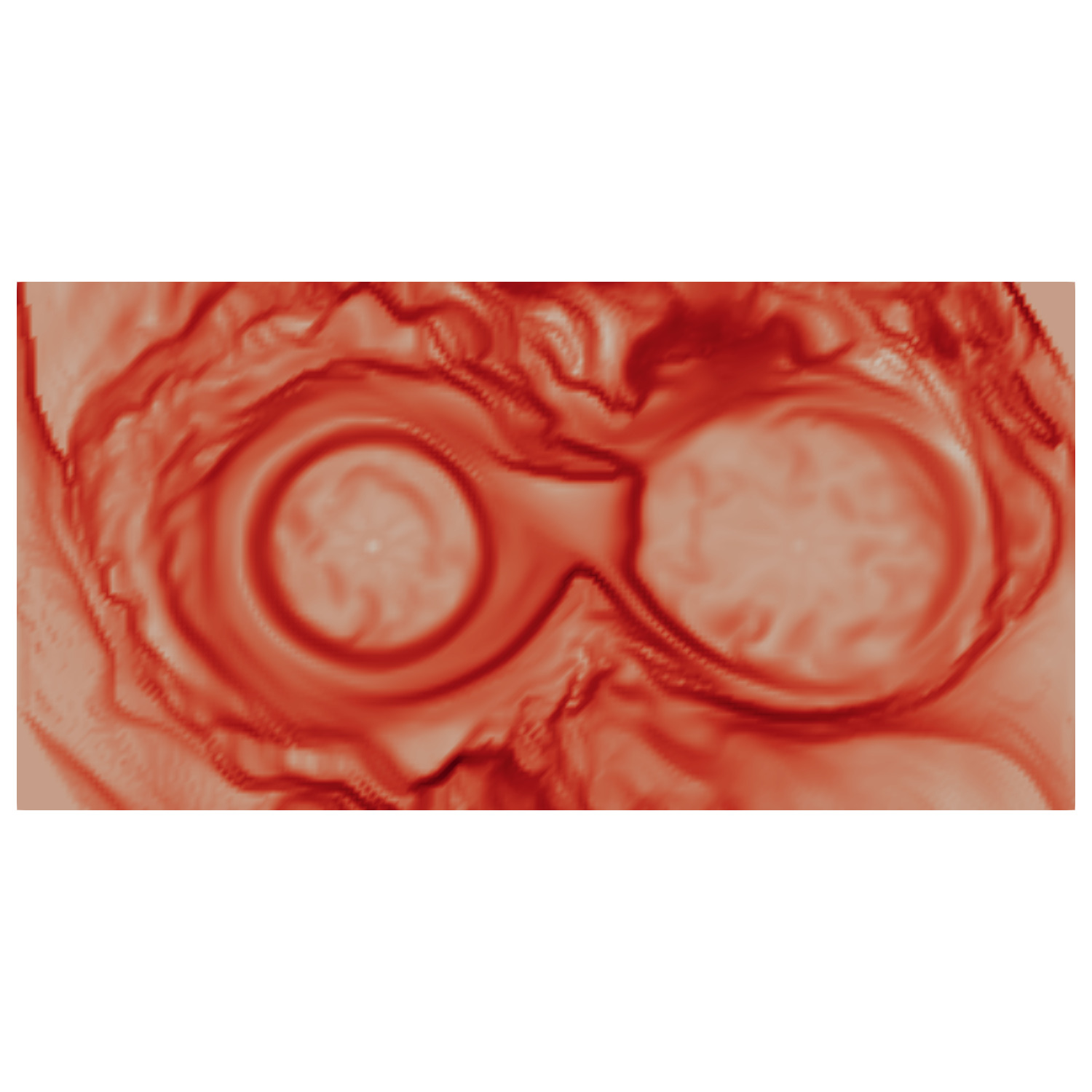}};%
            \draw[white,draw opacity=0.0,text opacity=1.0,fill=black,fill opacity=0.5] (5.40,2.25) circle (0.25) node {$\mathbf{a}$};%
            \draw[white,draw opacity=0.0,text opacity=1.0,fill=black,fill opacity=0.5] (4.95,3.23) circle (0.25) node {$\mathbf{a_1}$};%
            \draw[white,draw opacity=0.0,text opacity=1.0,fill=black,fill opacity=0.5] (4.50,2.55) circle (0.25) node {$\mathbf{a_2}$};%
            \draw[white,draw opacity=0.0,text opacity=1.0,fill=black,fill opacity=0.5] (6.05,1.45) circle (0.25) node {$\mathbf{a_0}$};%
            \draw[white,draw opacity=0.0,text opacity=1.0,fill=black,fill opacity=0.5] (2.90,2.85) circle (0.25) node {$\mathbf{b}$};%
            \draw[white,draw opacity=0.0,text opacity=1.0,fill=black,fill opacity=0.5] (4.13,1.65) circle (0.25) node {$\mathbf{c}$};%
            \draw[white,draw opacity=0.0,text opacity=1.0,fill=black,fill opacity=0.5] (4.85,1.50) circle (0.25) node {$\mathbf{c_0}$};%
            \draw[white,draw opacity=0.0,text opacity=1.0,fill=black,fill opacity=0.5] (6.70,0.50) circle (0.25) node {$\mathbf{d}$};%
        \end{tikzpicture}%
    }%
    \hfill%
    \subfloat[\label{fig:ftle-1072-slice-reverse}]{%
        \begin{tikzpicture}%
            \node[anchor=south west, inner sep=0] at (0,0)%
                {\includegraphics[width=0.49\linewidth,trim=50pt 400pt 50pt 400pt,clip]%
                {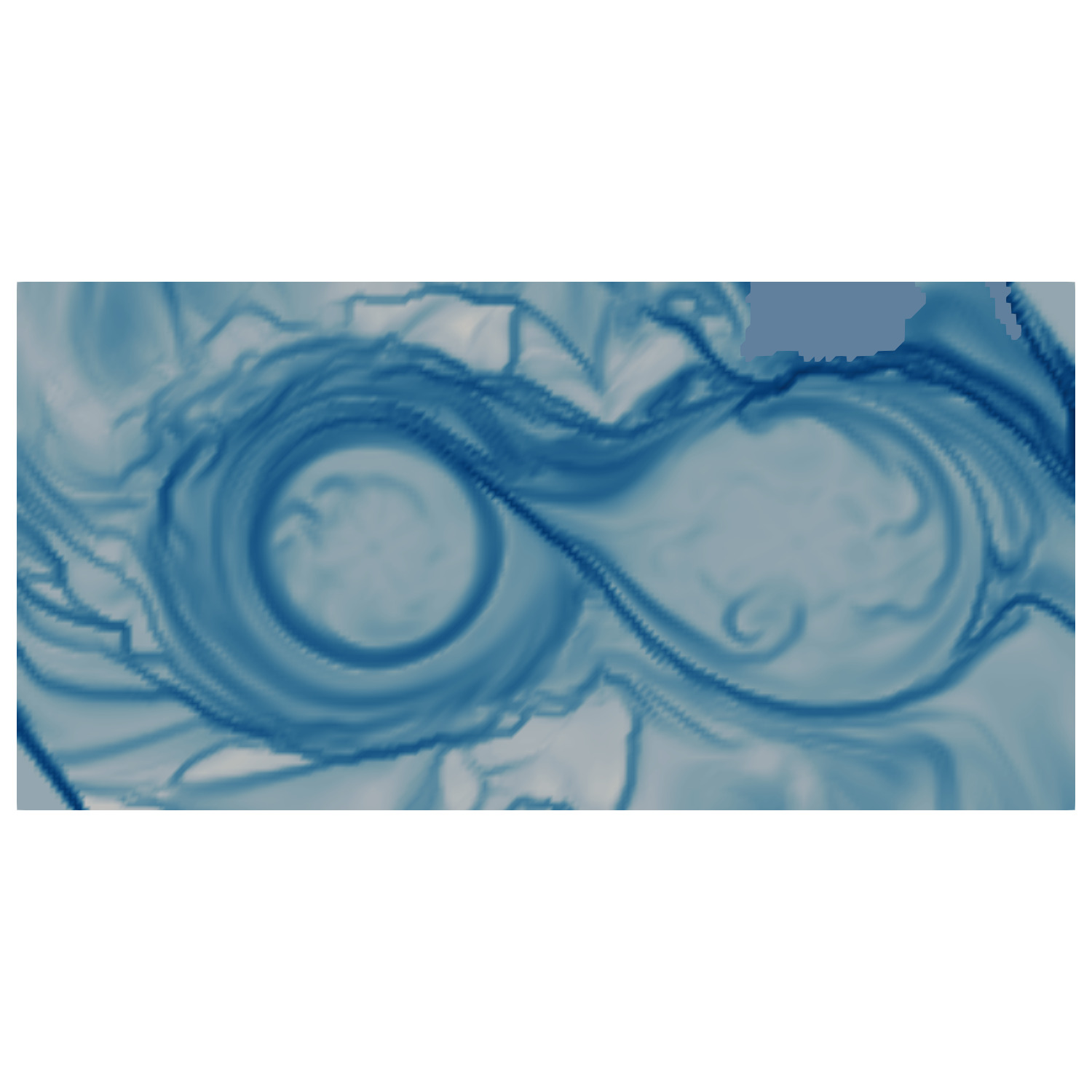}};%
            \draw[white,draw opacity=0.0,text opacity=1.0,fill=black,fill opacity=0.5] (4.90,2.30) circle (0.25) node {$\mathbfit{a}$};%
            \draw[white,draw opacity=0.0,text opacity=1.0,fill=black,fill opacity=0.5] (7.90,3.15) circle (0.25) node {$\mathbfit{d}$};%
            \draw[white,draw opacity=0.0,text opacity=1.0,fill=black,fill opacity=0.5] (2.90,2.85) circle (0.25) node {$\mathbfit{b}$};%
            \draw[white,draw opacity=0.0,text opacity=1.0,fill=black,fill opacity=0.5] (6.70,1.10) circle (0.25) node {$\mathbfit{c}$};%
        \end{tikzpicture}%
    }%
    \\%
    \subfloat[\label{fig:ftle-1072-pathlines}]%
        {\includegraphics[width=0.49\linewidth,trim=50pt 400pt 50pt 400pt,clip]%
        {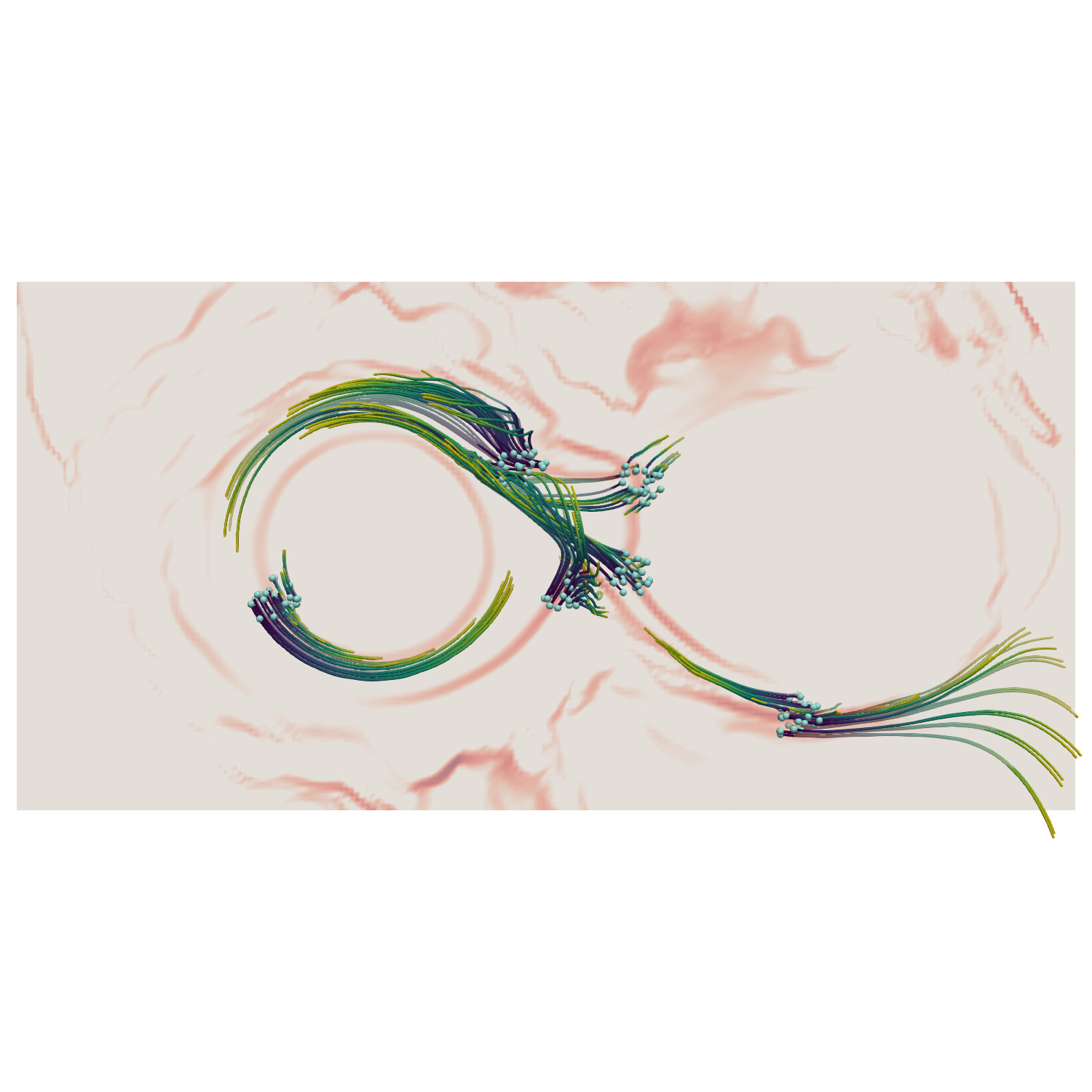}}%
    \\%
    \subfloat[\label{fig:ftle-1072-ridges}]%
        {\includegraphics[width=0.49\linewidth,trim=50pt 300pt 50pt 100pt,clip]%
        {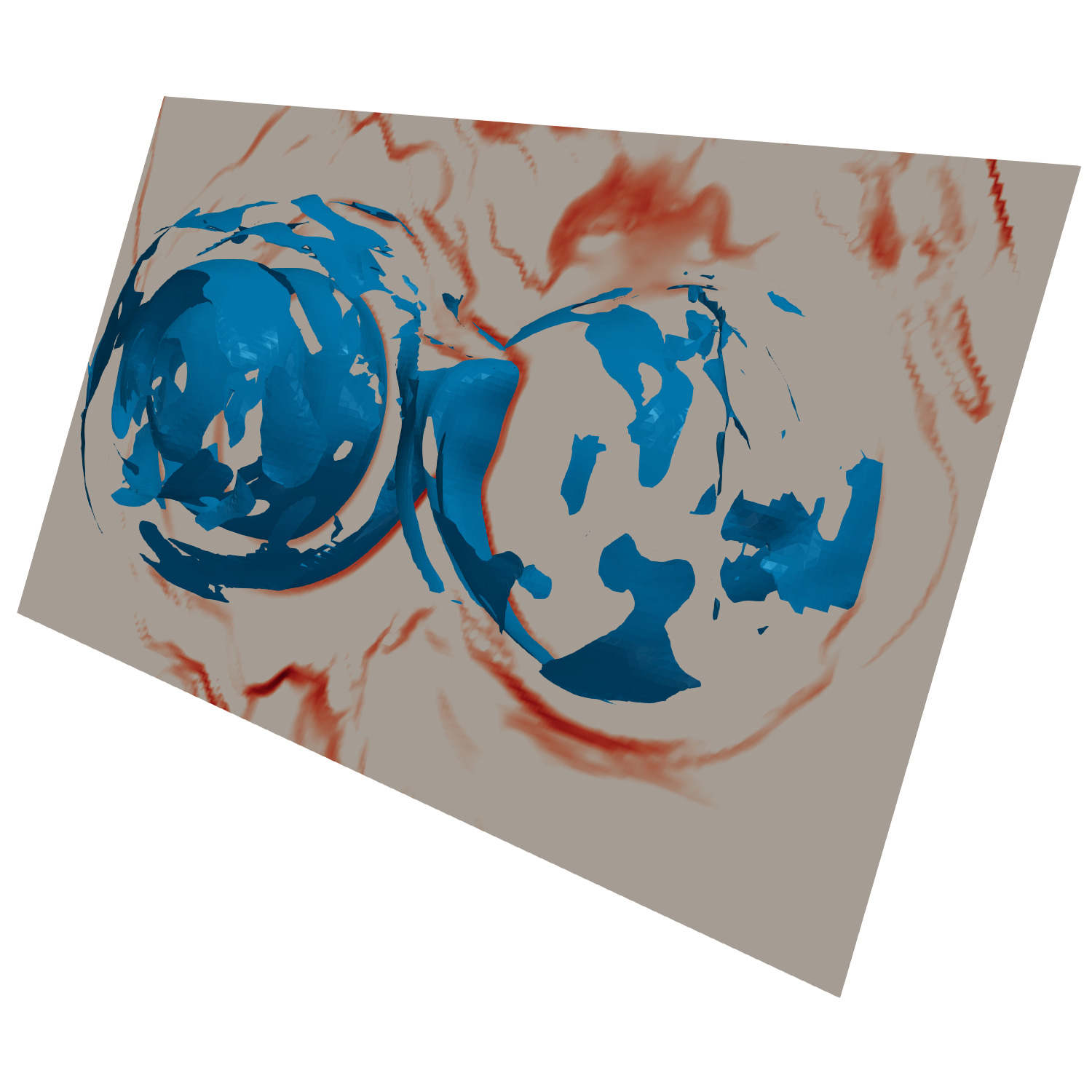}}%
    \hfill%
    \subfloat[\label{fig:ftle-1072-pathlines-side}]%
        {\includegraphics[width=0.49\linewidth,trim=50pt 200pt 50pt 200pt,clip]%
        {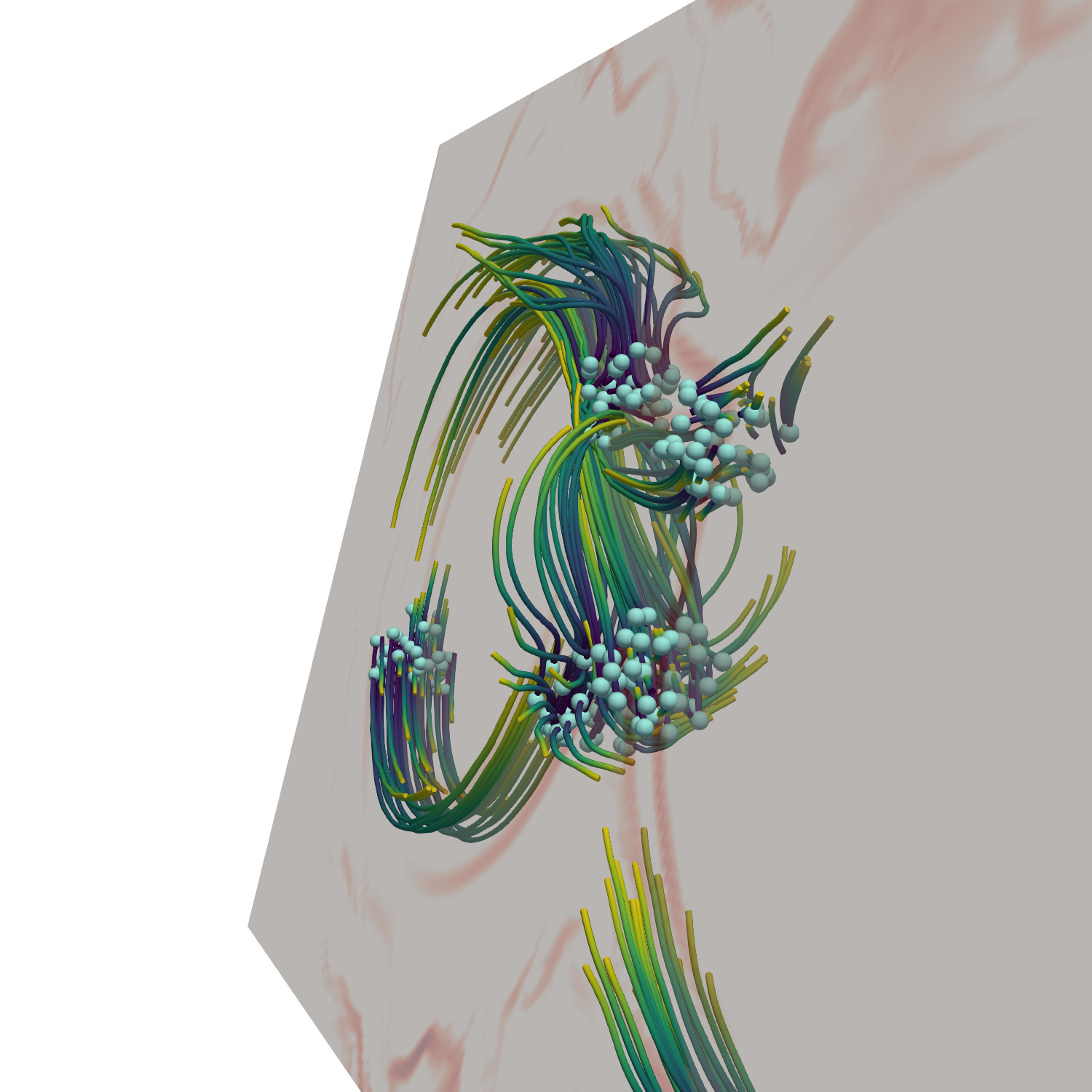}}%
    \caption{%
        Equatorial map of the Finite-Time Lyapunov Exponents~(FTLE) for rcb11 at $t = 1072.21 \seconds$ \protect\subref{fig:ftle-1072-slice} with forward integration for $40 \seconds$, and \protect\subref{fig:ftle-1072-slice-reverse} with backward integration for $40 \seconds$.
        \protect\subref{fig:ftle-1072-slice}~FTLE values mapped to color from white to red~(\includegraphics[width=6pt,height=32pt,angle=-90,origin=rB]{figures/legend_reds.jpeg}), \protect\subref{fig:ftle-1072-slice-reverse}~FTLE values mapped to color from white to blue~(\includegraphics[width=6pt,height=32pt,angle=-90,origin=rB]{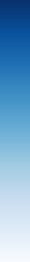}).
        \protect\subref{fig:ftle-1072-ridges}~Extracted ridges~(blue surface) to visualize the three-dimensional FTLE field.
        \protect\subref{fig:ftle-1072-pathlines},\protect\subref{fig:ftle-1072-pathlines-side}~ Pathlines seeded at both sides of FTLE ridges from $t=1072.21 \seconds$ to $t=1112.21 \seconds$ mapped to color~(\includegraphics[width=6pt,height=32pt,angle=-90,origin=rB]{figures/legend_viridis.jpeg}).
        Note that the equatorial slice is rendered semi-transparent in this case to highlight the symmetry of most of the pathlines. 
    }
    \label{fig:ftle-1072}
\end{figure*}

Here we refine the analysis of the equatorial flow making use of the techniques described in~\autoref{sec:vis-ftle}.
In order to facilitate the discussion, we label selected portions of the forward FTLE ridges on \autoref{fig:ftle-1072-slice}, and backward ridges on \autoref{fig:ftle-1072-slice-reverse}.
A forward FTLE ridge identifies a boundary of strong divergence: fluid elements at opposite sides of the ridge separate as the flow proceeds.
Such a ridge is therefore a repelling Lagrangian Coherent Structure (rLCS).
In contrast, a backward FTLE ridge identifies a convergence boundary, along which fluid elements come together with the flow.
Such a ridge is therefore an attracting LCS (aLCS).
In isolation, rLCS and aLCS can be loosely thought of as material ``separatrices" of the 2D flow on the equatorial plane since they are not crossed by the fluid.
Further background and details in \citet{Haller2015} and \citet{Li2022}.


The behavior of pathlines at suitably selected regions of the equatorial flow shown in \autoref{fig:ftle-1072-pathlines}, generated as described before by seeds straddling the LCS, helps to verify the role of the various LCS. It is worth emphasizing that the equatorial slices give us a cross section of the full LCS, which are truly 3D structures as illustrated by \autoref{fig:ftle-1072-ridges}
and the selected pathlines shown in \autoref{fig:ftle-1072-pathlines-side}. For clarity, only one half of the rLCS surfaces with $z>0$ are shown. Note the surface corresponding to 
ridge {\bf a} on the equatorial plane (see below), which separates the mass-transfer stream from the body of the donor.

Focusing on \autoref{fig:ftle-1072-slice}, the rLCS labeled {\bf a}, oriented parallel to the $y$-axis, in the vicinity of the inner
Lagrangian point (a saddle-point of the effective potential), could be considered as the origin of the mass transfer stream.

\begin{figure*}
    \centering%
    \rotatebox[origin=c]{90}{forward}%
    \hfill%
    \begin{minipage}{0.98\linewidth}%
        \subfloat[\label{fig:ftle-slices-1072-fwd}$t = 1072.21 \seconds$]%
            {\includegraphics[width=0.195\linewidth,trim=50pt 400pt 50pt 400pt,clip]%
            {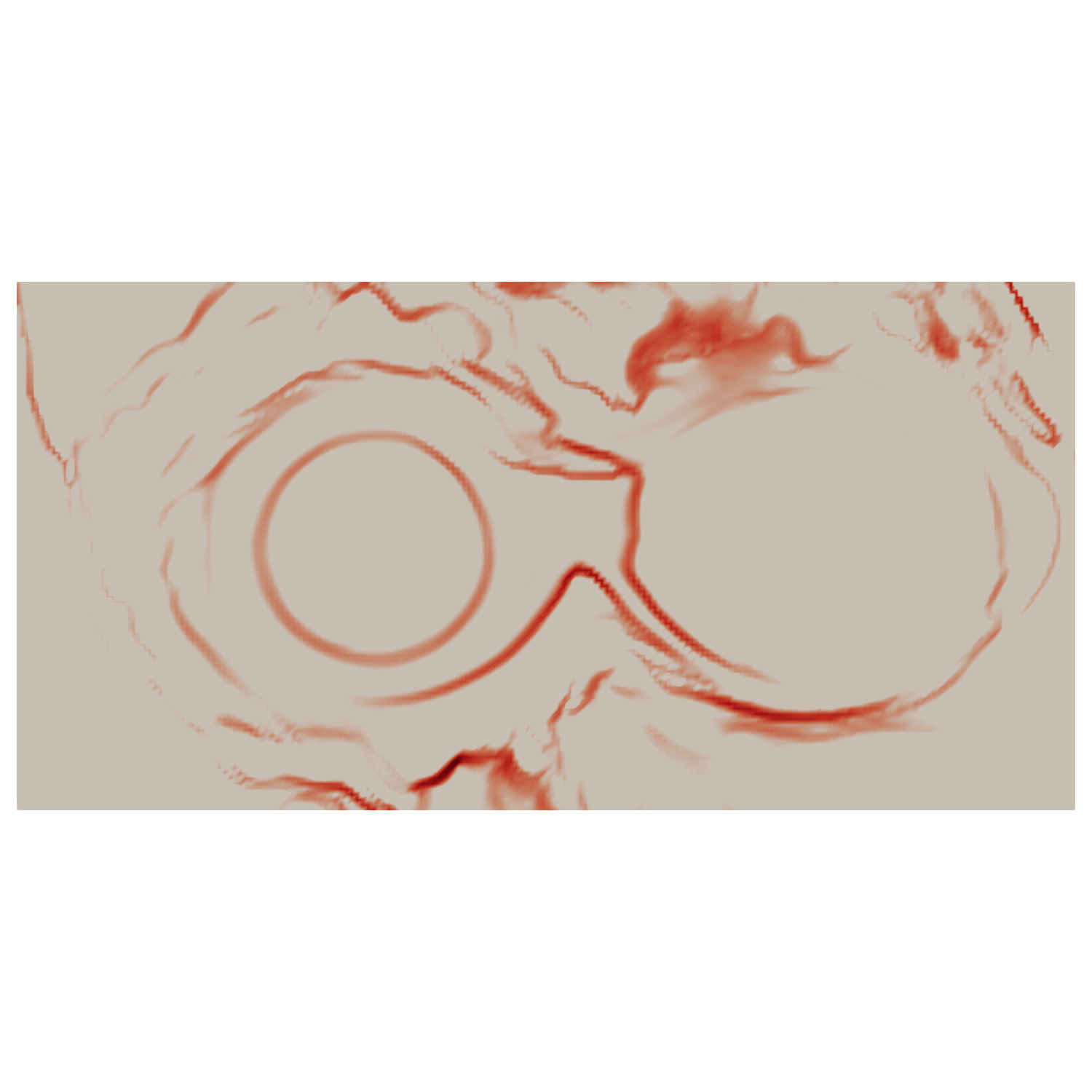}}%
        \hfill%
        \subfloat[\label{fig:ftle-slices-2166-fwd}$t = 2166.96 \seconds$]%
            {\includegraphics[width=0.195\linewidth,trim=50pt 400pt 50pt 400pt,clip]%
            {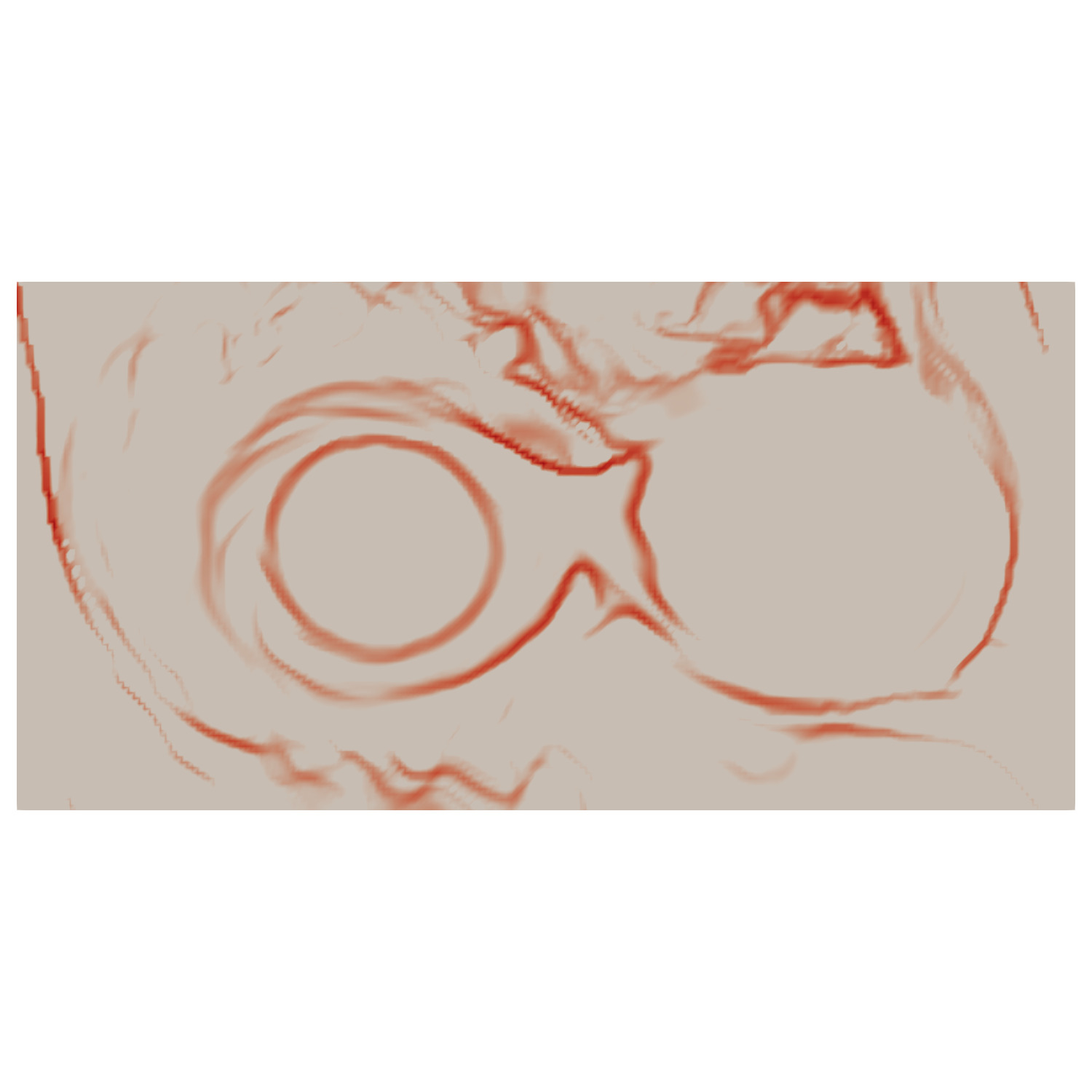}}%
        \hfill%
        \subfloat[\label{fig:ftle-slices-2737-fwd}$t = 2737.1 \seconds$]%
            {\includegraphics[width=0.195\linewidth,trim=50pt 400pt 50pt 400pt,clip]%
            {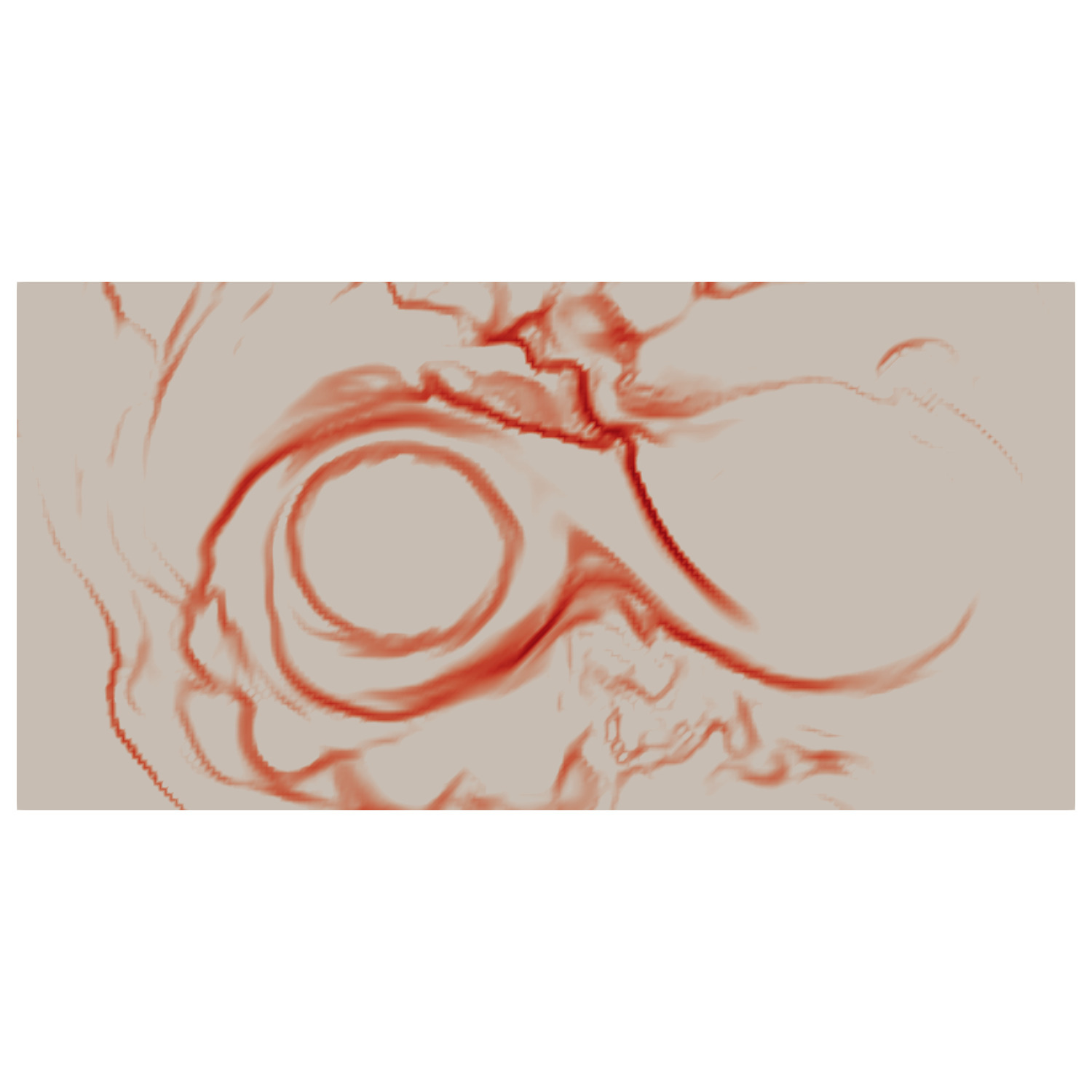}}%
        \hfill%
        \subfloat[\label{fig:ftle-slices-2814-fwd}$t = 2814.64 \seconds$]%
            {\includegraphics[width=0.195\linewidth,trim=50pt 400pt 50pt 400pt,clip]%
            {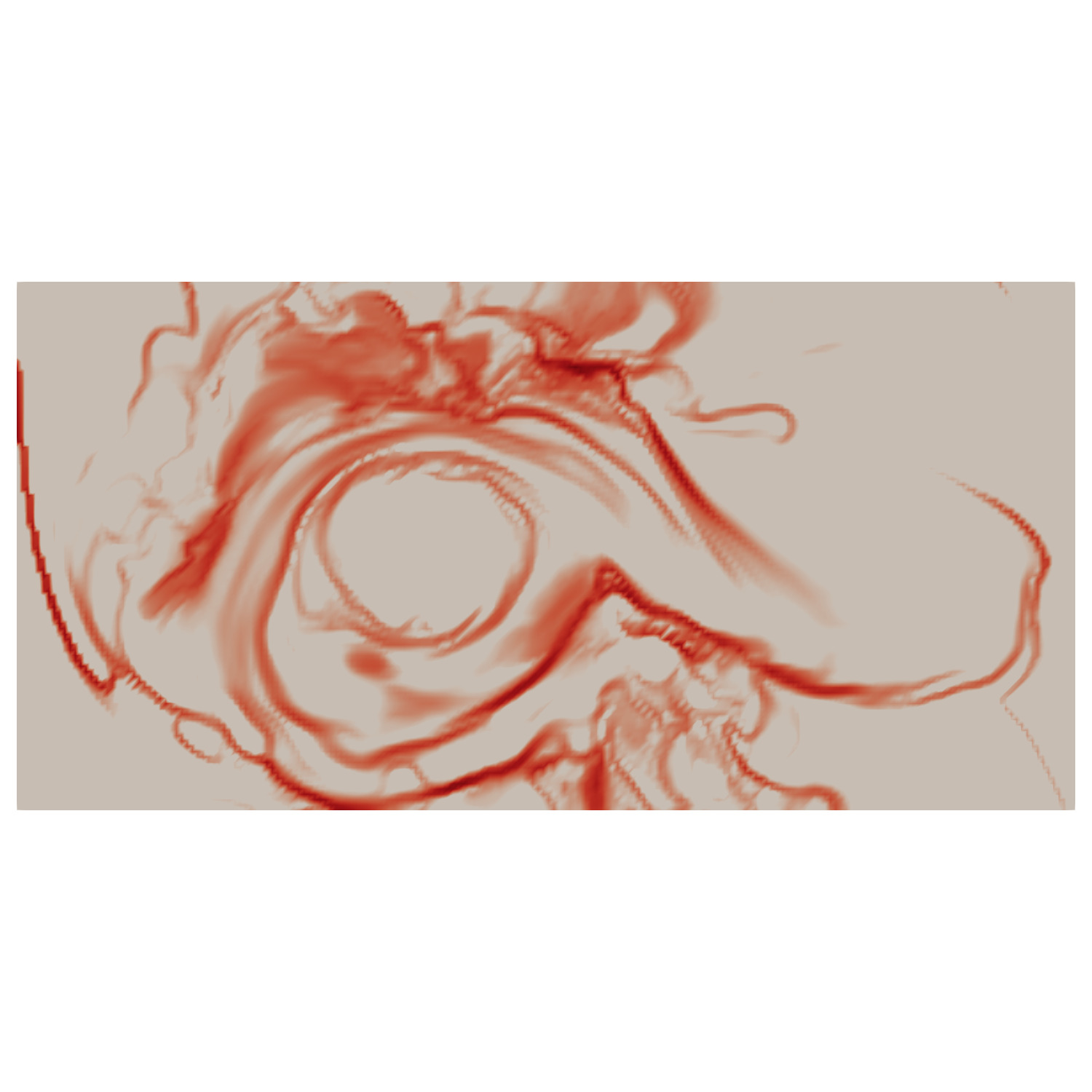}}%
        \hfill%
        \subfloat[\label{fig:ftle-slices-3193-fwd}$t = 3193.17 \seconds$]%
            {\includegraphics[width=0.195\linewidth,trim=50pt 400pt 50pt 400pt,clip]%
            {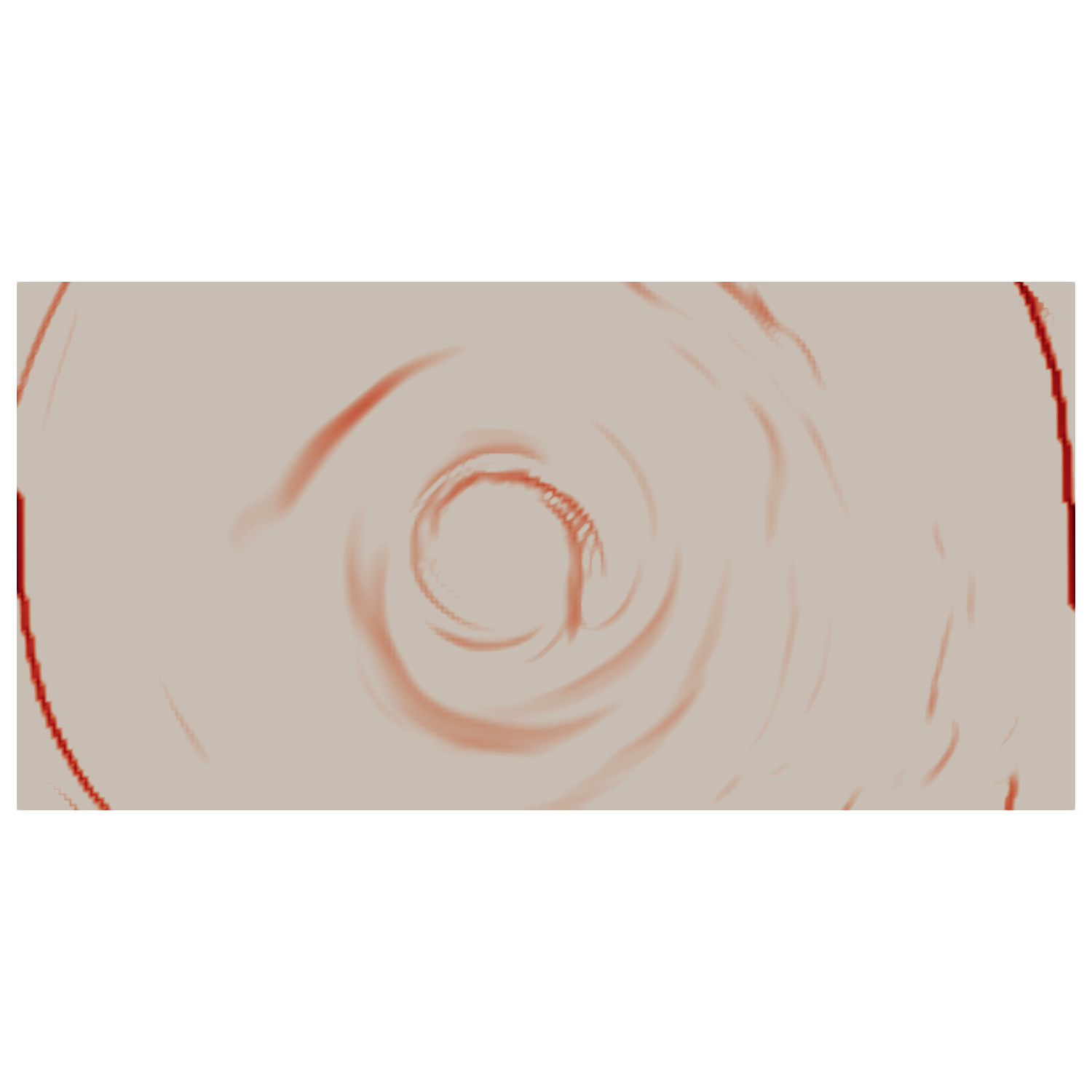}}%
    \end{minipage}%
    \\%
    \rotatebox[origin=c]{90}{backward}%
    \hfill%
    \begin{minipage}{0.98\linewidth}%
        \subfloat[\label{fig:ftle-slices-1072-bwd}$t = 1072.21 \seconds$]%
            {\includegraphics[width=0.195\linewidth,trim=50pt 400pt 50pt 400pt,clip]%
            {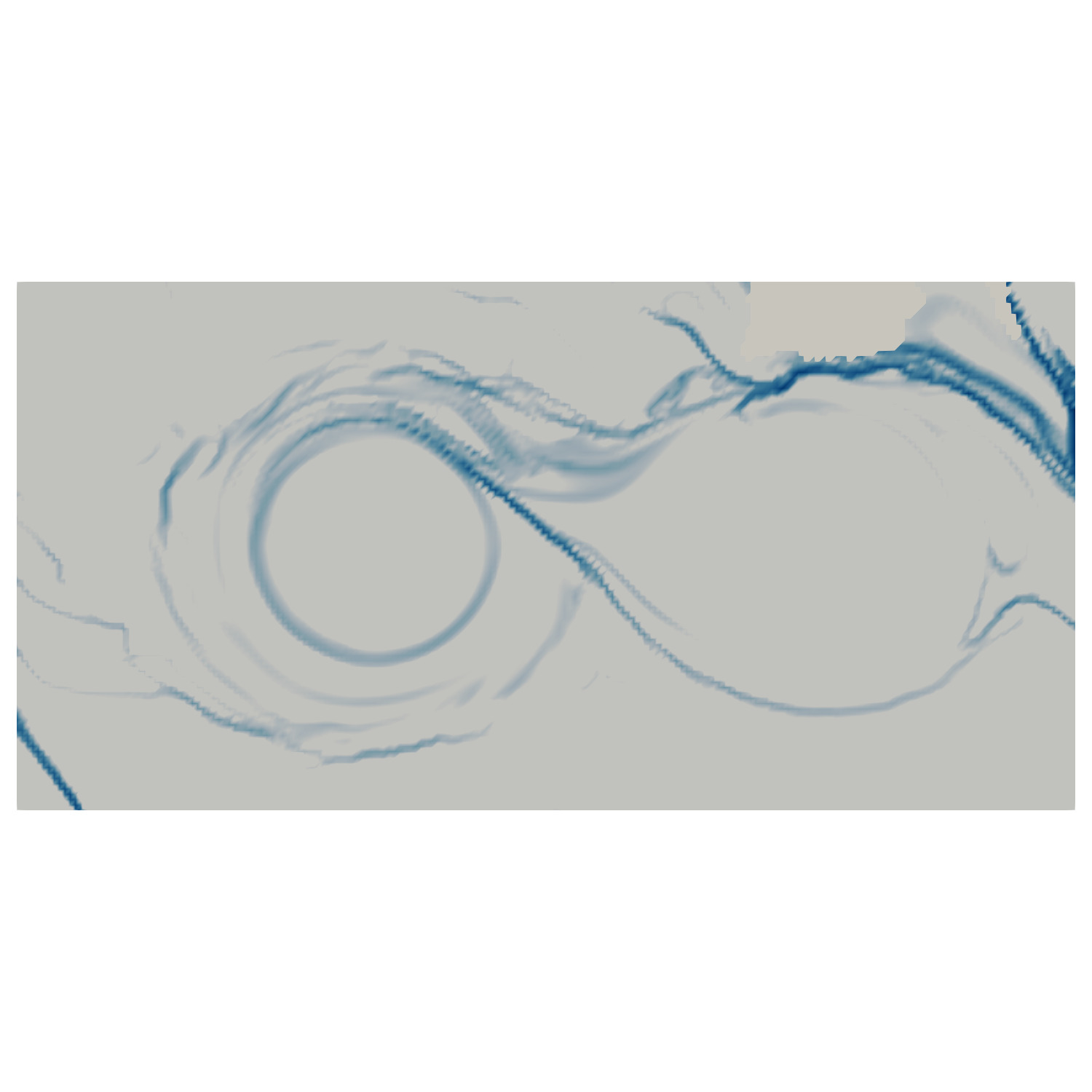}}%
        \hfill%
        \subfloat[\label{fig:ftle-slices-2166-bwd}$t = 2166.96 \seconds$]%
            {\includegraphics[width=0.195\linewidth,trim=50pt 400pt 50pt 400pt,clip]%
            {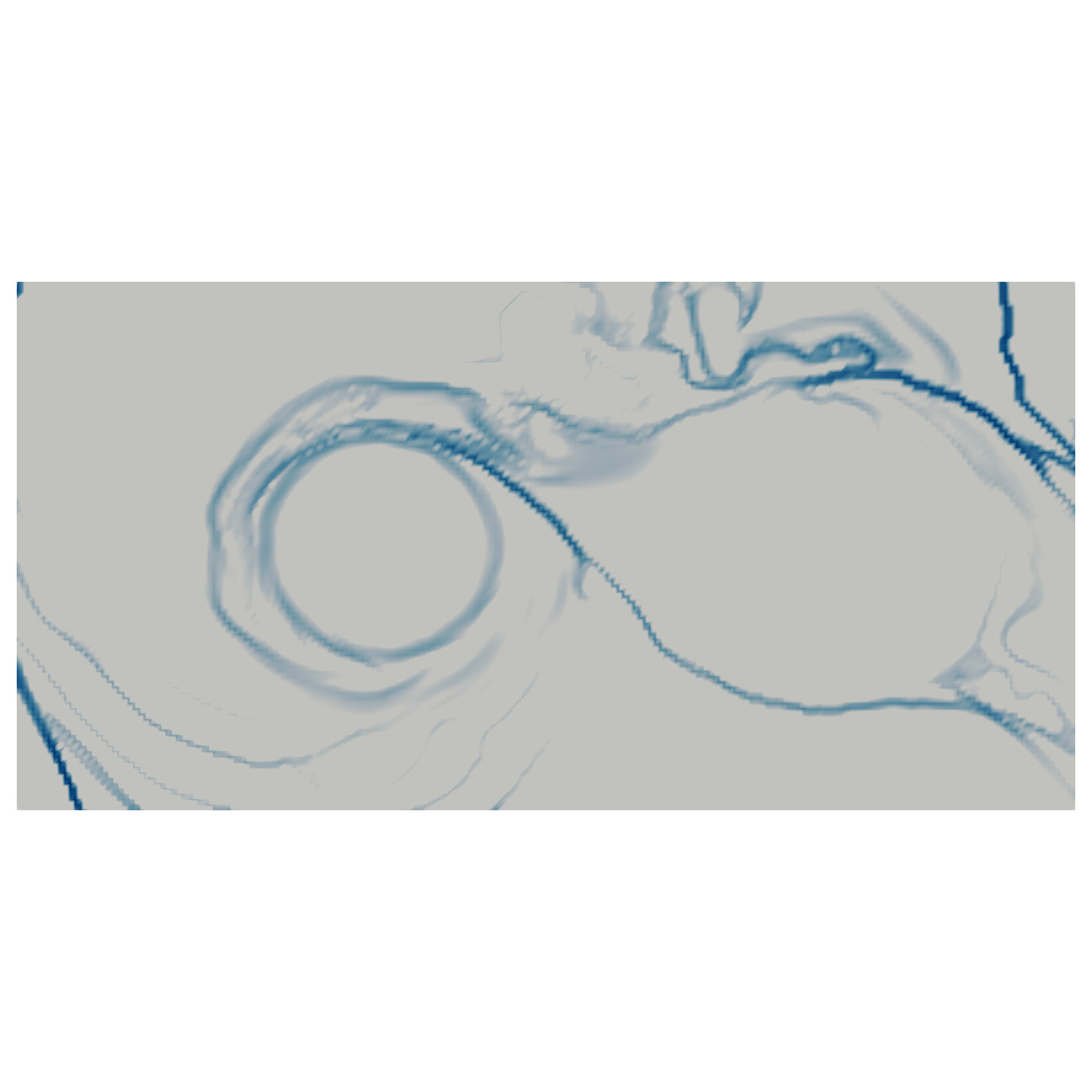}}%
        \hfill%
        \subfloat[\label{fig:ftle-slices-2737-bwd}$t = 2737.1 \seconds$]%
            {\includegraphics[width=0.195\linewidth,trim=50pt 400pt 50pt 400pt,clip]%
            {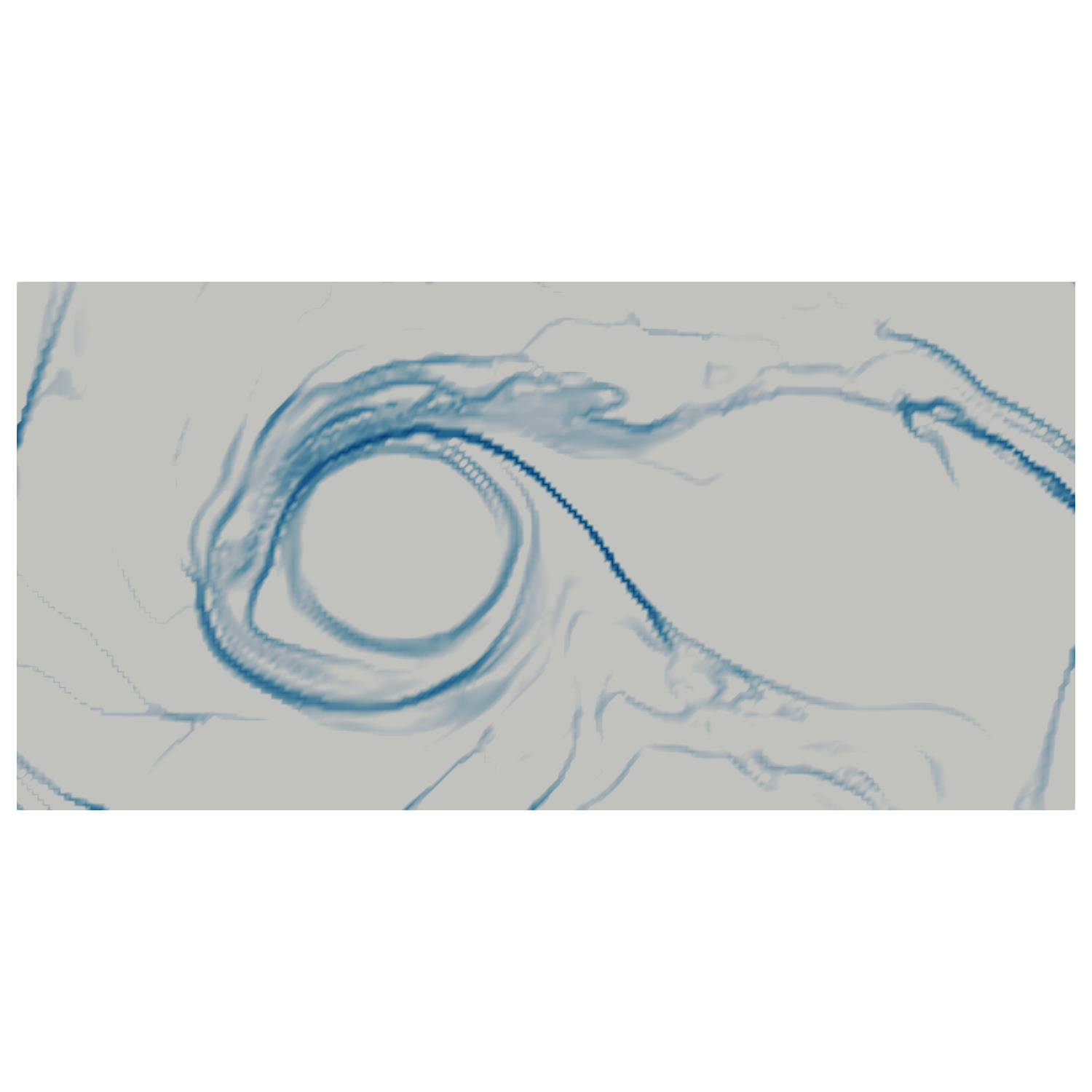}}%
        \hfill%
        \subfloat[\label{fig:ftle-slices-2814-bwd}$t = 2814.64 \seconds$]%
            {\includegraphics[width=0.195\linewidth,trim=50pt 400pt 50pt 400pt,clip]%
            {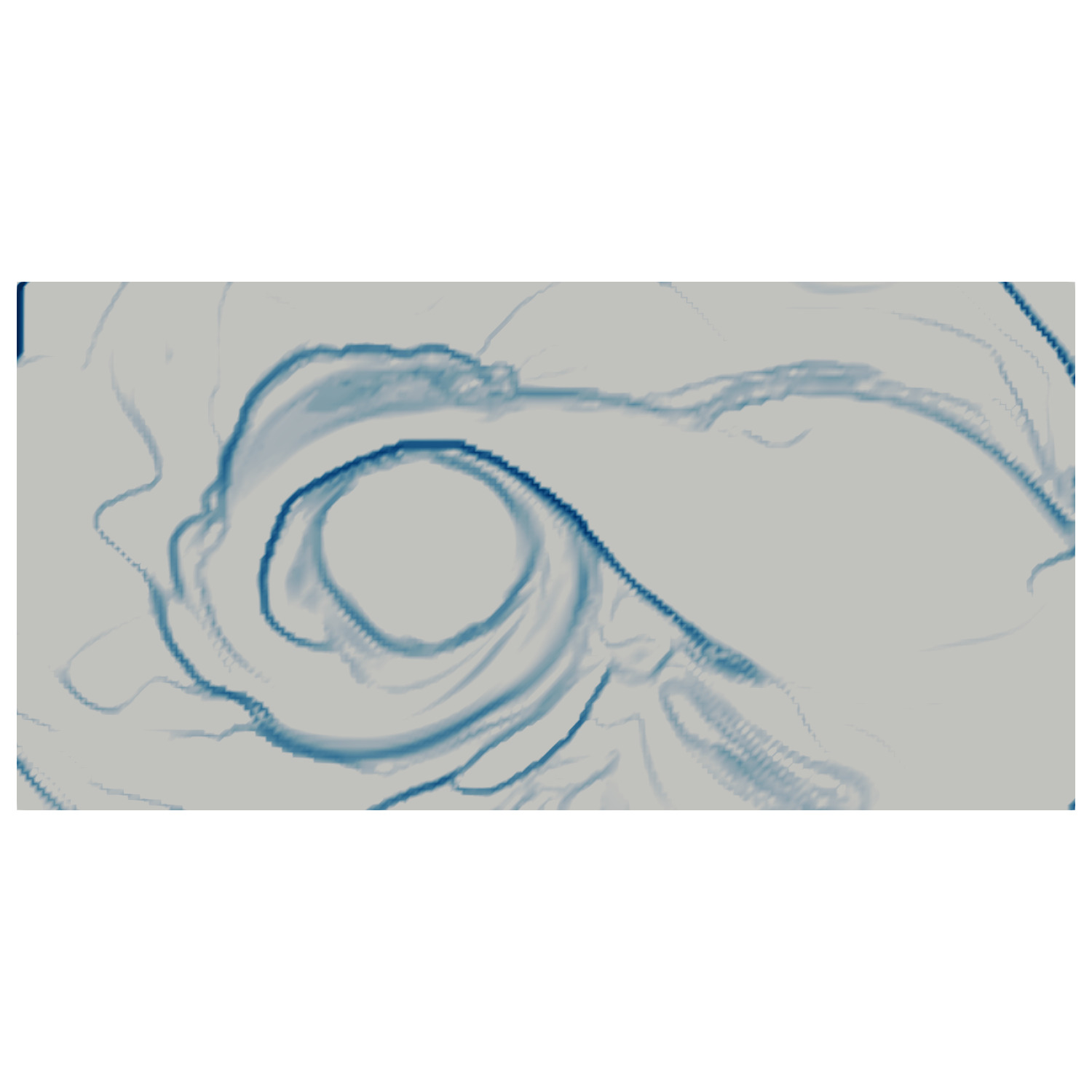}}%
        \hfill%
        \subfloat[\label{fig:ftle-slices-3193-bwd}$t = 3193.17 \seconds$]%
            {\includegraphics[width=0.195\linewidth,trim=50pt 400pt 50pt 400pt,clip]%
            {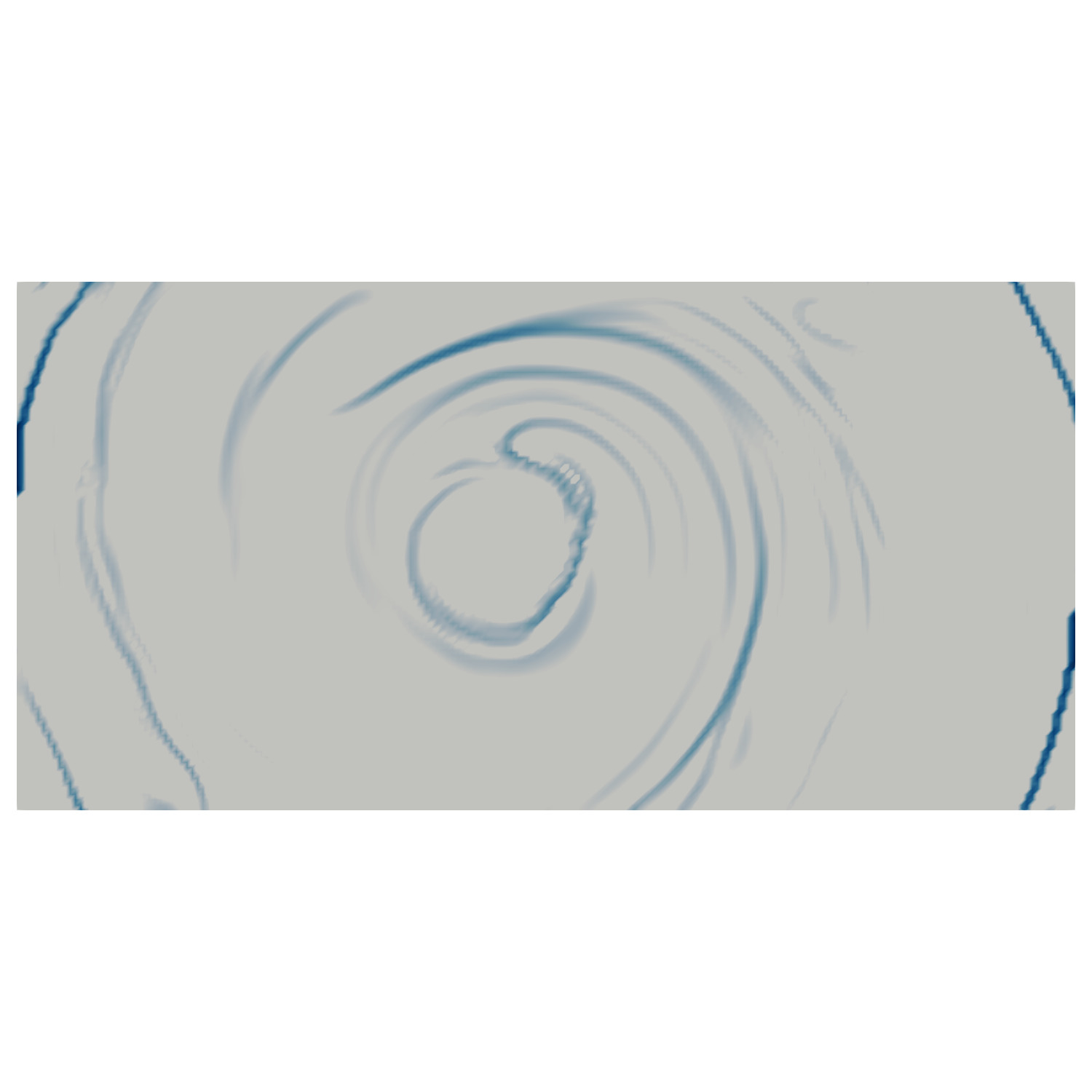}}%
    \end{minipage}%
    \\%
    \rotatebox[origin=c]{90}{fwd/bwd}%
    \hfill%
    \begin{minipage}{0.98\linewidth}%
        \subfloat[\label{fig:ftle-slices-1072-comp}$t = 1072.21 \seconds$]%
            {\includegraphics[width=0.195\linewidth,trim=12pt 96pt 12pt 96pt,clip]%
            {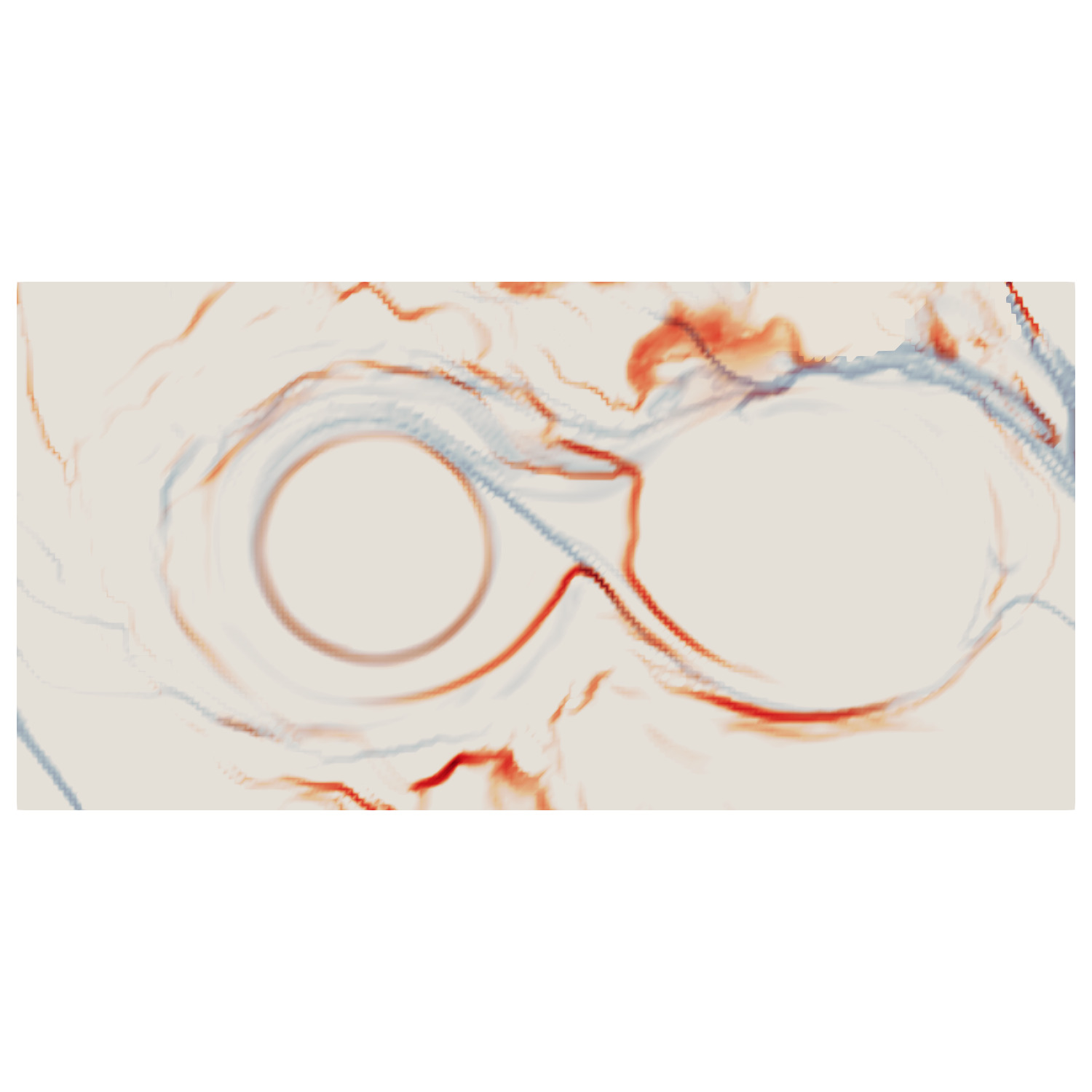}}%
        \hfill%
        \subfloat[\label{fig:ftle-slices-2166-comp}$t = 2166.96 \seconds$]%
            {\includegraphics[width=0.195\linewidth,trim=12pt 96pt 12pt 96pt,clip]%
            {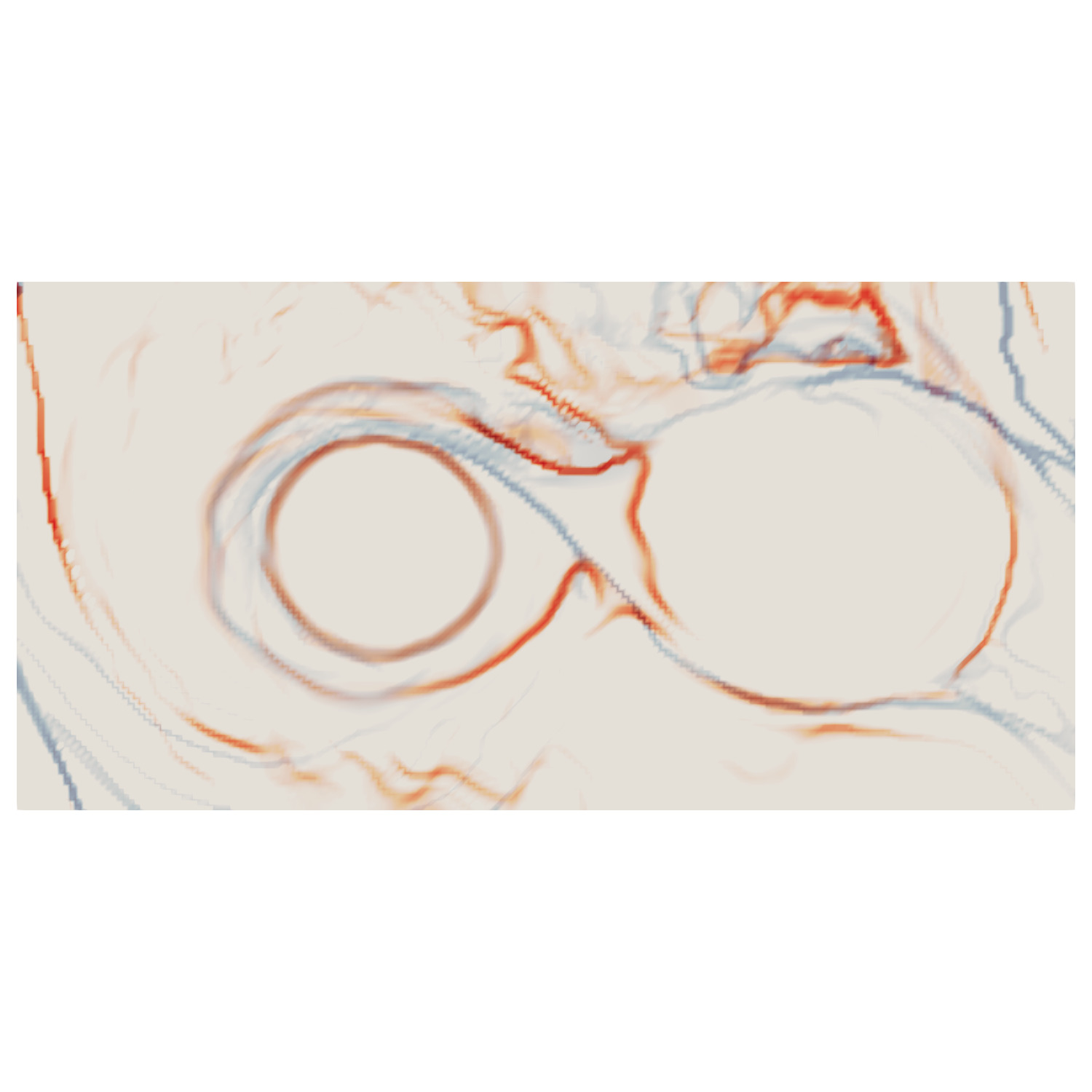}}%
        \hfill%
        \subfloat[\label{fig:ftle-slices-2737-comp}$t = 2737.1 \seconds$]%
            {\includegraphics[width=0.195\linewidth,trim=12pt 96pt 12pt 96pt,clip]%
            {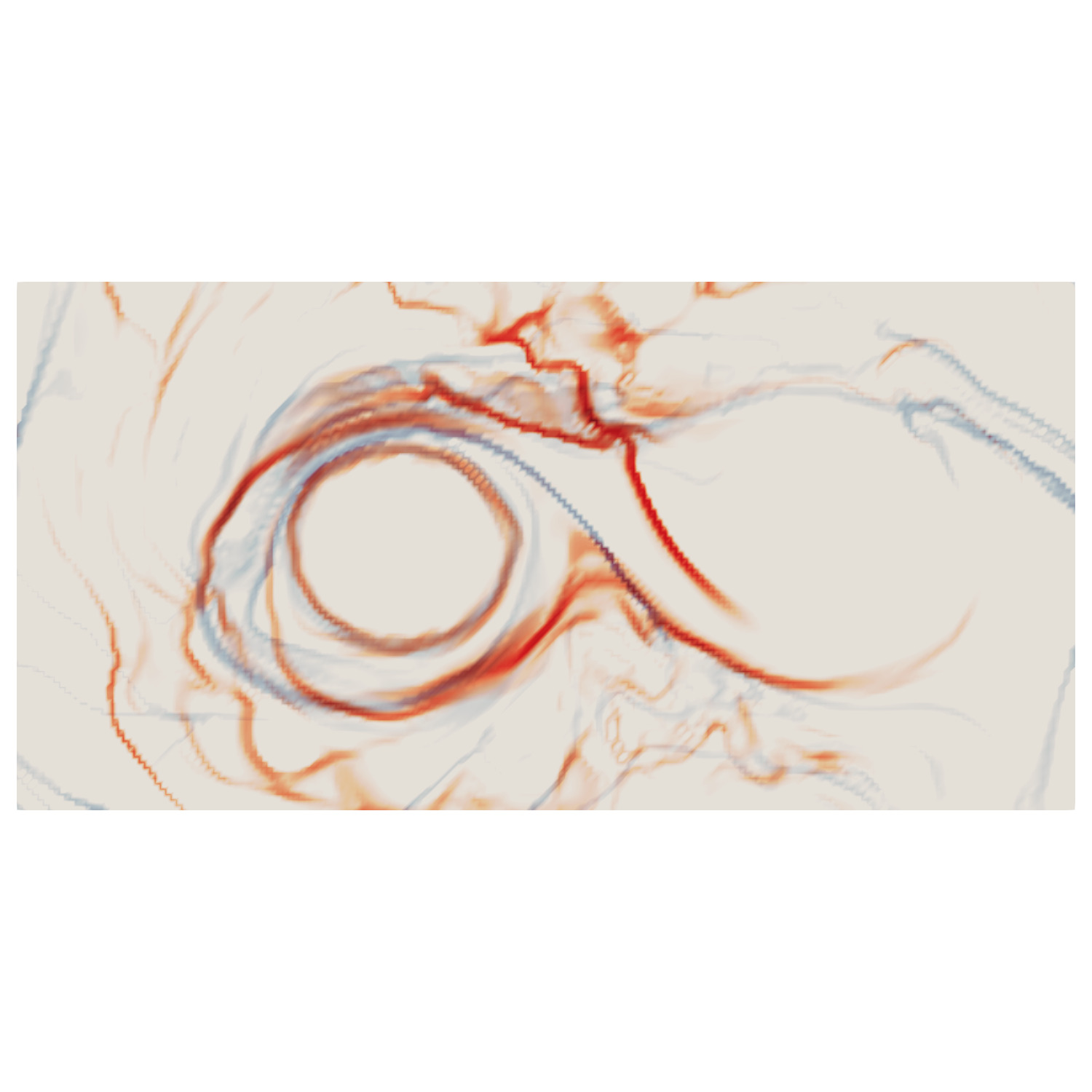}}%
        \hfill%
        \subfloat[\label{fig:ftle-slices-2814-comp}$t = 2814.64 \seconds$]%
            {\includegraphics[width=0.195\linewidth,trim=12pt 96pt 12pt 96pt,clip]%
            {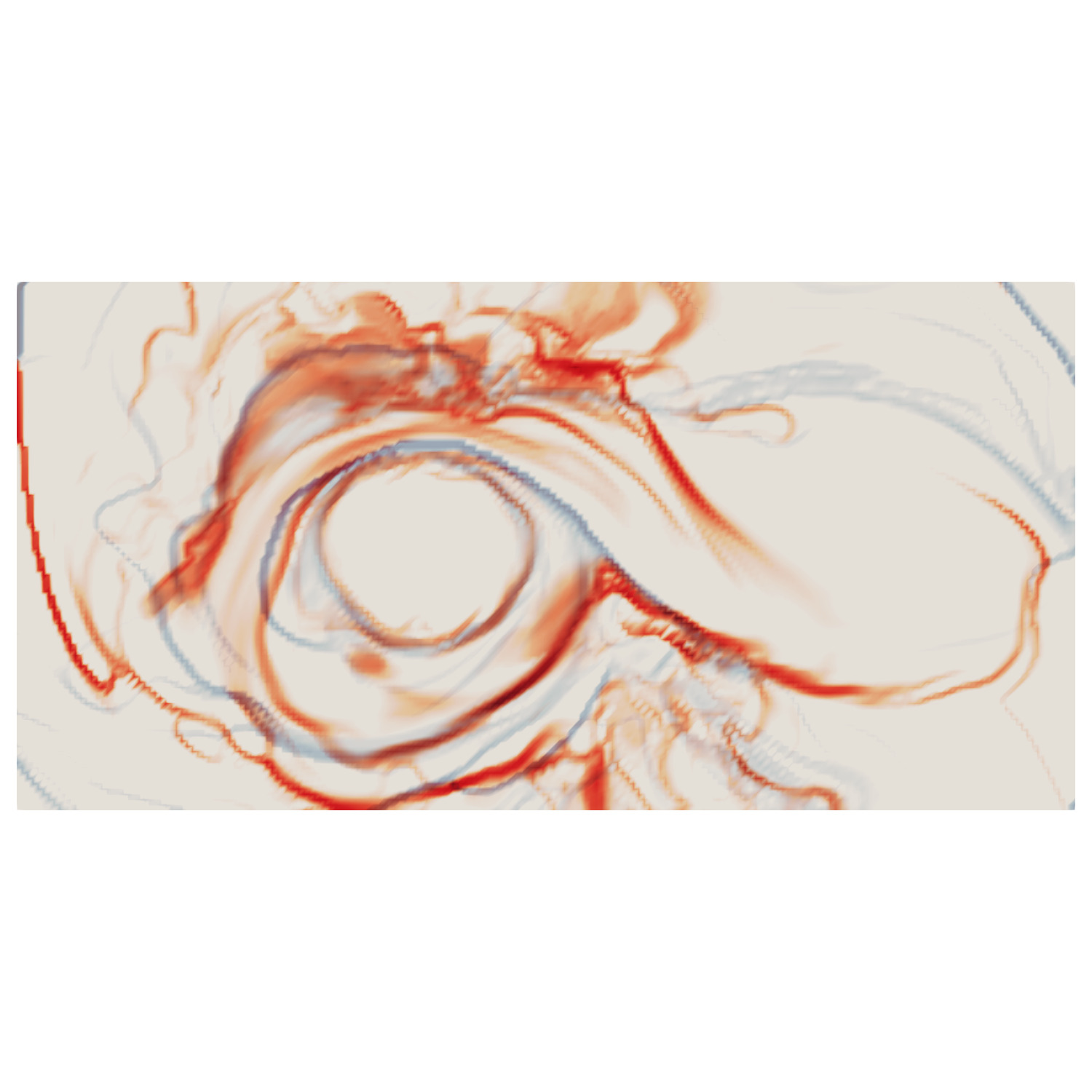}}%
        \hfill%
        \subfloat[\label{fig:ftle-slices-3193-comp}$t = 3193.17 \seconds$]%
            {\includegraphics[width=0.195\linewidth,trim=12pt 96pt 12pt 96pt,clip]%
            {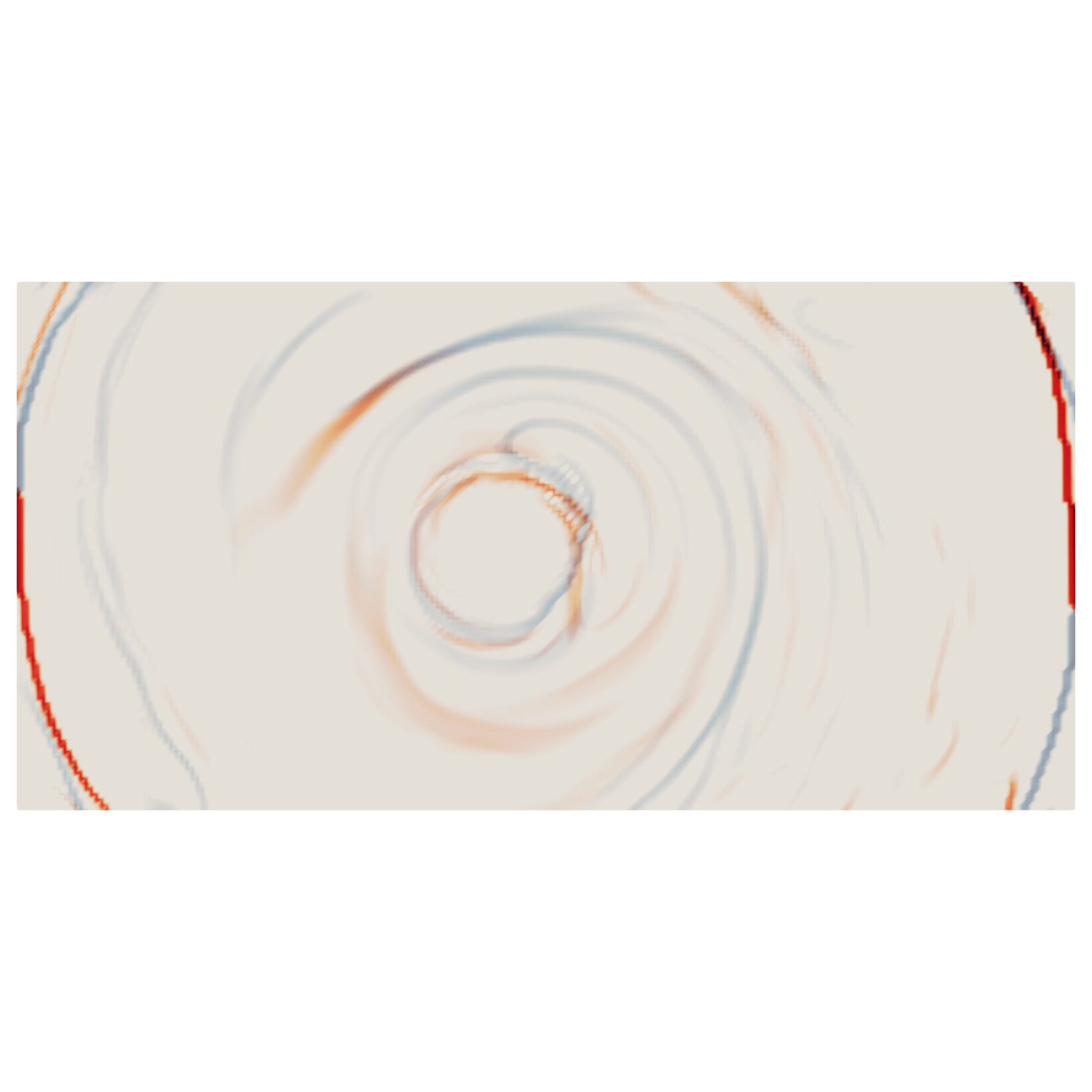}}%
    \end{minipage}%
    \\%
    \rotatebox[origin=c]{90}{velocities}%
    \hfill%
    \begin{minipage}{0.98\linewidth}%
        \subfloat[\label{fig:velocity-slices-1072}$t= 1072.21 \seconds$]%
            {\includegraphics[width=0.195\linewidth,trim=50pt 400pt 50pt 400pt,clip]%
            {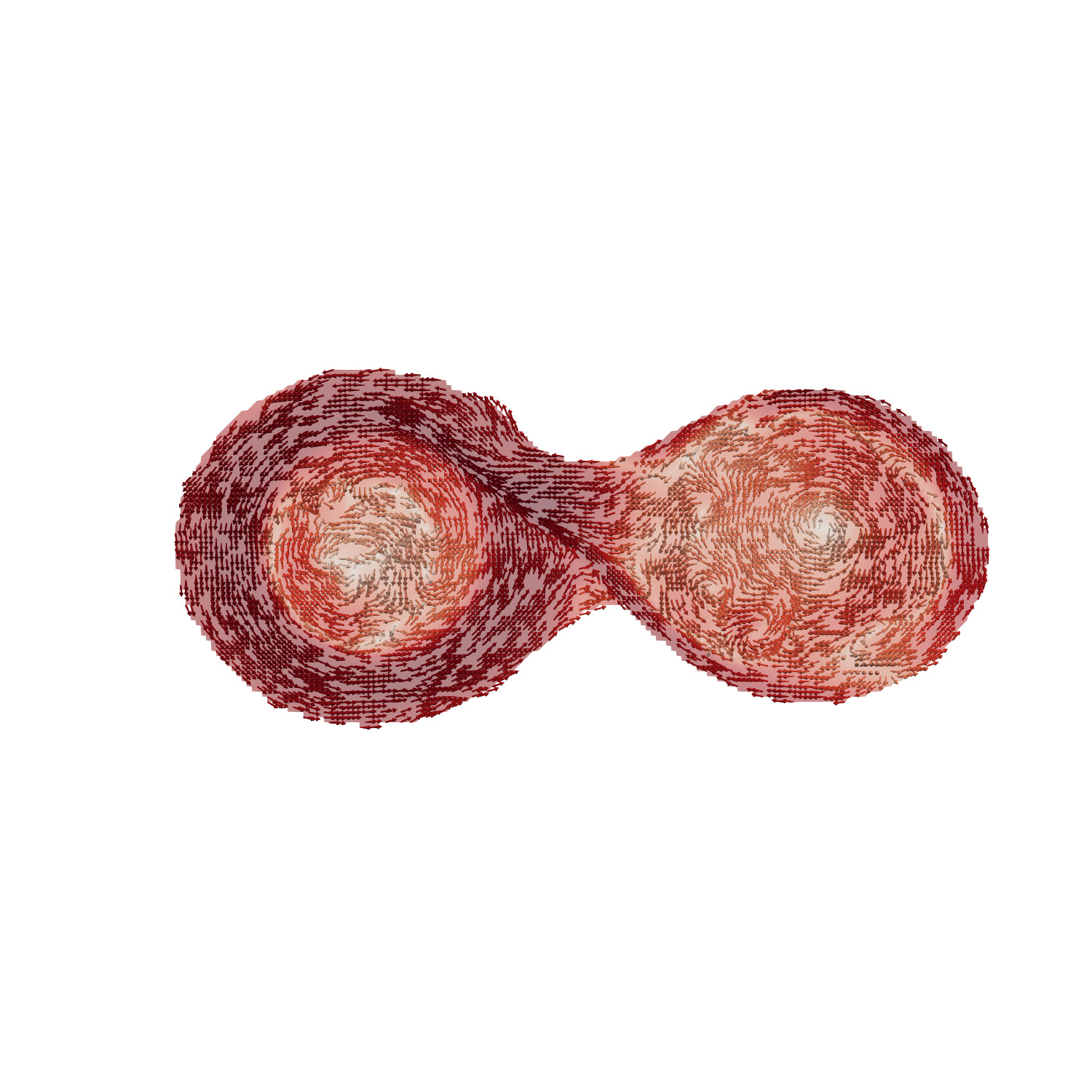}}%
        \hfill%
        \subfloat[\label{fig:velocity-slices-2166}$t= 2166.96 \seconds$]%
            {\includegraphics[width=0.195\linewidth,trim=50pt 400pt 50pt 400pt,clip]%
            {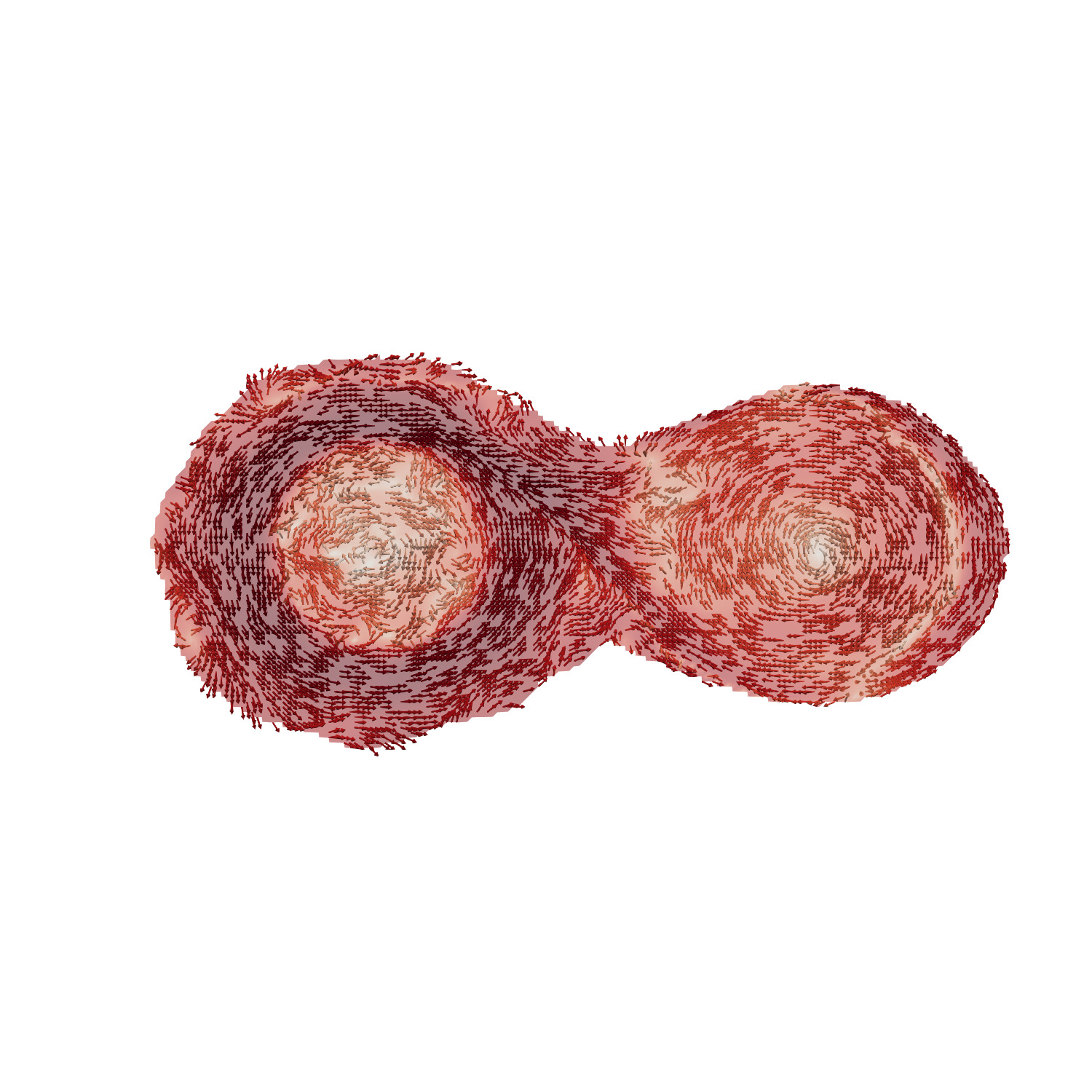}}%
        \hfill%
        \subfloat[\label{fig:velocity-slices-2737}$t= 2737.1 \seconds$]%
            {\includegraphics[width=0.195\linewidth,trim=50pt 400pt 50pt 400pt,clip]%
            {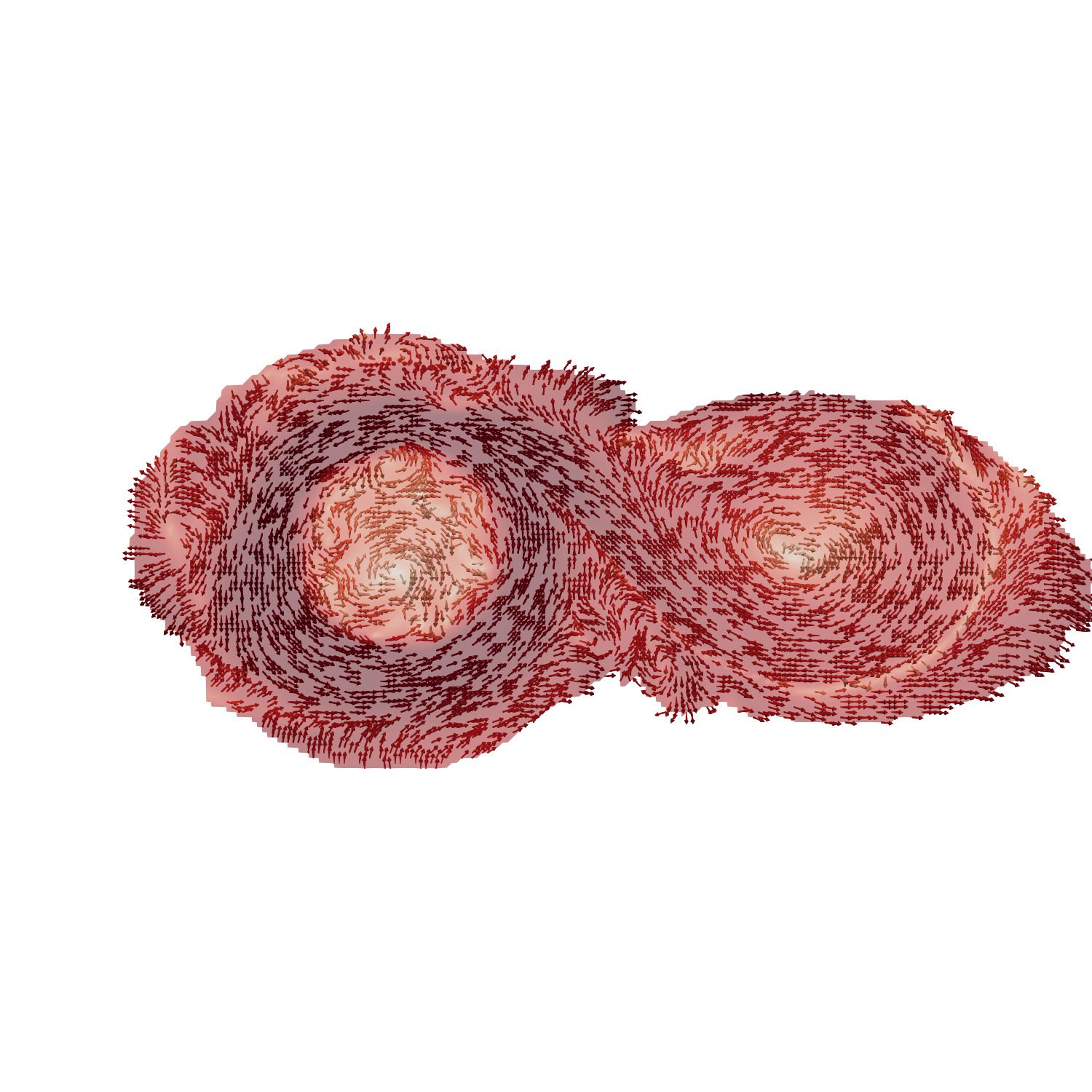}}%
        \hfill%
        \subfloat[\label{fig:velocity-slices-2814}$t= 2814.64 \seconds$]%
            {\includegraphics[width=0.195\linewidth,trim=50pt 400pt 50pt 400pt,clip]%
            {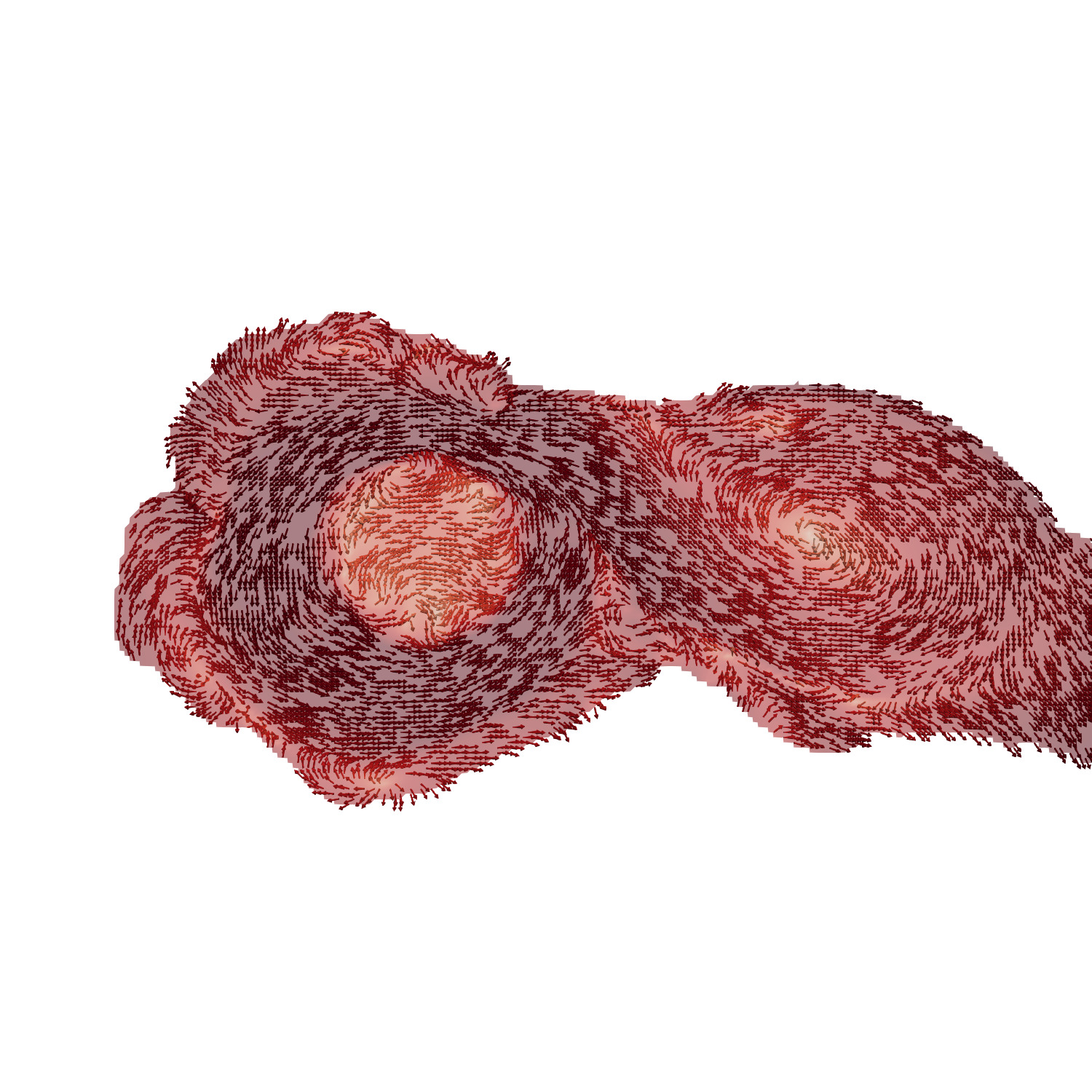}}%
        \hfill%
        \subfloat[\label{fig:velocity-slices-3193}$t= 3193.17 \seconds$]%
            {\includegraphics[width=0.195\linewidth,trim=50pt 400pt 50pt 400pt,clip]%
            {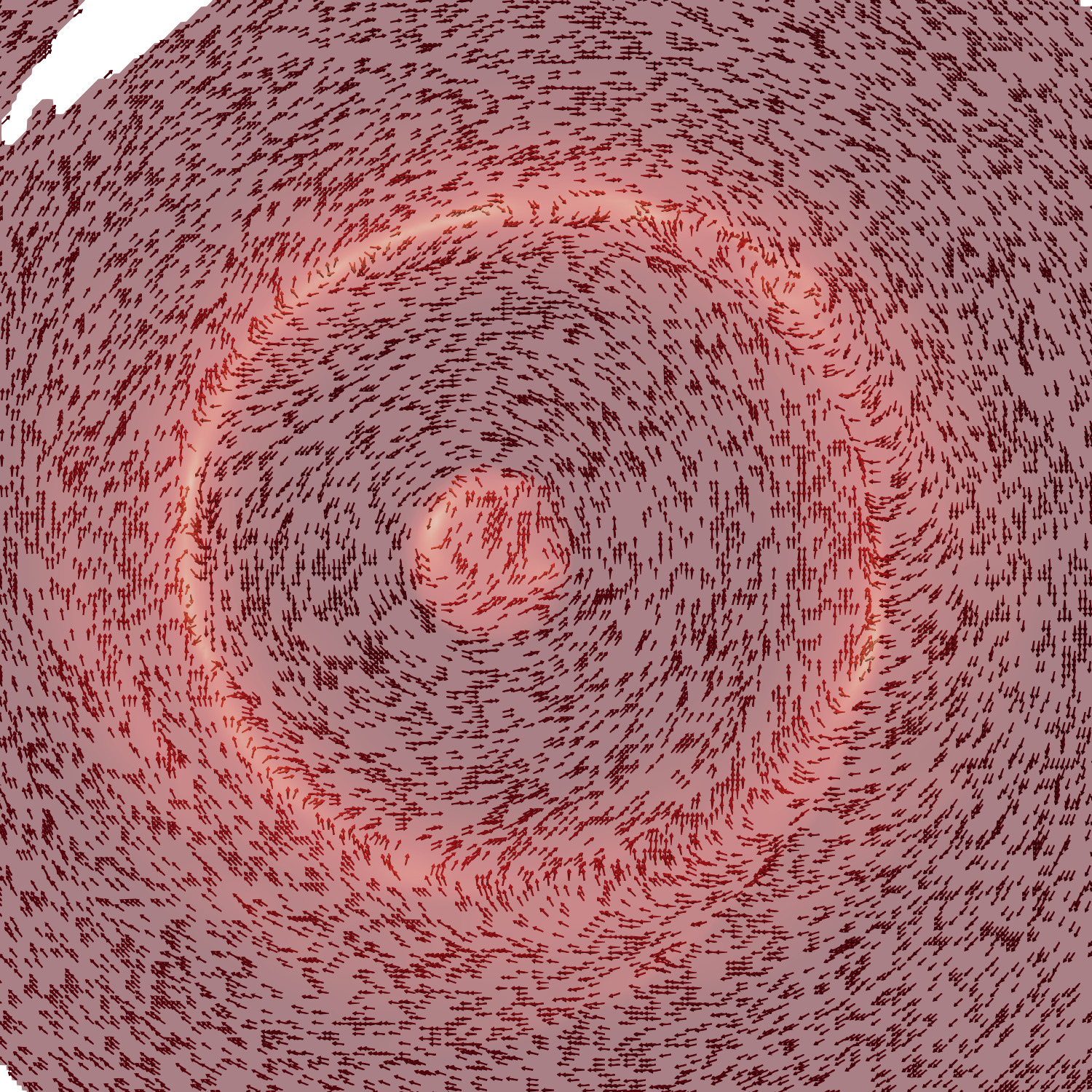}}%
    \end{minipage}%
    \caption{%
        Equatorial slices for different time steps showing only the large values of the FTLE field for rcb11.
        The first two steps show time steps for the well developed stream, the third shortly before the tidal disruption, the fourth at the tidal disruption, and the last after the merger.
        The FTLE values for \protect\subref{fig:ftle-slices-1072-fwd}--\protect\subref{fig:ftle-slices-3193-fwd}~forward time integration are mapped to color from white to red~(\includegraphics[width=6pt,height=32pt,angle=-90,origin=rB]{figures/legend_reds.jpeg}), and for reverse time integration~\protect\subref{fig:ftle-slices-1072-bwd}--\protect\subref{fig:ftle-slices-3193-bwd} from white to blue~(\includegraphics[width=6pt,height=32pt,angle=-90,origin=rB]{figures/legend_blues.jpeg}).
        \protect\subref{fig:ftle-slices-1072-comp}--\protect\subref{fig:ftle-slices-3193-comp}~Compositions of forward and backward integrated FTLE fields.
        \protect\subref{fig:velocity-slices-1072}--\protect\subref{fig:velocity-slices-3193}~Velocity magnitudes are mapped logarithmically to color~(\includegraphics[width=6pt,height=32pt,angle=-90,origin=rB]{figures/legend_reds.jpeg}) in the range $[1 \cdot 10^5,~1 \cdot 10^8] \cmps$.
    }
    \label{fig:ftle-slices}
\end{figure*}

Note how {\bf a} joins its extension {\bf a$_0$} near the trailing surface of the donor, and how, at the other end, {\bf a} bifurcates into {\bf a}$_1$ and {\bf a}$_2$ near the stream bounce.
    This set of LCS  
    {\bf a}, 
    {\bf a}$_0$,
    {\bf a}$_1$, and 
    {\bf a}$_2$,
    persists throughout the evolution, only changing gradually up to the merger. 
These LCS         
separate the
relatively slow internal flow in the donor body from the near-surface flow feeding the mas transfer stream.
Similarly LCS {\bf b} separates the internal flow in the accretor body from the equatorial accretion belt.
Finally LCS {\bf c}, {\bf c}$_0$ and {\bf d} define the separation between accretion-belt flow and a return flow carrying some mass back to the donor and eventually leaking out around the  $L_2$ external Lagrange point. 
These four main LCS shift in position and change shape during the evolution as illustrated in the top row of panels in \autoref{fig:ftle-slices-1072-fwd}--\ref{fig:ftle-slices-3193-fwd}. 

As described in \cite{Shiber2024}, the mass transfer rate, after an initial fast rise, settles on a value of $\approx 10^{-3}~M_{\rm \odot}$/$P_0 \approx 8\times 10^{-6}~M_{\rm \odot}~{\rm s^{-1}}$, and then continues to rise slowly throughout the evolution up until close to the merging time when the donor star is tidally disrupted. During this time, most of the mass that the donor loses is being accreted on to the accretor star and is added to the accretion belt. The rest of the mass is lost from the system and can be calculated using our diagnostics method (see \autoref{sec:vis-slice}), which identifies each cell as a part of a star only if it is inside the star’s Roche lobe. In \cite{Shiber2024} they calculated the mass lost as described and have found that only $10^{-3}~M_{\rm \odot}$ (out of the total system mass of $0.9~M_{\rm \odot}$) is lost at a time of two orbits before the merger, where the mass lost increases sharply to $7.5\times 10^{-3}~M_{\rm \odot}$ at the time of the merger (see their Figure 7). However, a fraction of this mass remains bound and is eventually expected to be accreted by the merged object at late times. The total unbound mass found at the end of the simulation is $\lesssim 5\times 10^{-3}~M_{\rm \odot}$.

Focusing next on the backward-integrated FTLE ridges or aLCS, shown with labels in \autoref{fig:ftle-1072-slice-reverse}, and their evolution on \autoref{fig:ftle-slices-1072-bwd}--\ref{fig:ftle-slices-3193-bwd},
we can think of them as lines of convergence of the equatorial flow. The strongest ridge labeled {\bf{\em d}} is where the leading
side of the donor interacts with the external medium through shocks. For the purpose of understanding the internal flow,
the most important convergence ridges are {\bf\em a}, where the stream flow and the return flow along the accretion belt converge and interact; {\bf\em b}, separating internal accretor flow from the accretion belt; {\bf\em c}, along the return flow along the trailing side of the donor, and exiting the binary near $L_2$. In general one expects convergence ridges to fall between divergence ridges, and to outline the main direction of the local flow. These 
backward-integrated LCS, also change in shape and shift in position as the flow evolves, but it is easy to see that they persist during the evolution. Throughout most of the evolution the {\bf b} and {\bf\em b} ridges coincide capturing a strong shearing layer  at the interface between the CW eddies on the accretor side and the accretion belt encircling it. This strong shear layer  marks the location of the inevitable mixing of donor and accretor material.

The overall evolution to tidal disruption of the donor and merger could be described as in increasing transfer of mass from the trailing side
of the donor, accompanied by a growth of the equatorial belt around the accretor, until eventually the disrupted donor is wrapped around the accretor.
The lowest row of equatorial slices during the entire evolution, showing velocity direction and magnitude in \autoref{fig:velocity-slices-1072}--\ref{fig:velocity-slices-3193}, confirms the above interpretation, especially when one focuses on the highest velocities shown in dark red.

\begin{figure*}
    \centering%
    \subfloat[\label{fig:quasi-steady1-rcb12-645}]%
        {\includegraphics[width=0.49\linewidth,trim=0 400pt 0 400pt,clip]%
        {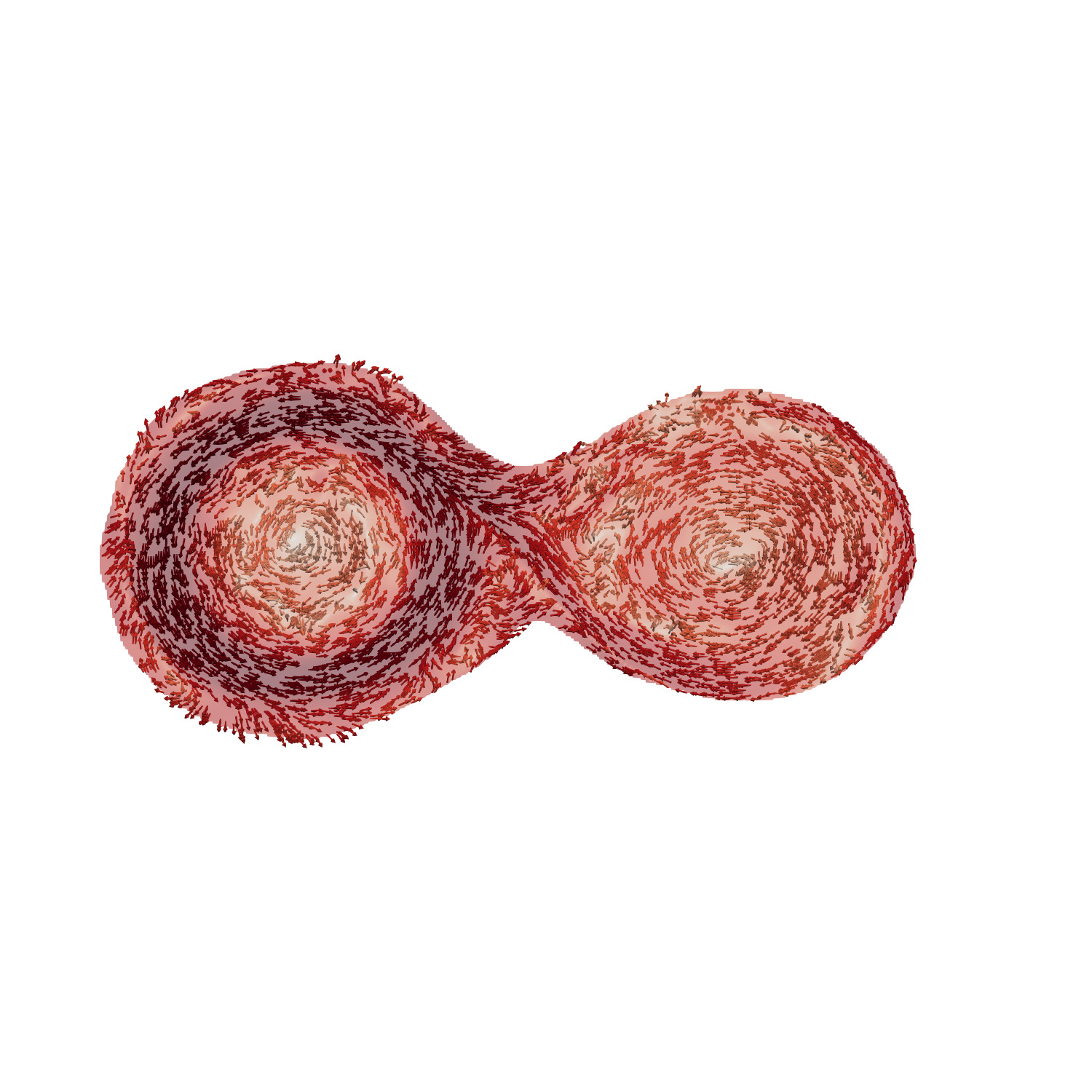}}%
    \hfill%
    \subfloat[\label{fig:quasi-steady1-rcb12-645-lic}]%
        {\includegraphics[width=0.49\linewidth,trim=0 400pt 0 400pt,clip]%
        {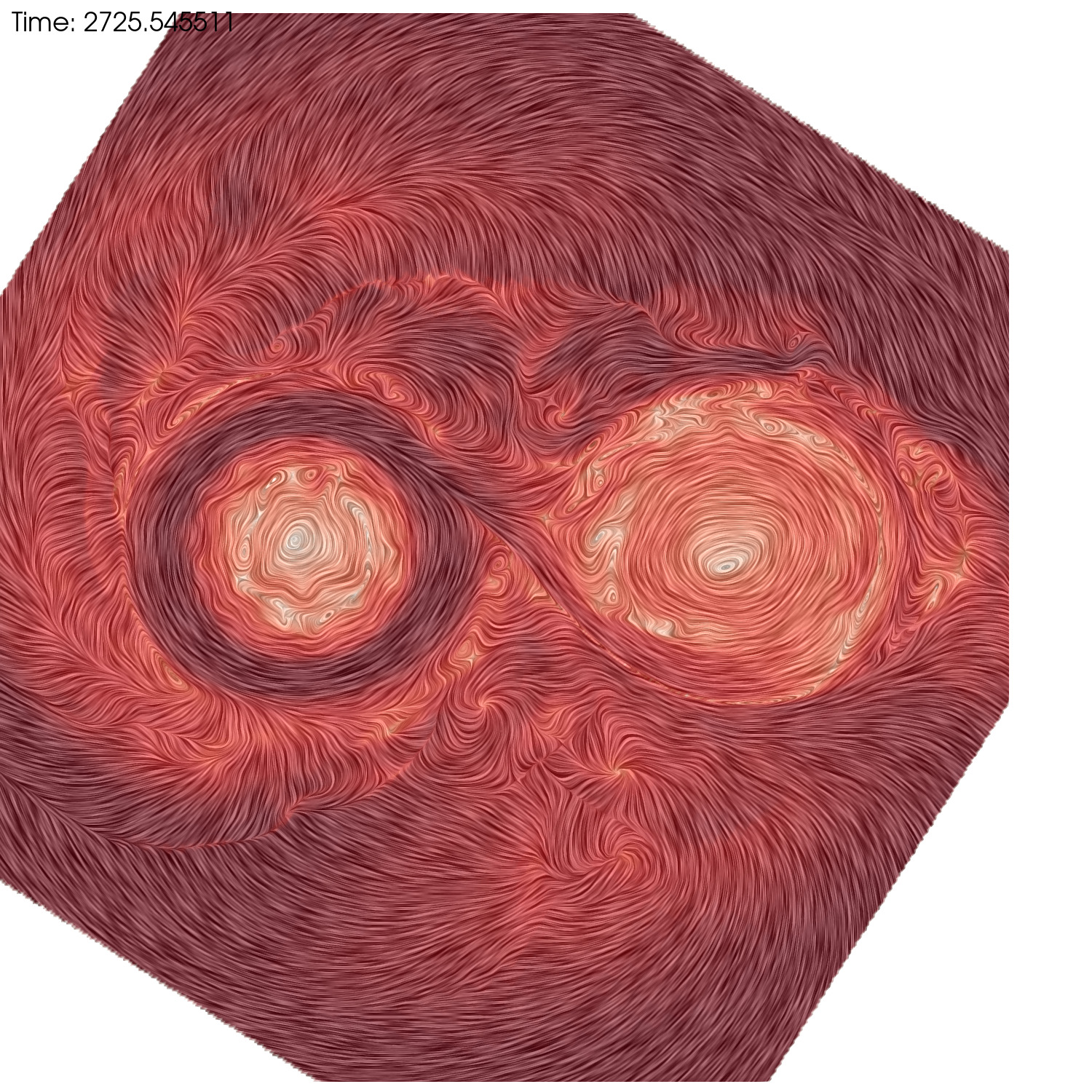}}%
    \caption{%
        Orbital plane and equatorial flows at a representative time $t= 2725.55 \seconds$ during the quasi-steady phase for rcb12.
        Velocity magnitudes are mapped logarithmically to color~(\includegraphics[width=6pt,height=32pt,angle=-90,origin=rB]{figures/legend_reds.jpeg}) in the range $[1 \cdot 10^5,~1 \cdot 10^8] \cmps$.
    }
    \label{fig:quasi-steady1}    
\end{figure*}

The evolution in both simulations goes
through a quasi-steady phase lasting 10-15
orbits during which the flow structure remains the same, with the rate of 
mass-flow in the stream and accretion belt increasing monotonically. As illustrated in \autoref{fig:quasi-steady1}, the behavior of rcb12
is consistent with rcb11:
the fastest flows during this stage are the stream and the accretion belt. A varying number (7-10) eddies develop between the accretion belt and the slower-rotating body of the accretor. The interior of the donor, lagging behind the orbital evolution, and a CW spiraling flow feeds and joins the mass transfer stream.

\section{Discussion and conclusions}
\label{sec:discussion}

In this paper we describe briefly some visualization techniques that have proven useful for the analysis of data pertaining to complex 3D fluid flows in several fields of science and engineering, hoping that they also find application in astrophysical fluid simulations.
In tight collaboration between astrophysicists, simulation scientists, and visualization experts, we apply these techniques to visualize and analyze the interior and surface flows in a double white dwarf binary in a merger process. The simulation datasets we selected are the same as in \citet{Shiber2024}, where the authors investigated DWD mergers as progenitors of R Coronae Borealis stars (RCB), and concluded that some dredge-up of accretor material into the partial helium-burning shell helped to produce the observed range of O$^{16}$/O$^{18}$ ratios. Shear-driven mixing was potentially the cause of the dredge-up. Here, by visualizing the flow in the shear layer between the accretion belt and the body of the accretor, we confirm that expectation (see further details below). We also touch briefly on possible applications to the binary progenitors of other types of astrophysical objects of interest.

\subsection{Advantages of the visualization methods}
\label{sec:advantages}

Two of the current state-of-the-art flow visualization techniques described in
\autoref{sec:vis} helped us to gain a refined understanding of the interior and the surface flows in a double white dwarf binary in a merger process: the Line-Integral Convolution and the Lagrangian Coherent Structures.

For example, when attempting to visualize the 2D flow on a surface or slice using velocity vectors, it is usually necessary to limit the number of arrows or glyphs displayed to a select subset of the cells to prevent overlap and visual clutter. 
In visualization, approaches that employ glyphs such as arrows are accordingly classified as \text{sparse} as they only show flow at select positions.
A more fine-grained \textit{dense} visualization of the flow is obtained by calculating surface LIC (\autoref{sec:vis-isolic}), which makes use of all velocity values on the grid, albeit at the cost of omitting directionality information. Then one must correlate the LIC flows with maps of velocity glyphs or selectively drop glyphs onto the LIC flow where needed.

Another example of the advantages of the techniques employed here is the use of 
LCS, a robust approach to generate trajectory patterns and to yield a skeleton depicting the major dynamics of the system. In this work, we used it to identify and classify the most important physically distinct regions of the flow and their evolution during disruption and merger.
For example, with this technique one can locate and visualize the properties of the boundary between internal donor flow and mass-transfer stream, the strong shear layer between the accretion belt and the body of the accretor, how the stream is fed, and by which layers, etc.


\subsection{Astrophysical implications}
\label{sec:astrophysical}

The binary parameters, initial configurations, and physical processes included in the simulations visualized in this paper were designed to investigate the origin of RCB stars and to match their observational properties \citep{Shiber2024}. Although not optimized for applications to other types of close binaries, it is nevertheless possible to draw some tentative conclusions about processes thought to occur in other interacting binaries by linking them to specific stages of the evolutions described here.
However, one must remain aware of the limitations of this approach, remembering that our simulations do not include a realistic equation of state, radiation transport, cooling, nor energy generation by nuclear burning, and that the timescales of the mergers are likely influenced by interaction with the low-density ``vacuum" fluid, numerical diffusion, and dissipation.

\emph{RCB progenitors.}
In the late stages of evolution, right up to the tidal disruption of the donor, the binary orbital period is shorter than the internal rotation of both component stars. In the case of the accretor, most of the angular momentum transferred is in the equatorial belt, while the core is still rotating slower. 

The internal flow of both components, viewed from the frame of reference corotating CCW with the orbit, will therefore appear to counter-rotate CW. 
As \autoref{fig:quasi-steady1-rcb12-645-lic} shows, a ring-like corotation boundary separates the CW interior from the CCW outer layers.
In the equatorial slice of the accretor, the faster CCW flow of the accretion belt is separated from the CW interior by 7-10 CCW eddies. The strong shearing in this layer contributes to the dredging up of accretor core material into the shell of fire where partial helium burning will eventually cause the expansion of the merger object into a luminous red giant with the spectroscopic properties and distinct isotopic ratios of the RCB stars \citep{Shiber2024}. 

\emph{AM Canum Venaticorum Binaries.} These are hydrogen-poor short-period binaries with $5\, {\rm min} \leq P_{\rm orb}< 1\, {\rm hr}$ \citep{Solheim2010,Ramsay2018,Chakrabortyetal2024}, with more than a dozen being LISA verification sources \citep{Amaro-Seoane2023}. Two of these, HM Cnc ($P_{\rm orb} =5.36 \minutes$) and V407 Vul ($P_{\rm orb} =9.49 \minutes$) are thought to be direct impact accretors. A third binary ES Ceti ($P_{\rm orb} = 10.33 \minutes$) \citep{WarnerWoudt2002} may be an intermediate case in which the accretion belt has developed into a small accretion disk \citep{Bakowska2021}. The recent discovery of three ultra-compact binaries by
\citet{Chakrabortyetal2024} has added two disk accretors with orbital periods below 10 min. In Section 4.2 of their paper they set firm limits on the mass ratio based on avoiding direct impact. As in the case of ES Ceti,
even if grazing direct impact occurs, given enough time, especially if the accretor has been spun up, the belt may grow radially and develop a disk due to viscosity.
This alters both the gravitational radiation and electromagnetic signals and the orbital evolution due to the re-arrangement of mass and angular momentum.
The evolution of ultra-close binaries (UCBs) is expected to be driven by a combination of gravitational radiation and astrophysical fluid effects acting in concert. For LISA verification sources for which we have distance and mass information, \emph{e.g.} from Gaia or devoted follow-up campaigns (for example \citet{Roelofs2010, Burdge2020, Brown2020}), it may be possible to disentangle astrophysical effects from purely gravitational ones, thus contributing to elucidate both \citep{Demarchi2023}. 
For a more careful investigation of mass and angular momentum flows in these UCBs, numerical hydrodynamic simulations of DWD with mass ratios $q\approx 0.2$ or less, and a more careful treatment of the donor non-degenerate atmosphere are necessary.

\emph{Cataclysmic Variables.}
In cataclysmic variables (CVs) a near-main-sequence star transfers mass by Roche lobe overflow onto a white dwarf. The orbital periods of these secularly stable binaries are much longer than in the case studied here; and the rates of mass transfer are orders of magnitude lower. It is currently not possible to simulate numerically such low rates of mass transfer which are typical of cataclysmic variables. Nevertheless, one is tempted to speculate about which aspects of the model studied here, if any, are applicable to CVs.  
For example, in our case, material with the chemical composition of the donor can move by a combination of advection and diffusion to higher latitudes in a relatively small number of orbits.
However, as already mentioned, the mass-transfer rates in our simulations are orders of magnitude above realistic values. In CVs the donor material gradually arrives at the accretor through a boundary layer and accretion belt. The spreading of this material has been investigated analytically by \citet{Piro2004} who found that higher accretion rates result in larger spread in latitude. It would be desirable to tackle this question numerically without the simplifying assumptions required for the analytic treatment. Unfortunately, we are unable to simulate the realistic quasi-steady flow in between nova explosions with our current tools. We conjecture that high shear would make the accretion belt and boundary layer prone to instabilities. In that case, there should be ample time for the donor material to cover the entire WD surface with hydrogen-rich gas before the next nova explosion occurs. 

\emph{Short-Period Algol Binaries.}
These binaries consist of an evolved low-mass companion transferring mass to a main-sequence primary \citep{Richards1992}. When $P_{\rm orb}\leq 5-6\, {\rm d}$, \emph{e.g.}, in Algol itself, U CrB and RS Vul, the transfer stream impacts the primary and in some cases a partial disk is present and in others there is a transition between a disk and no-disk. \citet{Blondin1995} carried out a 2D hydrodynamic study concluding that the transition was probably due to variations in the mass transfer rate. The most recent investigation using 3D hydrodynamic
simulations \citep{Raymer2012} considered magnetic effects that tilted the stream and formed a warped disk with significant emission from out-of-plane structures. Our simulations, while not suited to these systems, suggest that the accretion belt could easily develop into a tidally truncated disk filling much of the primary's Roche lobe.

\emph{Type Ia Supernovae.} One of the evolutionary channels leading to type Ia supernovae is unstable mass transfer in a DWD whose total mass is near or exceeds the Chandrasekhar limit. Direct impact accretion is likely to occur if the mass ratio is not too small. The mass transfer flow prior to detonation is very similar to the situation discussed in this paper, even if the masses and mass ratios are different. See \citet{Munday2025} for the recent discovery of a super-Chandrasekhar DWD,
and the details of the simulations that predict a double detonation in 23 Gyr.

\section*{Acknowledgments}
\label{sec:acknow}
The numerical work was carried out using the computational resources (QueenBee2) of the Louisiana Optical Network Initiative (LONI) and Louisiana State University’s High Performance Computing (LSU HPC).
Juhan Frank thanks the Lorraine and Leon August Endowment to LSU for support.
This work was partially funded by Deutsche Forschungsgemeinschaft (DFG, German Research Foundation) under Germany’s Excellence Strategy in EXC 2075 ‘‘SimTech’’ (390740016), International Research Training Group GRK 2160 ‘‘DROPIT’’ (270852890), and Collaborative Research Centre SFB 1313 (327154368).
This work was supported by the U.S. Department of Energy through the Los Alamos National Laboratory. Los Alamos National Laboratory is operated by Triad National Security, LLC, for the National Nuclear Security Administration of U.S. Department of Energy (Contract No. 89233218CNA000001). LA-UR-25-25740. We thank the referee for comments that improved the presentation of our results.

\section{Data Availability}

\octo is available on GitHub\footnote{\url{ https://github.com/STEllAR-GROUP/octotiger}}, and was built using the following build chain\footnote{\url{https://github.com/STEllAR-GROUP/OctoTigerBuildChain}}.
The visualization used ParaView, based on the Visualization Toolkit~(VTK).
With its modular approach, available filters can be combined with custom modules programmed in C\texttt{++}.
For this work, some filters in the Two-Phase Flow~(TPF) framework\footnote{\url{https://github.com/UniStuttgart-VISUS/tpf}}, and general ParaView plugins\footnote{\url{https://github.com/UniStuttgart-VISUS/ParaView-Plugins}} were developed and used. These are already publicly available under the permissive MIT license.
Other visualizations are provided directly by ParaView, or by the Visual Computing Group at Heidelberg University\footnote{\url{https://vcg.iwr.uni-heidelberg.de/plugins}}.

\clearpage
\bibliographystyle{mnras}
\bibliography{bibliography}
\bsp
\label{lastpage}

\end{document}